\documentclass{article}

\usepackage{amsmath}
\usepackage{enumitem}

\usepackage{arxiv}

\usepackage[utf8]{inputenc} 
\usepackage[T1]{fontenc}    
\usepackage{hyperref}       
\usepackage{url}            
\usepackage{booktabs}       
\usepackage{amsfonts}       
\usepackage{nicefrac}       
\usepackage{microtype}      
\usepackage{cleveref}       
\usepackage{lipsum}         
\usepackage{graphicx}
\usepackage{doi}
\usepackage{subfig}

\title{Enforcing asymptotic behavior with DNNs for approximation and regression in finance}

\newif\ifuniqueAffiliation
\uniqueAffiliationtrue

\ifuniqueAffiliation 
\author{Hardik Routray \\
	Wells Fargo Bank, N.A. \\
	\texttt{hardik.routray@wellsfargo.com} \\
	\And
	Bernhard Hientzsch\\
	Wells Fargo Bank, N.A.\\
	\texttt{bernhard.hientzsch@wellsfargo.com} \\
}
\else
\usepackage{authblk}

\newbox{\orcid}\sbox{\orcid}{\includegraphics[scale=0.06]{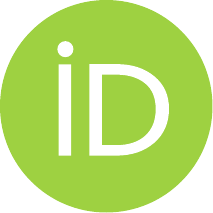}} 
\author[1]{%
	\href{https://orcid.org/0000-0000-0000-0000}{\usebox{\orcid}\hspace{1mm}David S.~Hippocampus\thanks{\texttt{hippo@cs.cranberry-lemon.edu}}}%
}
\author[1,2]{%
	\href{https://orcid.org/0000-0000-0000-0000}{\usebox{\orcid}\hspace{1mm}Elias D.~Striatum\thanks{\texttt{stariate@ee.mount-sheikh.edu}}}%
}
\affil[1]{Department of Computer Science, Cranberry-Lemon University, Pittsburgh, PA 15213}
\affil[2]{Department of Electrical Engineering, Mount-Sheikh University, Santa Narimana, Levand}
\fi

\renewcommand{\headeright}{} 
\renewcommand{\undertitle}{} 
\renewcommand{\shorttitle}{\textit{DNNs with asymptotic forms in finance}}

\hypersetup{
pdftitle={Enforcing asymptotic behavior with DNNs for approximation and regression in finance},
pdfsubject={q-fin.Cp, q-fin.MF,math.NA},
pdfauthor={Hardik.~Routray, Bernhard.~Hientzsch},
pdfkeywords={Differential Machine Learning, Black-Scholes},
}

\begin{document}
\maketitle

\begin{abstract}
    We propose a simple methodology to approximate functions with given asymptotic behavior by specifically constructed 
    terms and an unconstrained deep neural network (DNN). 
    The methodology we describe extends to various asymptotic behaviors and multiple dimensions
    and is easy to implement. 
    In this work we demonstrate it for linear asymptotic behavior in one-dimensional examples.
    We apply it to function approximation and regression problems where we measure approximation of 
    only function values (``Vanilla Machine Learning''-VML) or also approximation of function 
    and derivative values (``Differential Machine Learning''-DML) on several examples. 
    We see that enforcing given asymptotic behavior leads to better approximation and faster convergence. 

\end{abstract}

\keywords{Asymptotic behavior and forms \and Differential Machine Learning \and Function Approximation \and Regression \and Approximation of conditional expectations} 

\section{Introduction}

Often solutions to scientific problems and financial problems have known asymptotic behavior. In particular pricing functions
in quantitative finance often have or are assumed to have linear asymptotes \cite[Section 2 in Appendix 4]{huge2020differential}.
If one represents pricing functions with a deep neural network (DNN) with activation functions with certain properties, these pricing functions 
will be asymptotically linear, but training the asymptotic part implicitly as part of the DNN presents challenges and 
requires special treatment \cite[Section 2 in Appendix 4]{huge2020differential}. Often, unless specifically addressed in
the training and setup, convergence of the DNN to the true asymptotic behavior can be slow or not even assured. 
Thus, one is interested in ways to use DNN to represent solutions while ensuring a given asymptotic form. 

\cite{antonov2020neural,antonov2020deeprisk} capture asymptotic behavior by a spline interpolator while in the interior, the solution 
is approximated by a linear combination of Gaussian kernels. The centers and widths of these kernels are given by 
artificial neural networks (\cite{antonov2020neural} calls this construction 
``Constrained Radial Layer'' or ``Gaussian Radial Layer'').
They are constrained such that the resulting linear combination has zero asymptotes and is numerically zero
in the asymptotic region beyond a certain distance. 

However, a simpler construction without a special layer and without constraints might be preferred. 

One such construction is the following: one represents the solution inside the nonasymptotic domain as the 
sum of a function which matches the asymptotic behaviors and of a product of a polynomial term that enforces
zero asymptotes (``{\em zasymptotic}'') and a DNN: 
\begin{equation}
f(x) = asymptotic(x) + DNN(x) * zasymptotic(x).
\end{equation}
Outside the domain, $f(x)$ is just $asymptotic(x)$ and the DNN only influences the solution within the 
domain. 

Here, we test this construction in one dimension for function approximation and regression, with two kinds of 
loss functions, one only involving value approximation (VML - vanilla machine learning) and the other 
one involving both value and derivative approximation (DML - differential machine learning). 

We will show on examples that adding asymptotic behavior with fixed or trainable coefficients improves the
approximation and convergence for both function approximation and regression.

The paper is organized as follows - we first introduce the approach and setting(s) in section \ref{sec:setup},
then consider some first simple function approximation in section \ref{sec:firstapprox} to demonstrate the methodology.
In section \ref{sec:bsfuncapprox}, we progress to approximating a Black-Scholes function with its asymptotic behavior. 
In section \ref{sec:bsfunccondex}, we learn a Black-Scholes function with asymptotic form from payoff and potentially
payoff derivative samples generated by simulation. We conclude in section \ref{sec:conclusion}.

\section{Approach and Set-Up}
\label{sec:setup}

We are trying to learn a function $f$ from function values (and possibly additional derivative values),  or to learn a conditional expectation function $f$ 
with $f(x)=E(Y|X=x)$ from samples of the target random variate $Y$ and the conditioning $X$ (and possibly sample-wise derivatives of the target random variate with 
respect to the conditioning). In addition, through observation, limiting behavior, or asymptotic methods, we know the asymptotic behavior or form of $f$,
either with free parameters or with already given fixed parameters. Standard deep learning approaches with generic DNN do not take advantage of that form 
and thus, standard DNN approximations resulting from such approaches do not satisfy or approximate this asymptotic behavior well. 

In general, this availability of asymptotic forms splits the domain of $f$ into an asymptotic domain where the asymptotic form is exact or a good approximation 
of the function to be sought and the non-asymptotic domain where the function needs to be learned otherwise. One possible approach is to extend, if possible, 
the asymptotic form(s) from the asymptotic domain to a function defined on the entire domain and to only learn the difference to the asymptotic form. 
For general domains, finding such extensions is not trivial and might require both mathematical and computational effort. 
Often, asymptotic domains are given by the complement of tensor products of intervals (``boxes'') or by the complement of a hypersphere. 
For the box case, spline interpolation (in particular univariate or multivariate cubic interpolation) is a natural starting point. 

In symbols, $f(x) = extension(asymptotic)(x) +nonasymptotic(x)$. $nonasymptotic(x)$ has to be such that it is zero in the asymptotic domain 
and smoothly blends into zero so as to not affect the asymptotic form. We will denote the extension of the asymptotic form by $asymptotic(x)$ as 
well and thus write:
\begin{equation}
    f(x) = \begin{cases} 
        asymptotic(x) & \mbox{in asymptotic domain} \\
        asymptotic(x)+DNN(x) \times zasymptotic(x) & \mbox{elsewhere}
     \end{cases}
\end{equation}
where $zasymptotic(x)$ ensures that its product with $DNN(x)$ does smoothly paste into the asymptotic form in the asymptotic domain. 

It is sometimes the case that the derivatives of the function to be approximated are given or can be approximated well. In such cases, approaches that 
take derivatives into account might be of interest, in particular if the improvement in efficiency from using derivatives outweighs the effort to 
compute or approximate the derivatives needed, or where the derivatives of the function are to be approximated also. Similarly, for the regression 
case, the process that generates the samples of the target variate and gives the values of the conditioning might allow one to efficiently 
generate sample-wise derivatives as well, making sample-wise derivatives easily accessible for regression as well. In these cases, 
the extension of the asymptotic form and the approximation in the nonasymptotic domain have to respect the derivatives as well and 
have to lead to a approximation where derivatives match as well. 

Here, we consider the one-dimensional case.
Then, 
\begin{equation}
f(x) = \begin{cases} 
    asymptotic(x) & x \leq LL \\
    asymptotic(x)+DNN(x) \times zasymptotic(x)  & LL\leq x\leq UL \\
    asymptotic(x)  & UL \leq x 
 \end{cases} \label{eq:fform}
\end{equation}

We will concentrate on the case of linear asymptotic forms.
\begin{equation}
asymptotic(x) = \begin{cases} 
    LS(x - LL) + LI & x \leq LL \\
    US(x - UL) + UI & UL \leq x 
\end{cases}.
\end{equation}
We use cubic interpolation to extend this to a continuous function with continuous
derivatives:
\begin{equation}
    asymptotic(x) = \begin{cases} 
        LS(x - LL) + LI & x \leq LL \\   
    (x-LL)(UL-x) (a_0 (x-LL) + \tilde{LS}) + (s_0(x-LL) + LI) & LL\leq x\leq UL \\
        US(x - UL) + UI & UL \leq x 
    \end{cases},\label{eq:asymptotic}
\end{equation}
with 
$s_0 := \frac{UI-LI}{UL-LL}$, $\tilde{LS} := \frac {LS-s_0}{UL-LL}$,
$\tilde{US} := \frac{s_0-US}{UL-LL}$, and $a_0 := \frac{\tilde{US}-\tilde{LS}}{UL-LL}$.
(\cite{antonov2020neural} present a more general version, but our form is enough for our purposes.)
For alternative asymptotic forms, we would use these forms below $LL$ and above $UL$,
and set $LS$ and $LI$ respective $US$ and $UI$ to the value and derivative of these forms at $LL$ and $UL$ and otherwise proceed in the 
same way. 
\cite{antonov2020neural} present an extension by multivariate splines for the multivariate box case 
and we could use their construction or alternative constructions for multivariate extensions of our 
approach as well. 

\cite{antonov2020neural} ensure smooth pasting to the asymptotic forms by representing 
$nonasymptotic(x)$ as a linear combination of Gaussian kernels, with a more involved domain decomposition
(asymptotic domain, "no-mans land", and inner domain). Here, we will use simpler forms.

For the $zasymptotic(x)$, we use a scaled polynomial:
\begin{equation}
zasymptotic(x) = \begin{cases} 
    0 & x \leq LL \\
    (x-UL)^2(x-LL)^2(\frac{1}{LL \cdot UL}) & LL\leq x\leq UL \\
    0 & UL \leq x 
 \end{cases}.\label{eq:zasymptotic}
\end{equation}
Appropriate scaling is important for normalization and easier learning, in particular when several functions and/or 
sets of parameters are to be learned simultaneously. 

The parameter are as follows: $UL$ - upper level or limit (beyond which the asymptotic form is used),
with $US$ and $UI$ being the upper slope and intercept of the linear asymptotic
form or the derivative and value of the asymptotic form in general, 
$LL$ - lower level or limit (beyond which the asymptotic form is used), 
with $LS$ and $LI$ being the lower slope and intercept of the linear asymptotic
form or the derivative and value of the asymptotic form in general. 

We consider applications in function approximation and in regression problems. 
For function approximation based on only function values (VML - vanilla machine learning), we are
given samples $x_i$ and $y_i$ and want to learn a function $f$ such that $y_i=f(x_i)$. We are using the VML loss function 
\begin{equation}
vmlloss = \sum_i \left( y_i - f(x_i)\right)^2 
\end{equation}
and are looking for a function $f$ that minimizes that loss function. In standard deep learning, we would represent the entire function $f$
by a deep neural network DNN, whereas for given asymptotic forms, we use the form of $f$ proposed above in (\ref{eq:fform}) with 
(\ref{eq:asymptotic}) and (\ref{eq:zasymptotic}) to enforce the given asymptotic behavior. 
In the function approximation case, the VML loss function would have a minimum of zero. 

For function approximation based on function and derivative values (DML - differential machine learning), we are 
given samples  $x_i$, $y_i$, $\frac{dy_i}{dx}$ and want to learn a function $f$ such that $y_i=f(x_i)$ and
$\frac{dy_i}{dx}=f'(x_i)$. We are using the DML loss function 
\begin{equation}
dmlloss = \sum_i \left( y_i - f(x_i)\right)^2 + \lambda  \sum_i \left( \frac{dy_i}{dx}- f'(x_i)\right)^2
\end{equation}
and are looking for a function $f$ that minimizes that loss function.\footnote{$\lambda$ 
allows the weighting of fitting of derivatives relative to fitting of function values as well as addressing 
different scales of function values and derivatives. In this paper and on our examples, we assume $\lambda=1$,
but other values of $\lambda$ might be appropriate for other settings. Our methods allow for general $\lambda$.} 
In standard deep learning, we would represent the entire function $f$
by a deep neural network DNN, whereas for given asymptotic forms, 
we use the form of $f$ proposed above in (\ref{eq:fform}) with 
(\ref{eq:asymptotic}) and (\ref{eq:zasymptotic}) to enforce the given asymptotic behavior. 
In the function approximation case, the DML loss function would have a minimum of zero, but will be larger than the VML
loss function since it measures both approximation of values and derivatives. 

For the regression setting where we are trying to approximate conditional expectations from samples, we are given 
sample $x_i \sim X$ and $y_i \sim Y$, and we are looking to learn a function that is equals to or approximates the conditional
expectation function $f(x) = E(Y|X=x)$. We are using the same VML loss function as for function approximation. The minimum loss now in general
will be larger than zero since in general there will be irreducible randomness, $Y$ is still random even conditioned on $X$. 

In many situations, $Y$ can be represented by a function of $X$ and some additional randomness (independent of $X$) which 
we will call $\Omega$, $Y=h(X,\Omega)$. For instance, $X$ could be random initial values of an SDE or random inputs
to some function. The function $h$ is often either explicitly given or given as some program or procedure. 
Appropriate derivatives of the function can then be computed by appropriate algorithmic differentiation of 
the program. We are given samples $x_i \sim X$, $y_i=h(x_i,\omega_i)$, and can efficiently compute 
or approximate sample-wise derivatives $\frac{dy_i}{dx}= \frac{dh}{dx}(x_i,\omega_i)$. We can then use DML and
the DML loss function as above, but the minimum loss will be not zero and larger than the VML loss since 
we also are approximating the conditional expectations of the derivatives. The approach also covers the
case where $X$ are intermediate values in some computation that are generated from some fixed initial values
and additional randomness $\Omega^X$ as in $X=g(\Omega^X)$.

For testing examples where we know the function or the conditional expectation to be approximated, 
we can compute the error by comparing against that function and approximate the minimum loss
by evaluating that loss function against generated samples. 
For the regression and conditional expectation setting, \cite{cheridito2021computation,cheridito2023computation}
offers ways to compute estimates and guarantees. 
 
In DML without asymptotic forms, standard deep learning frameworks implement computation of derivatives
by back propagation which is a special form of adjoint algorithm differentiation a.k.a. reverse 
mode so that $f'$ for the DNN function $f$ is easily and efficiently available. 
Including asymptotic forms, the $f$ is now defined piece-wise and somewhat more 
complicated. It can be differentiated either through the framework or explicitly 
as 
\begin{equation}
    f'(x) = \begin{cases} 
        asymptotic'(x) & x \leq LL \\
        asymptotic'(x)+DNN'(x) \times zasymptotic(x) + DNN(x) \times zasymptotic'(x) & LL\leq x\leq UL \\
        asymptotic'(x)  & UL \leq x 
     \end{cases}
\end{equation}
where $DNN'(x)$ is given by adjoint algorithmic differentiation or back propagation.

\section{A first function approximation example}
\label{sec:firstapprox}

Here, we study approximation of an example function 
\begin{equation}
    f(x) = \begin{cases} 
        asymptotic(x) & x \leq LL \\
        asymptotic(x)+x \times zasymptotic(x)  & LL\leq x\leq UL \\
        asymptotic(x)  & UL \leq x 
     \end{cases}
\end{equation}
with $LL=-5$, $LI=0$, $LS=50$, $UL=5$, $UI=0$, and $US=50$.

\begin{figure}%
    \centering
    \subfloat[\centering Example function]{{\includegraphics[width=7cm]{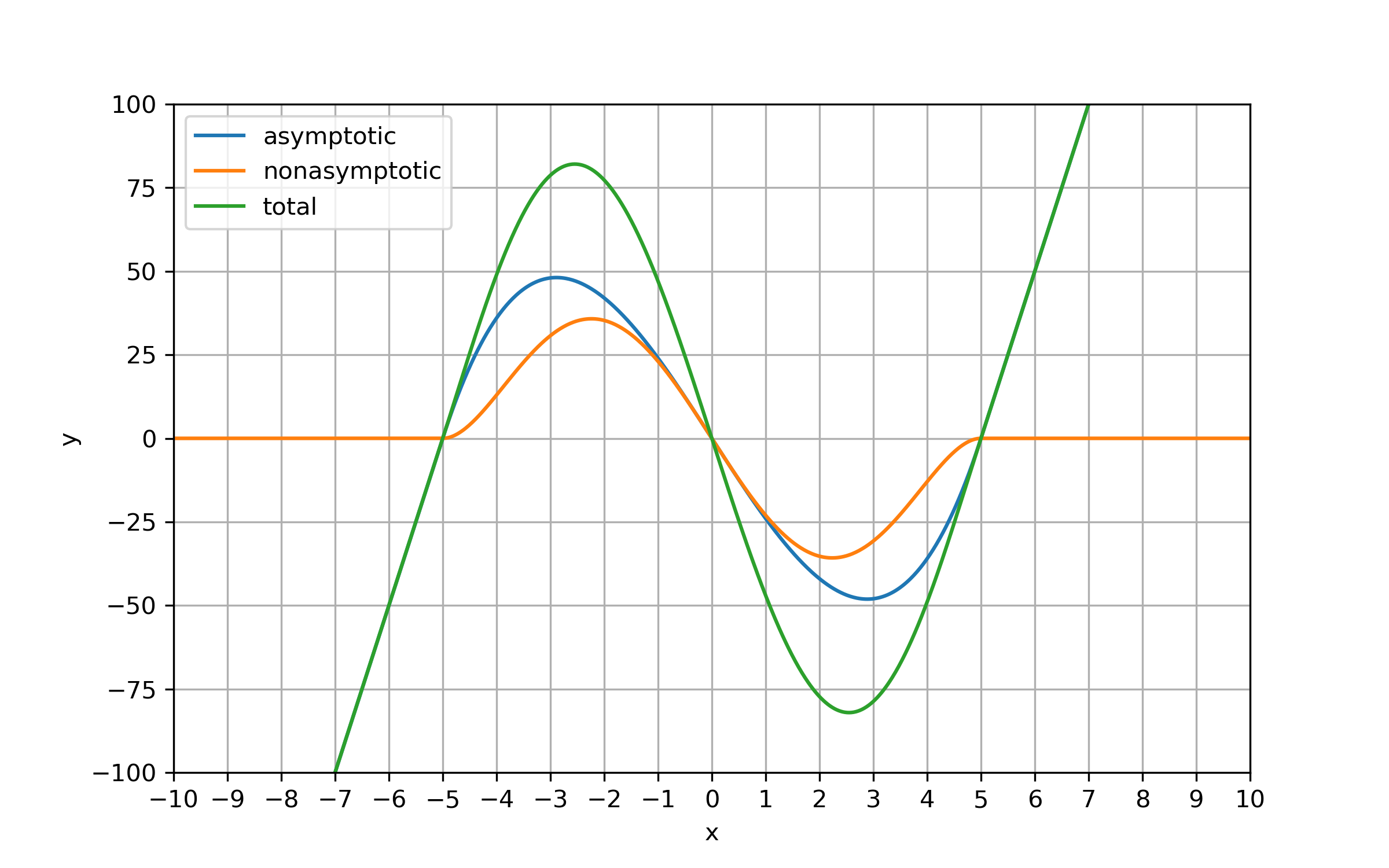} }}%
    \qquad
    \subfloat[\centering Derivative of example function]{{\includegraphics[width=7cm]{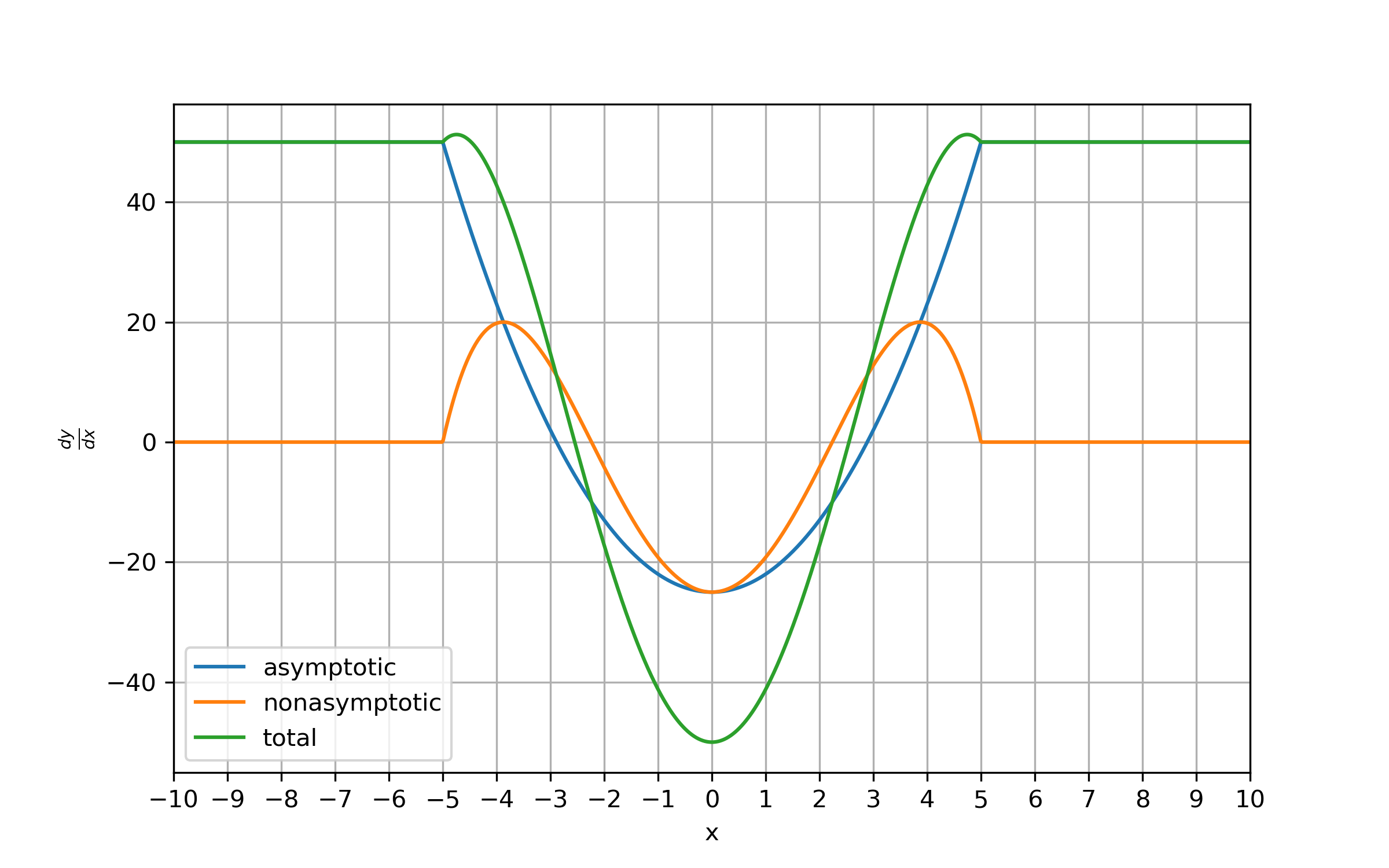} }}%
    \caption{First function approximation: The example function with associated derivative and its two components - asymptotic and non-asymptotic part.}%
    \label{fig:func_total}%
\end{figure}

\begin{figure}%
    \centering
    \subfloat[\centering Non-asymptotic component of example function]{{\includegraphics[width=7cm]{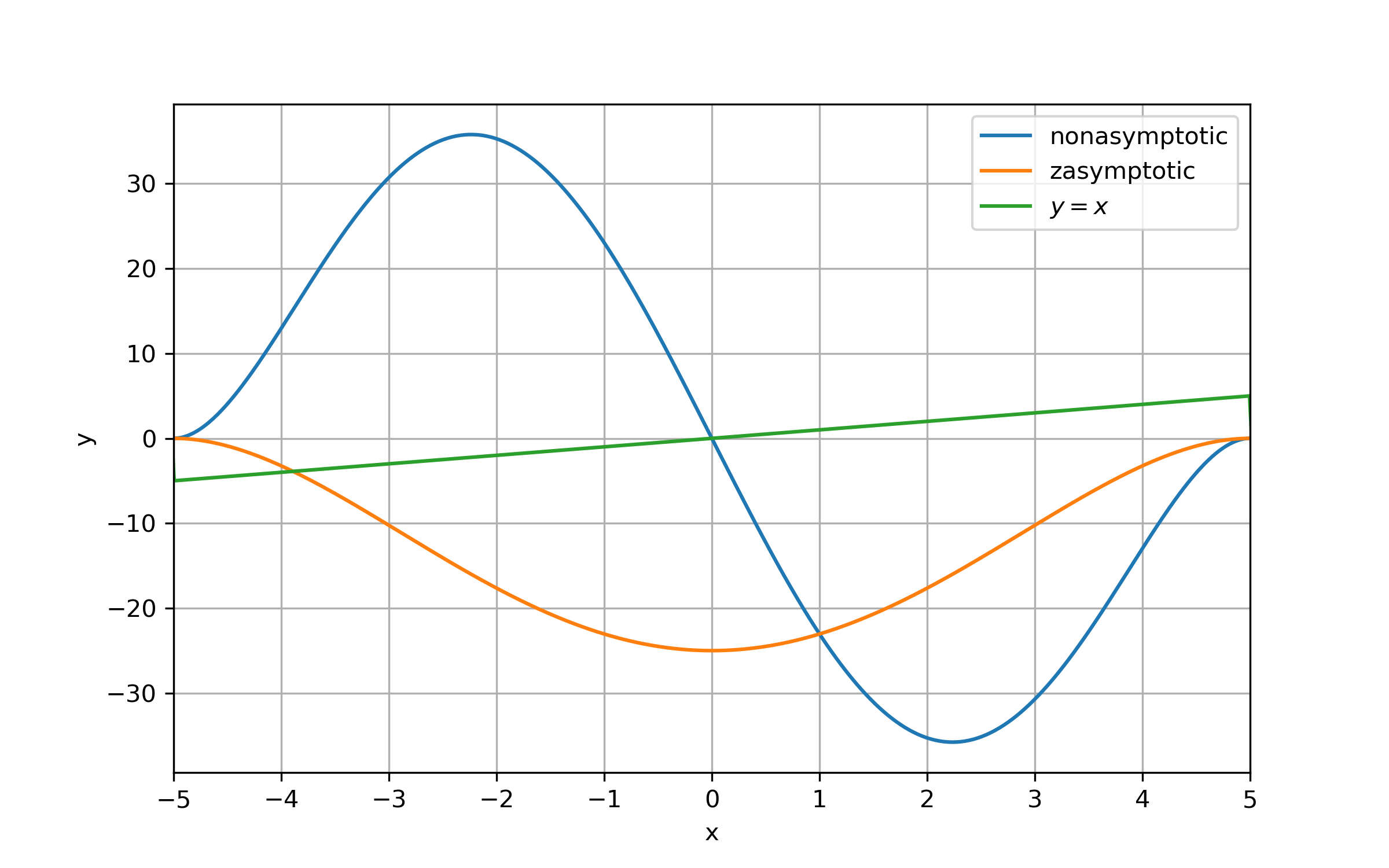} }}%
    \qquad
    \subfloat[\centering Derivative of non-asymptotic component of example function]{{\includegraphics[width=7cm]{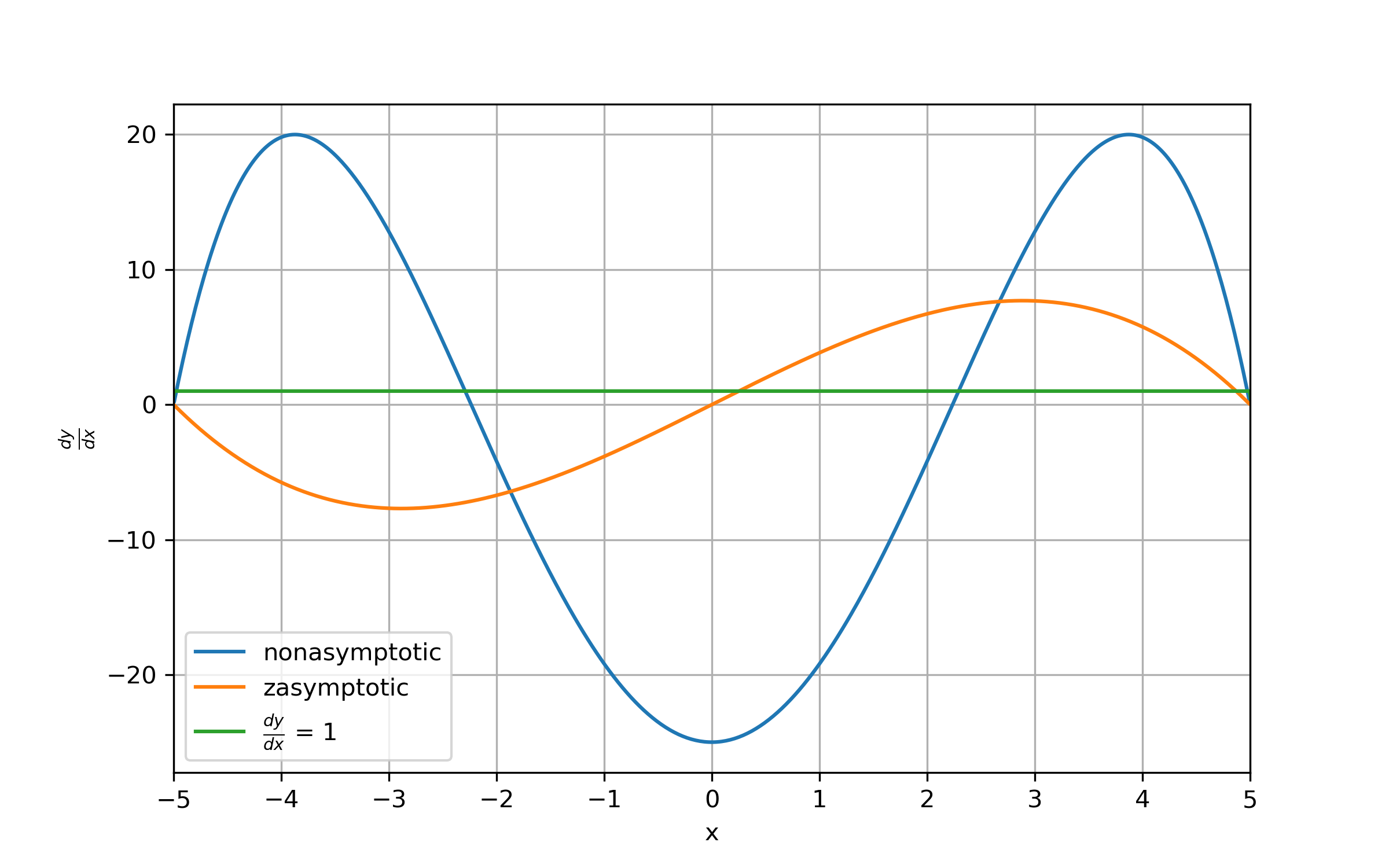} }}%
    \caption{First function approximation: The non-asymptotic component of the example function with associated derivative and its two components - NN and non-NN ({\em zasymptotic})}%
    \label{fig:func_nonasymp}%
\end{figure}

 Fig.~\ref{fig:func_total} shows the example function with associated derivative and its components 
 while Fig.~\ref{fig:func_nonasymp} shows the non-asymptotic component of the example function with associated 
 derivative and its components.

Now that we have defined the function, we generate input data for training using 50,000 randomly selected $x$  
from a uniform distribution between -10 and 10 along with associated true function and derivative values $y$ and $\frac{dy}{dx}$. 
In a first approach, we learn the parameters in the asymptotic form through stochastic gradient descent or Adam \cite{kingma2014adam}
at the same time the parameters of the neural network DNN are learned 
(``trainable parameters''). In a second approach, we use fixed given values for the parameters (``fixed parameters''). 
We tested using the true values of the asymptotic parameters (here, LS, LI, US, and UI) 
used to generate the samples as well as estimating the asymptotic parameters (LS, LI, US, and UI) through least squares
in the asymptotic regions separately, possibly constraining the intercept to ensure good fitting of function values at LL and UL. 
We found that for our examples, the difference between using true values 
and using values determined with least square estimation was negligible. Thus, we present the results 
with parameters determined by least square estimation in the asymptotic regions for all our examples,
since this would be also applicable if true parameters are not known.

The hyperparameters for the neural network used for training on normalized inputs are: input dimension = 1, output dimension = 1, hidden layers with nodes = [20, 20], 
activation function = softplus, optimizer = Adam, epochs = 200, and learning rate = 0.05. \\

\begin{figure}%
    \centering
    \subfloat[\centering Trained DNN along with true value]{{\includegraphics[width=7cm]{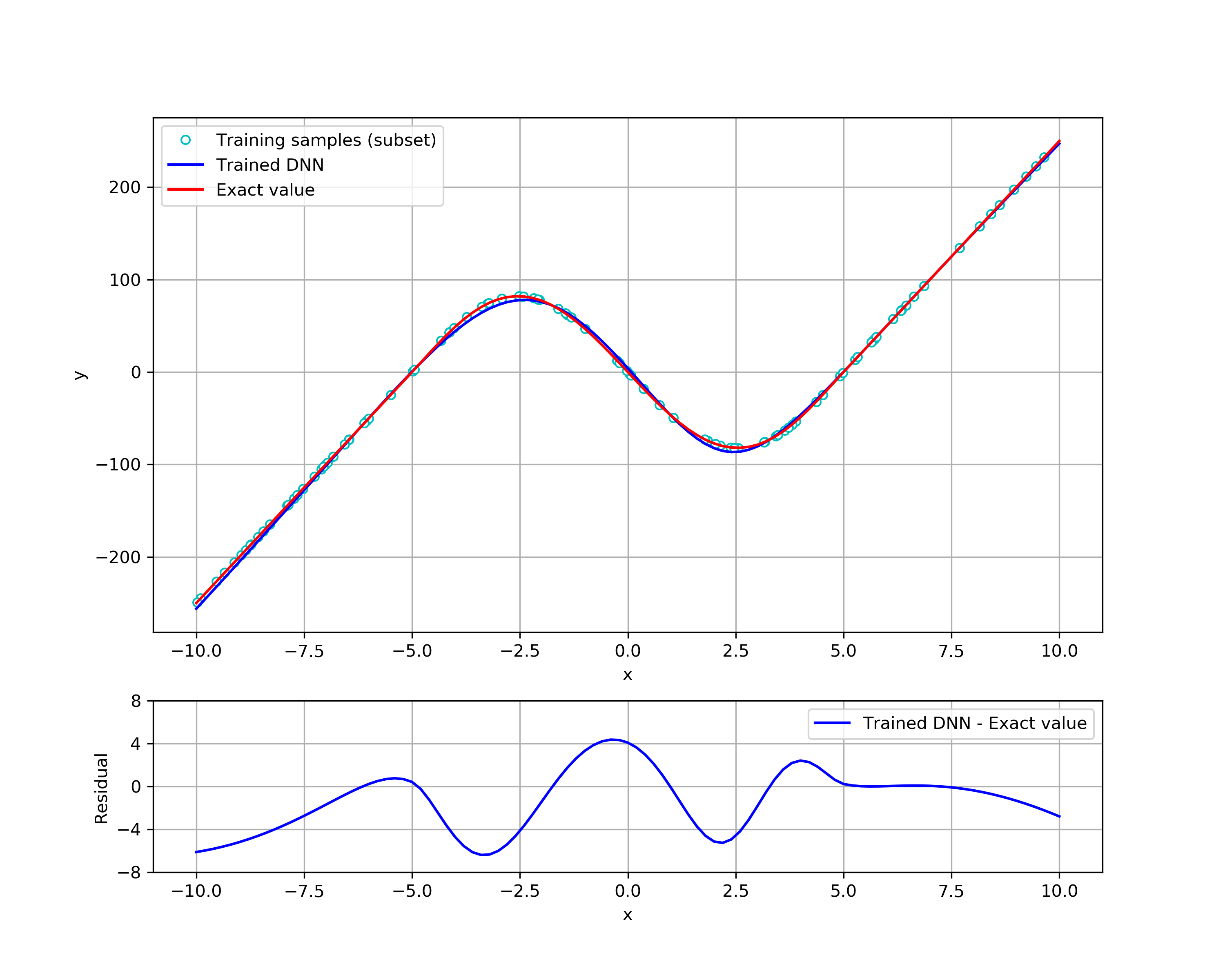} }}%
    \qquad
    \subfloat[\centering Gradient of trained DNN along with true  derivative]{{\includegraphics[width=7cm]{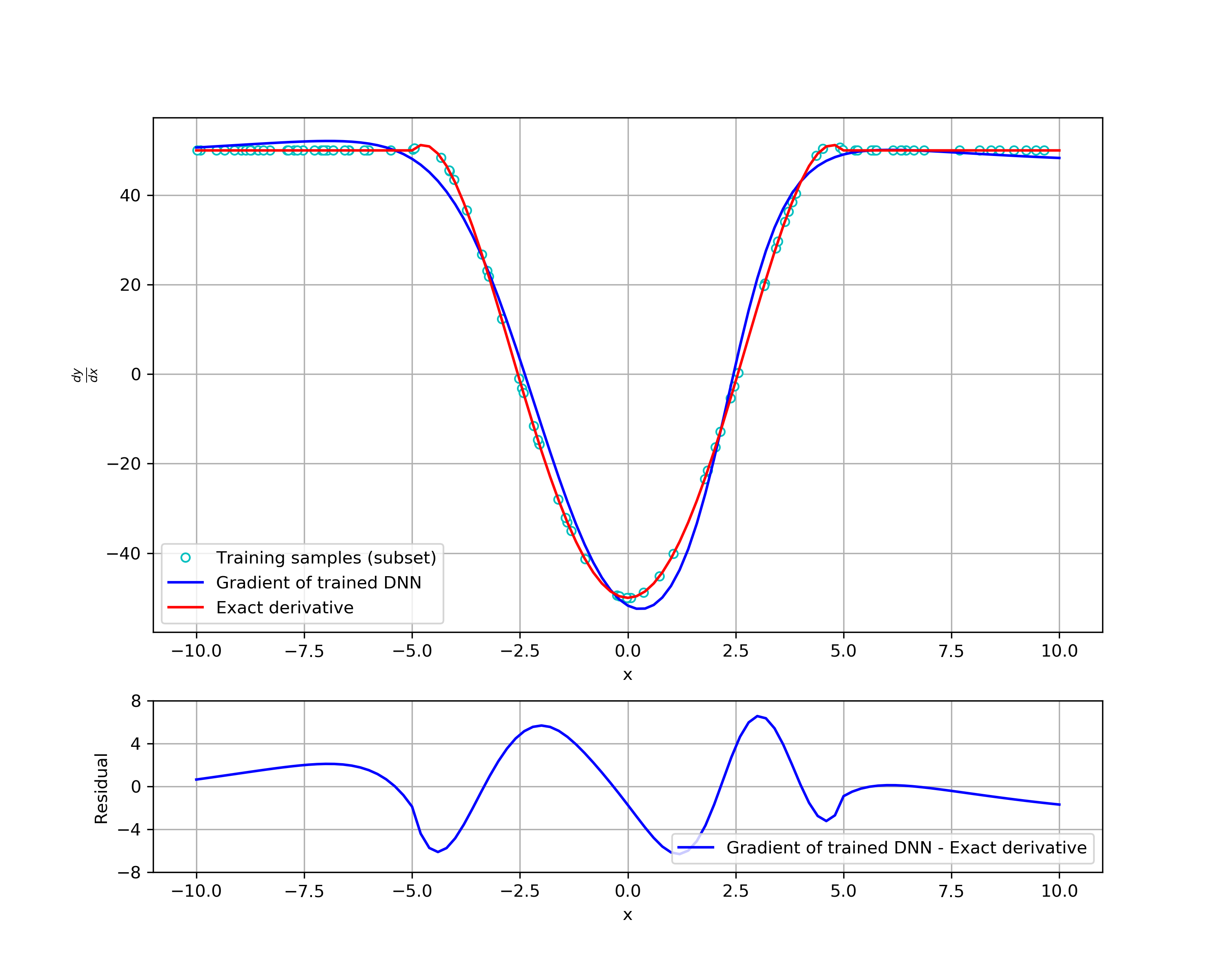} }}%
    \caption{First function approximation: Vanilla Machine Learning without asymptotic treatment where results from DNN are in solid blue, 
    true values are in solid red, and the blue scatter points represent a subset of samples used in training chosen randomly for visualization purposes only.}%
    \label{fig:results_vml}%
\end{figure}

\begin{figure}%
    \centering
    \subfloat[\centering Trained DNN along with true  value]{{\includegraphics[width=7cm]{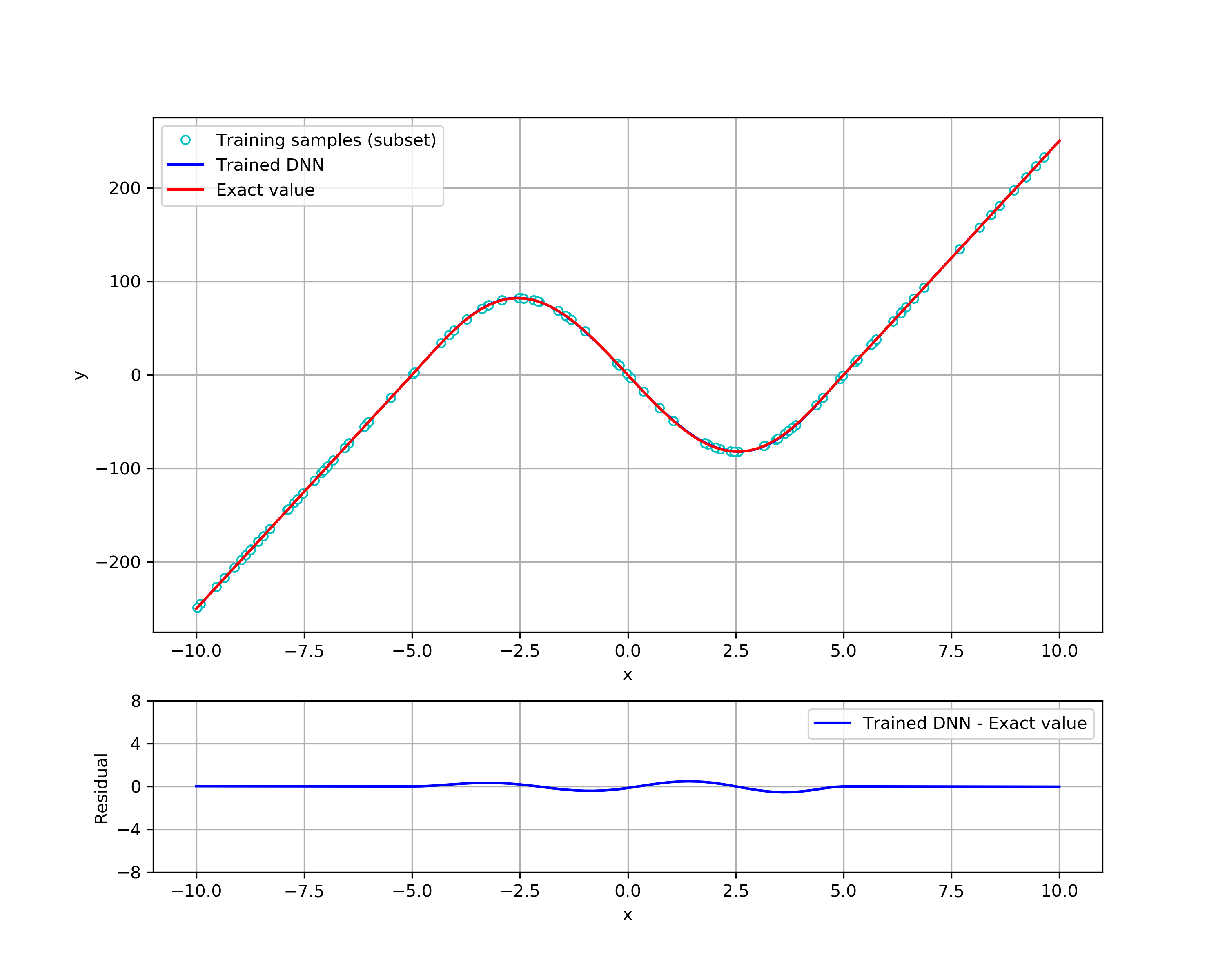} }}%
    \qquad
    \subfloat[\centering Gradient of trained DNN along with true  derivative]{{\includegraphics[width=7cm]{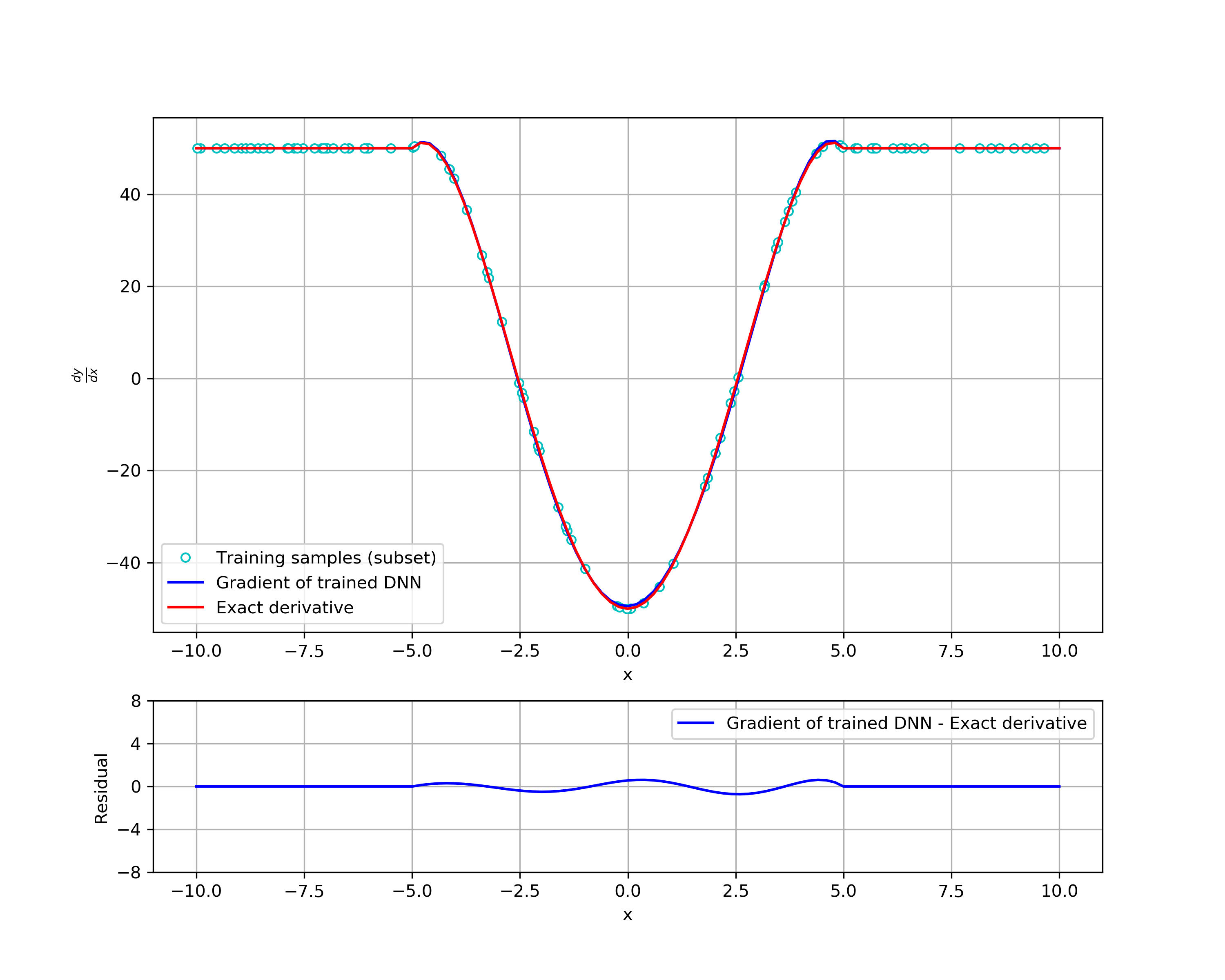} }}%
    \caption{First function approximation:  Vanilla Machine Learning with asymptotic treatment (trainable parameters) where results from DNN are in solid blue, 
    true values are in solid red, and the blue scatter points represent a subset of samples used in training chosen randomly for visualization purposes only.}%
    \label{fig:results_vml_trainasymp}%
\end{figure}

\begin{figure}%
    \centering
    \subfloat[\centering Trained DNN along with true  value]{{\includegraphics[width=7cm]{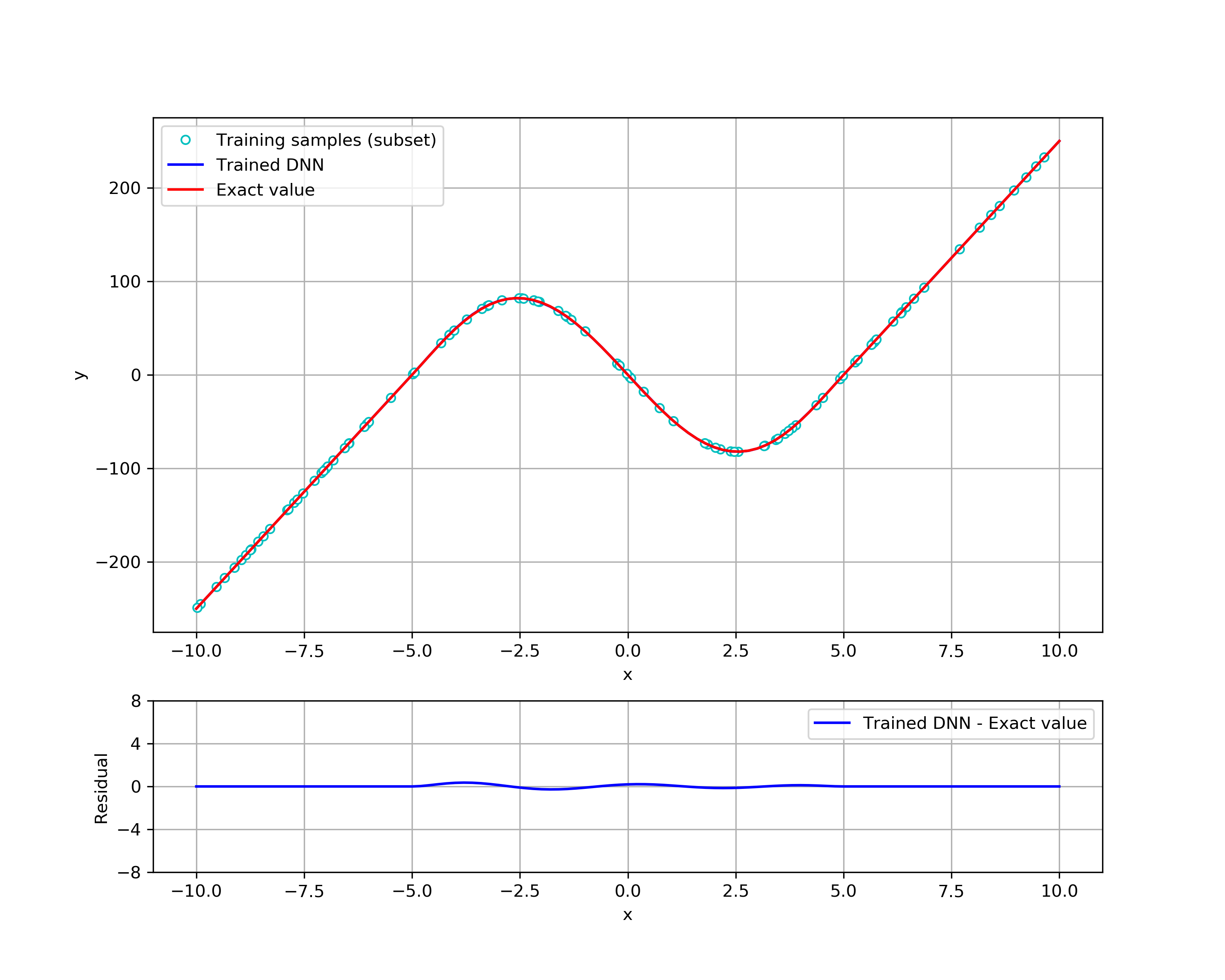} }}%
    \qquad
    \subfloat[\centering Gradient of trained DNN along with true  derivative]{{\includegraphics[width=7cm]{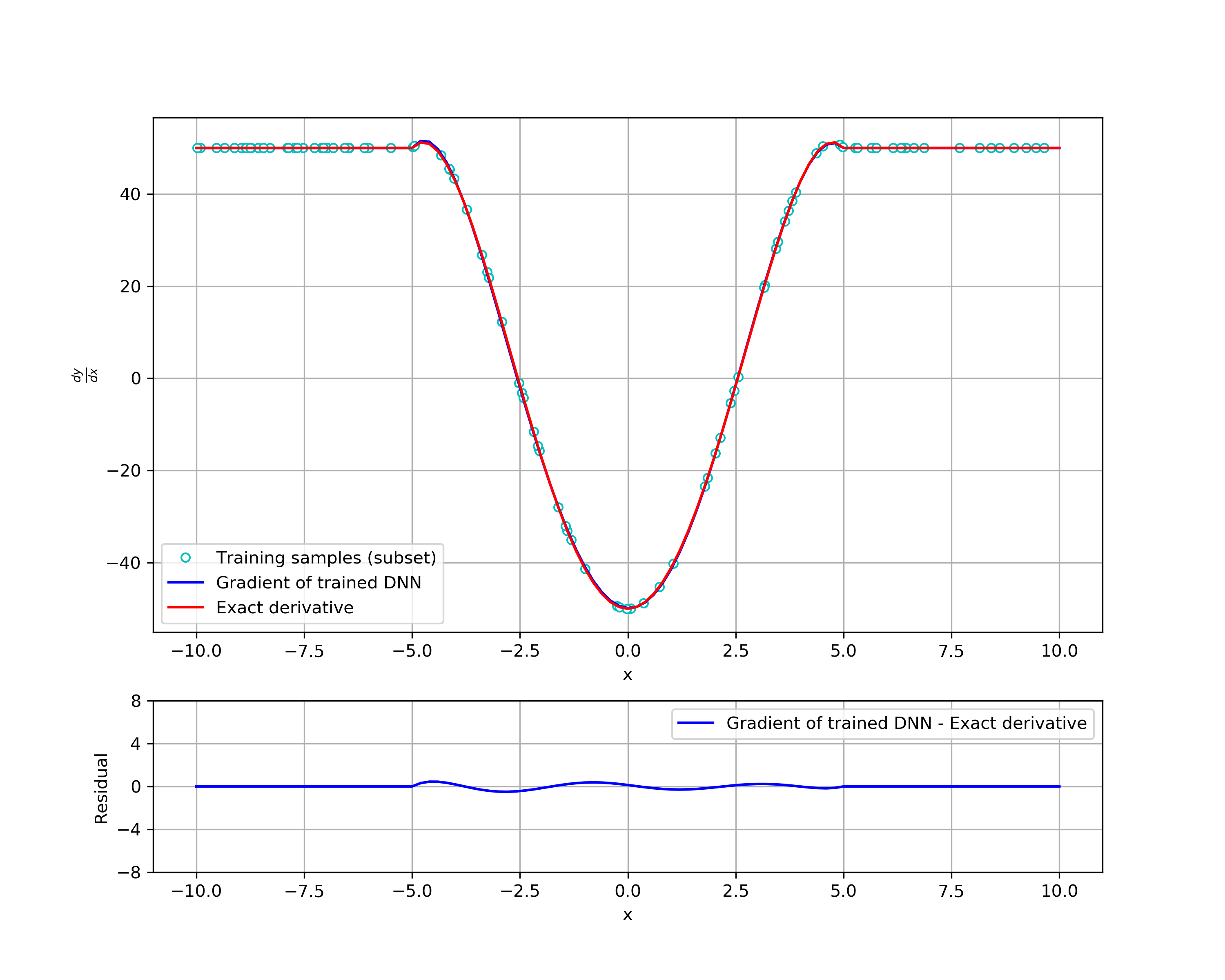} }}%
    \caption{First function approximation:  Vanilla Machine Learning with asymptotic treatment (fixed parameters) where results from DNN are in solid blue, 
    true values are in solid red, and the blue scatter points represent a subset of samples used in training chosen randomly for visualization purposes only.}%
    \label{fig:results_vml_fixedasymp}%
\end{figure}

\begin{figure}%
    \centering
    \subfloat[\centering Difference to true  value]{{\includegraphics[width=7cm]{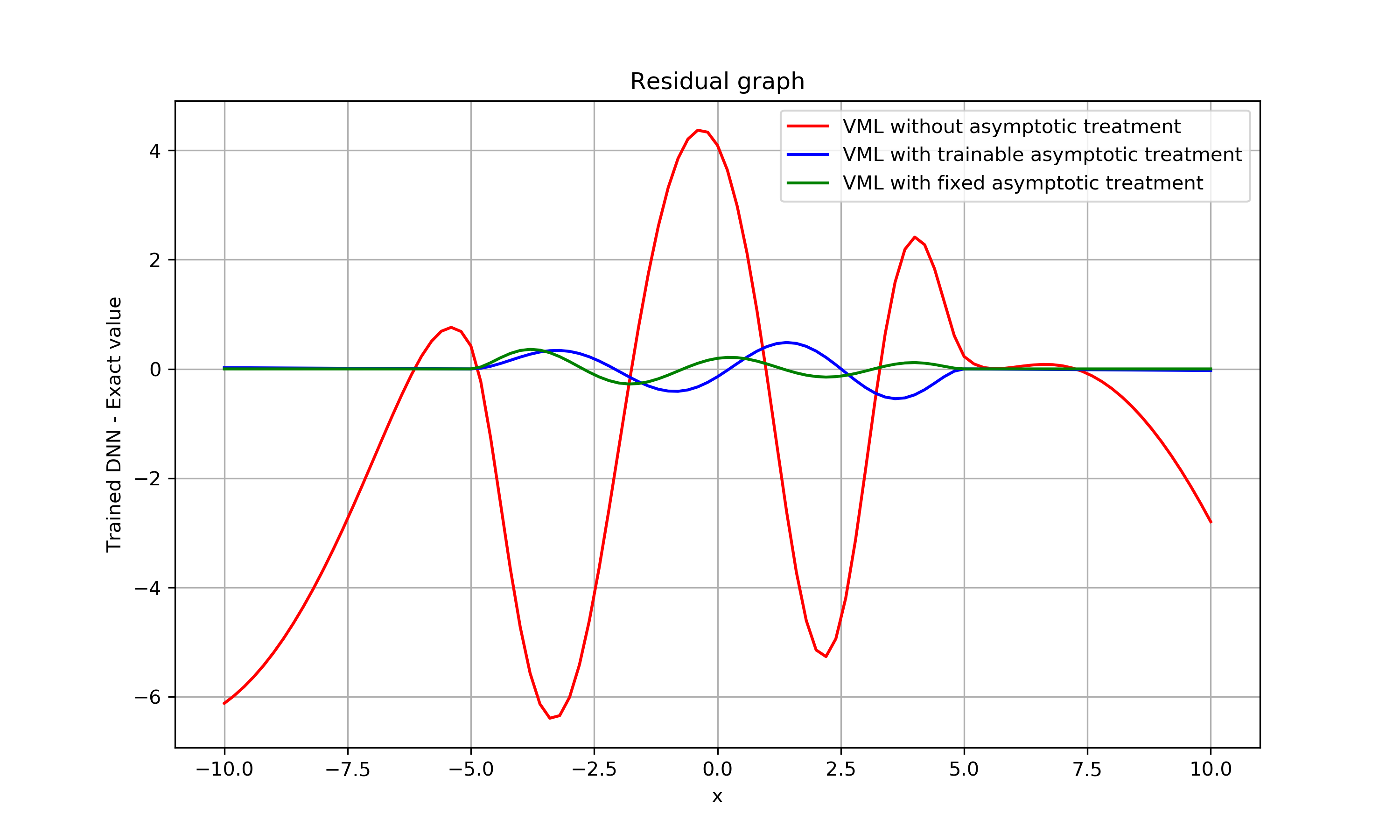} }}%
    \qquad
    \subfloat[\centering Difference to true derivative]{{\includegraphics[width=7cm]{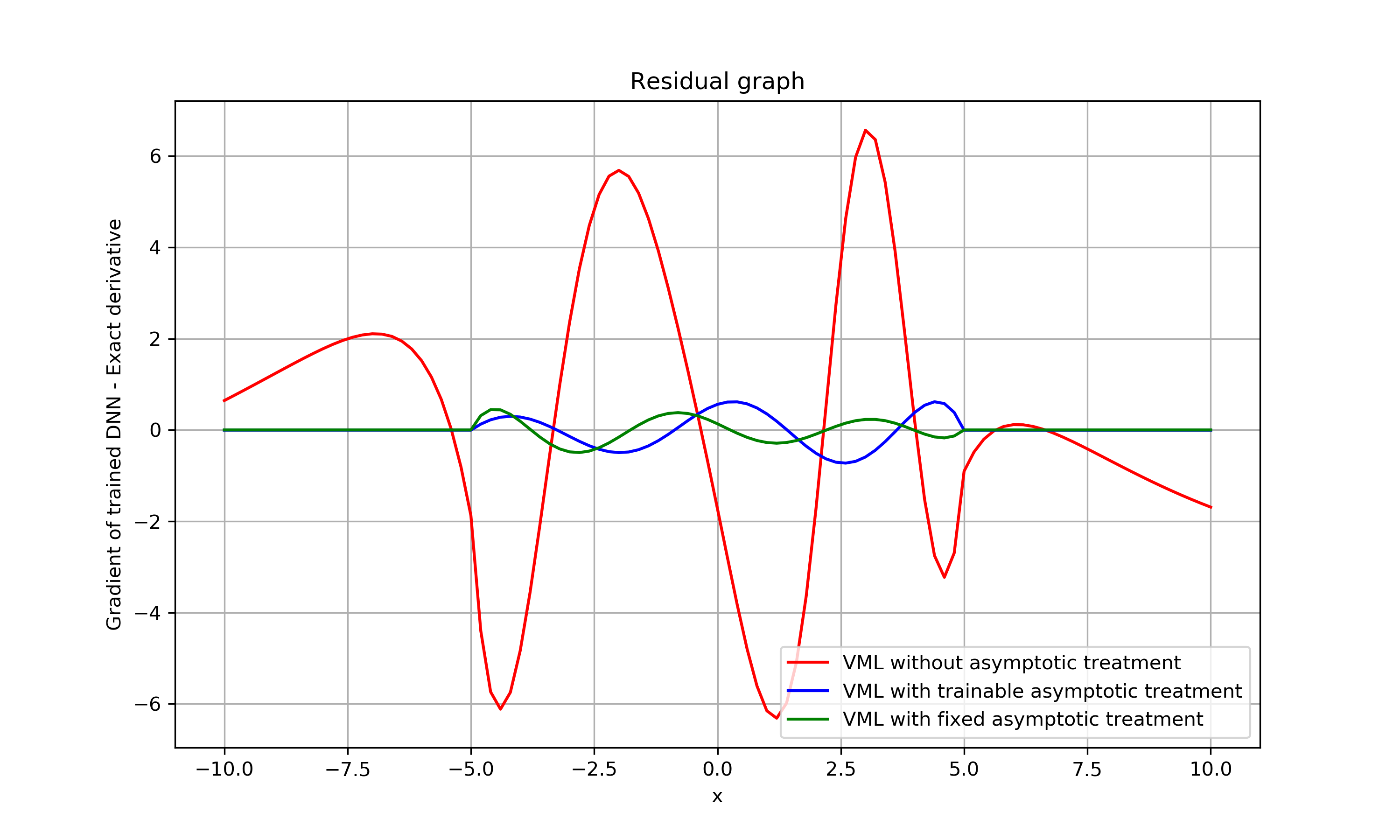} }}%
    \qquad
     \subfloat[\centering Training loss graph (VML loss)]{{\includegraphics[width=7cm]{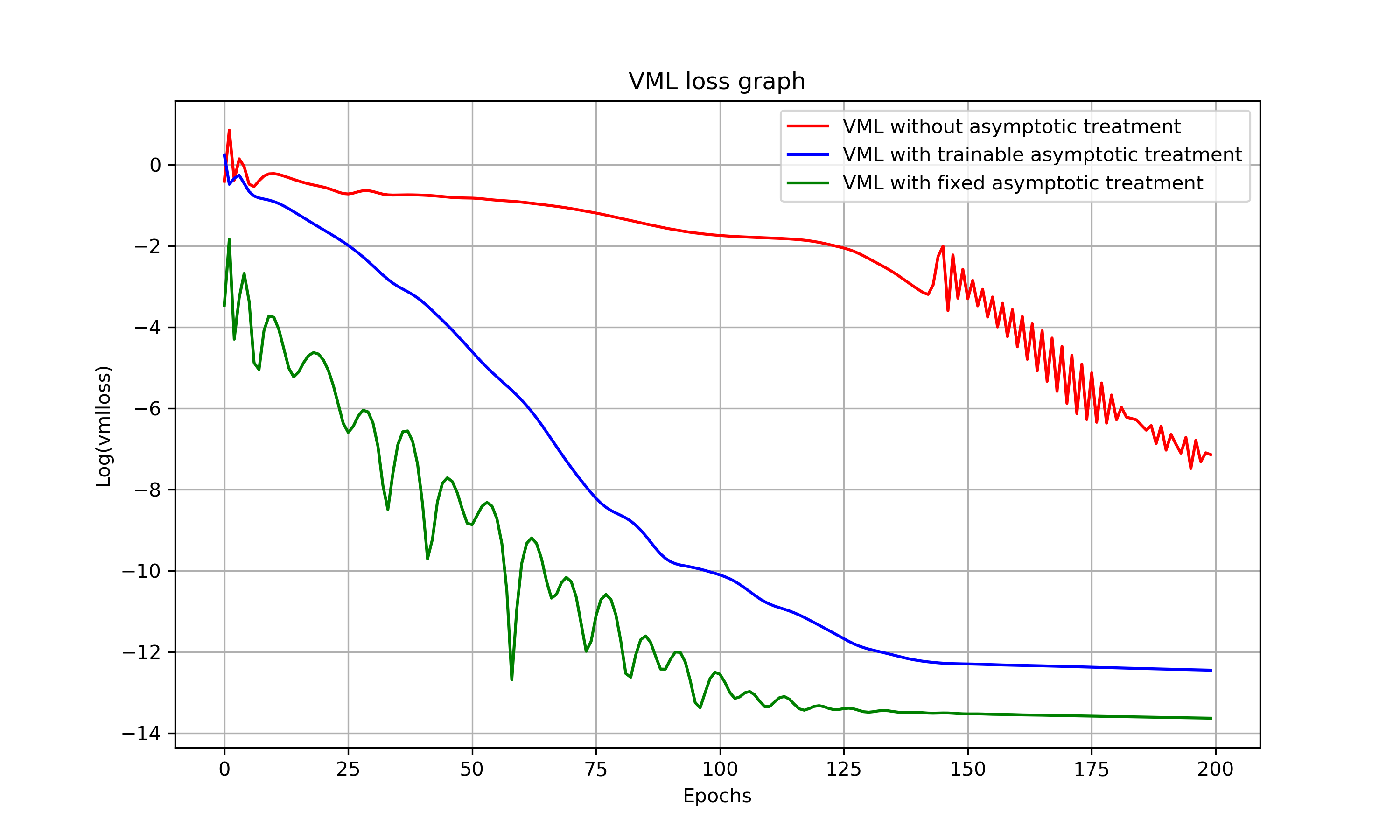} }}%
    \caption{First function approximation: Difference and loss graphs for VML without and with asymptotic treatment}%
    \label{fig:lossgraphs_vml}%
\end{figure}

\begin{figure}%
    \centering
    \subfloat[\centering Trained DNN along with true value]{{\includegraphics[width=7cm]{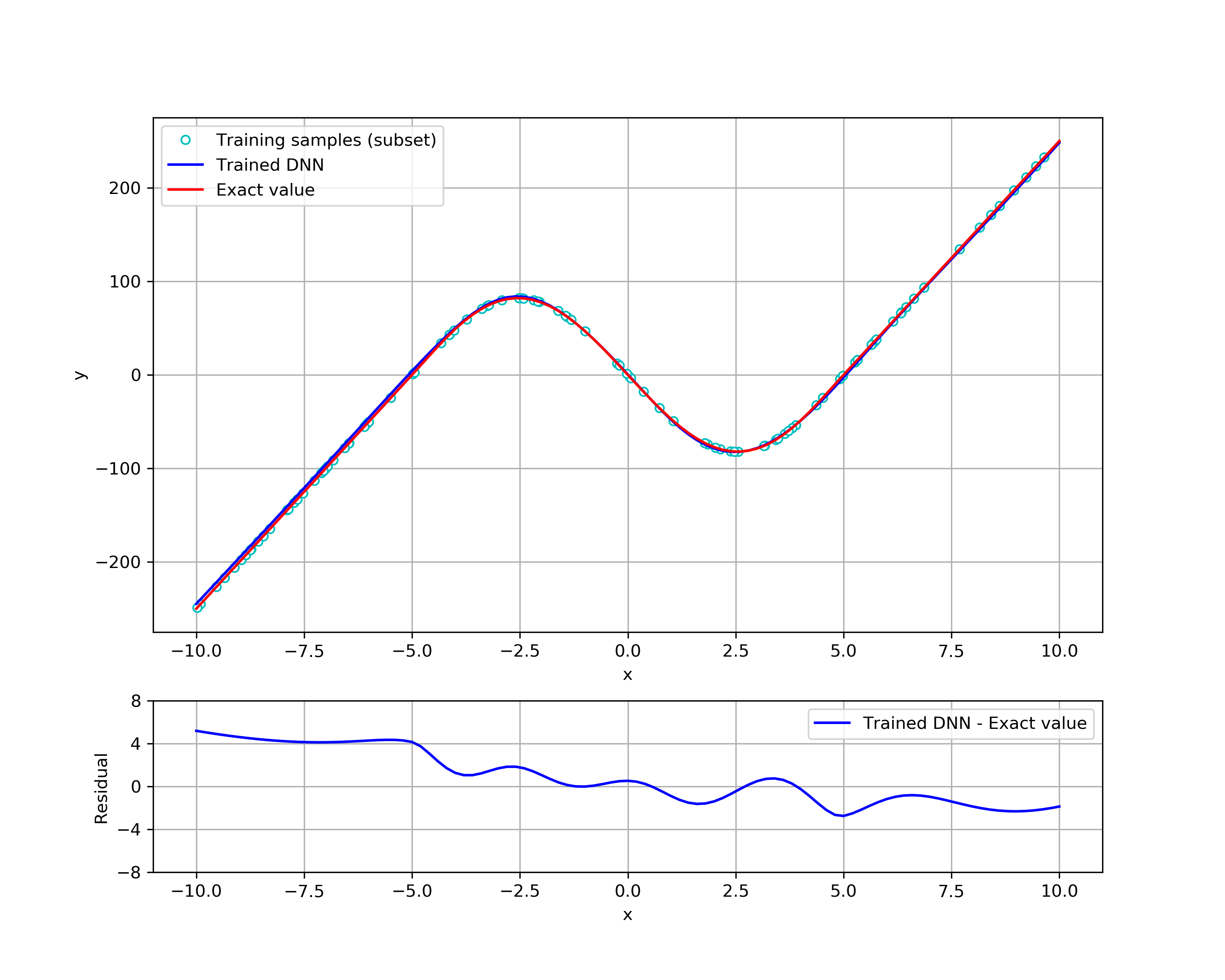} }}%
    \qquad
    \subfloat[\centering Gradient of trained DNN along with true derivative]{{\includegraphics[width=7cm]{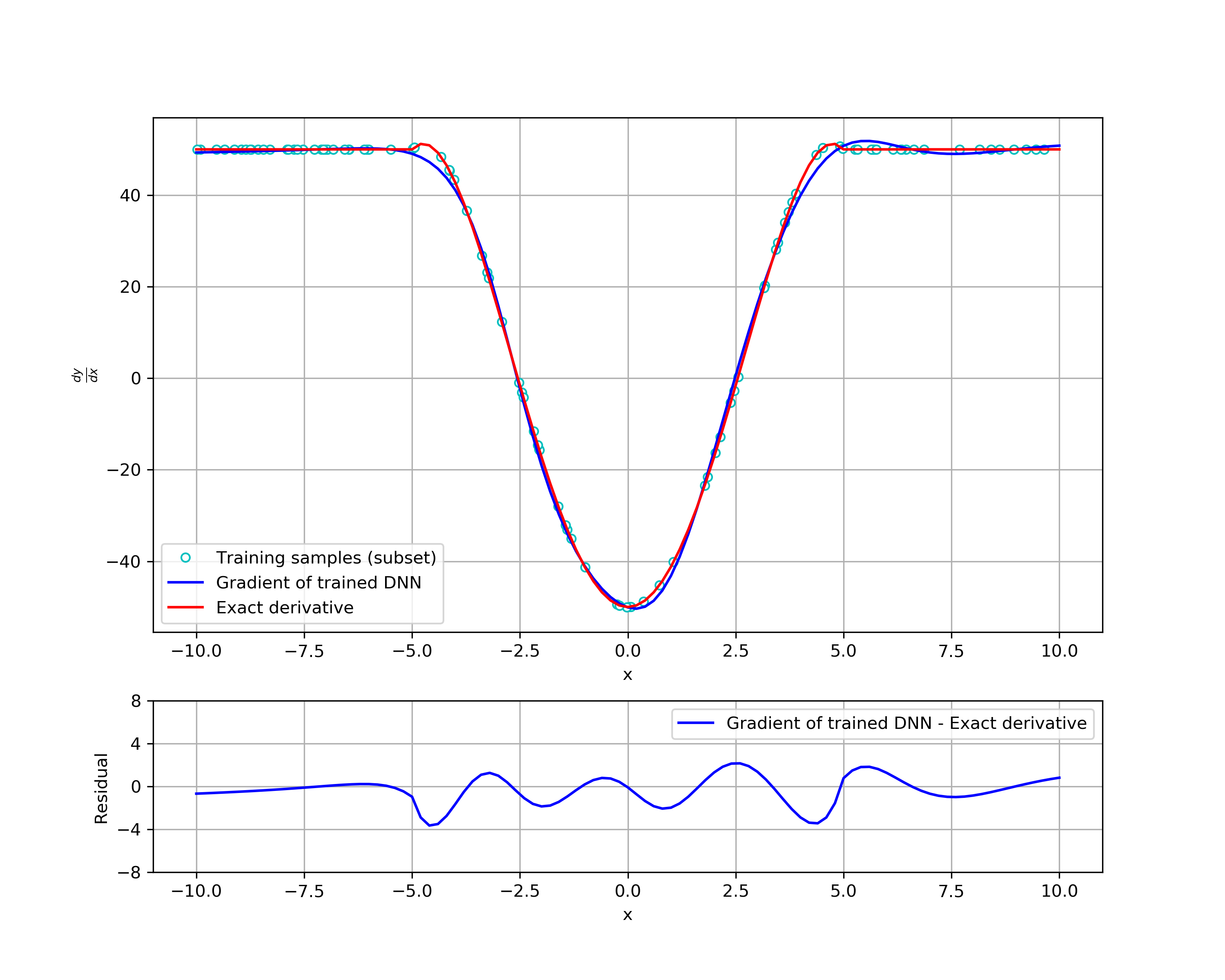} }}%
    \caption{First function approximation:  Differential Machine Learning without asymptotic treatment where results from DNN are in solid blue, 
    true values are in solid red, and the blue scatter points represent a subset of samples used in training chosen randomly for visualization purposes only.}%
    \label{fig:results_dml}%
\end{figure}

\begin{figure}%
    \centering
    \subfloat[\centering Trained DNN along with true value]{{\includegraphics[width=7cm]{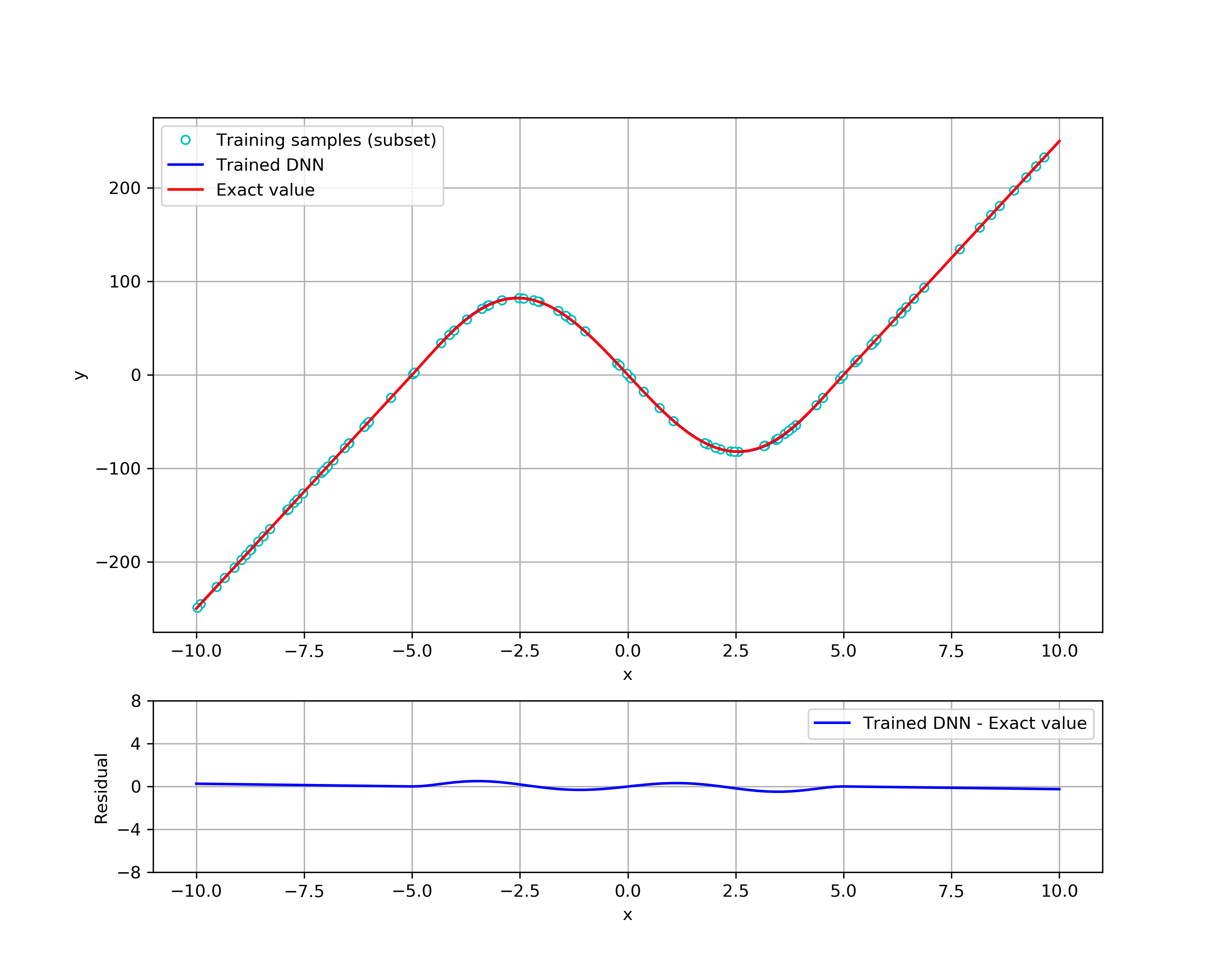} }}%
    \qquad
    \subfloat[\centering Gradient of trained DNN along with true  derivative]{{\includegraphics[width=7cm]{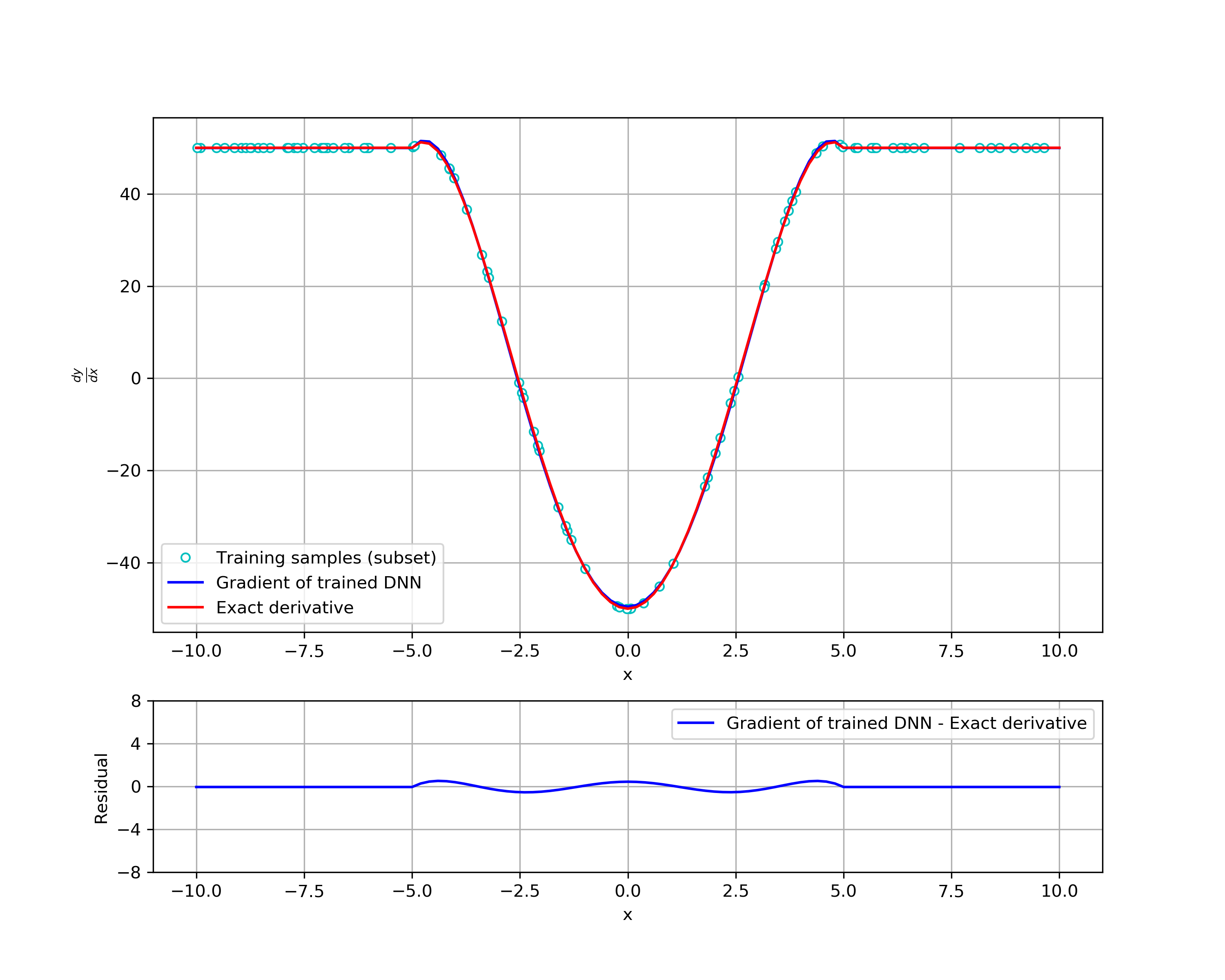} }}%
    \caption{First function approximation:  Differential Machine Learning with asymptotic treatment (trainable parameters) where results from DNN are in solid blue, 
    true values are in solid red, and the blue scatter points represent a subset of samples used in training.}%
    \label{fig:results_dml_trainasymp}%
\end{figure}

\begin{figure}%
    \centering
    \subfloat[\centering Trained DNN along with true  value]{{\includegraphics[width=7cm]{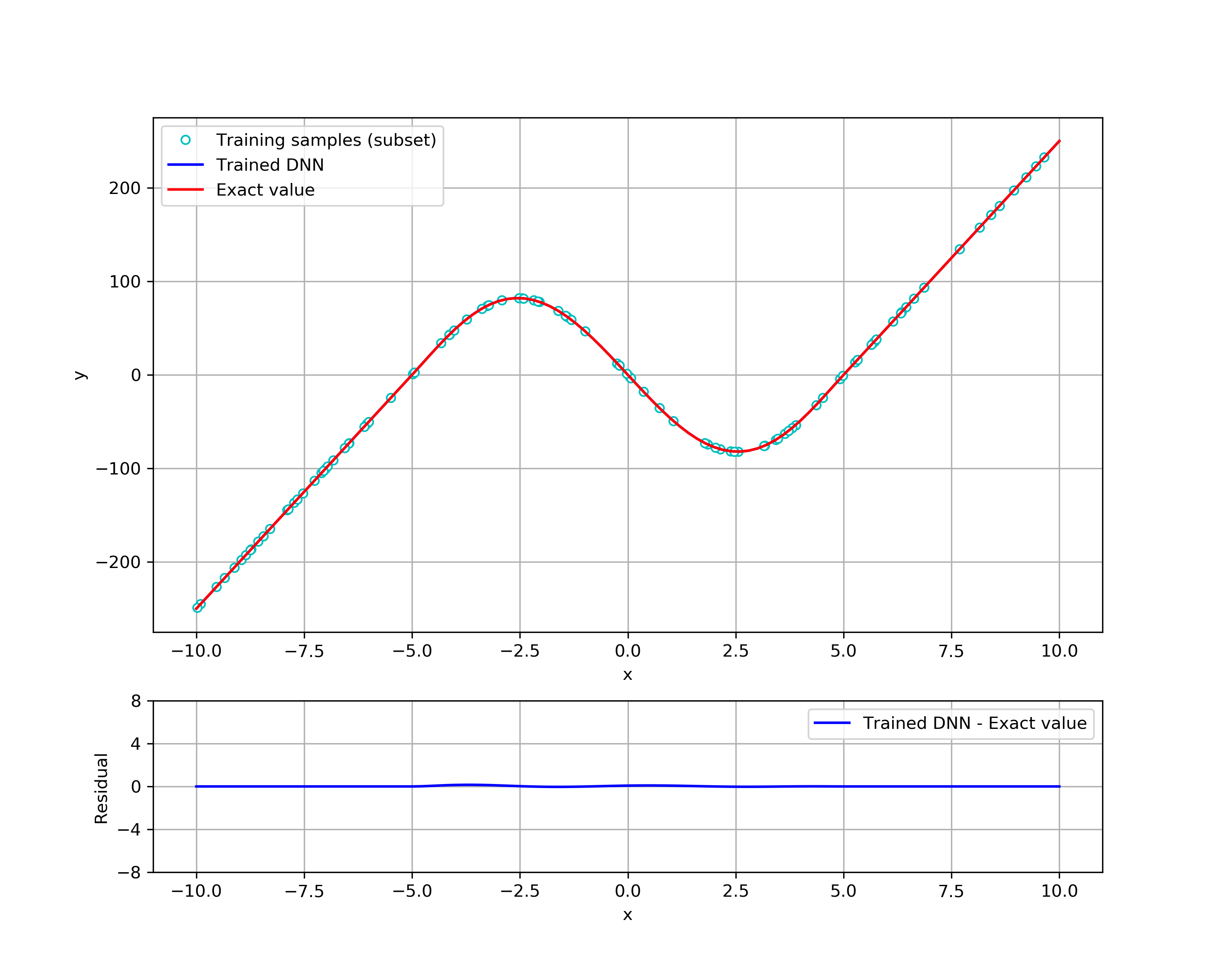} }}%
    \qquad
    \subfloat[\centering Gradient of trained DNN along with true derivative]{{\includegraphics[width=7cm]{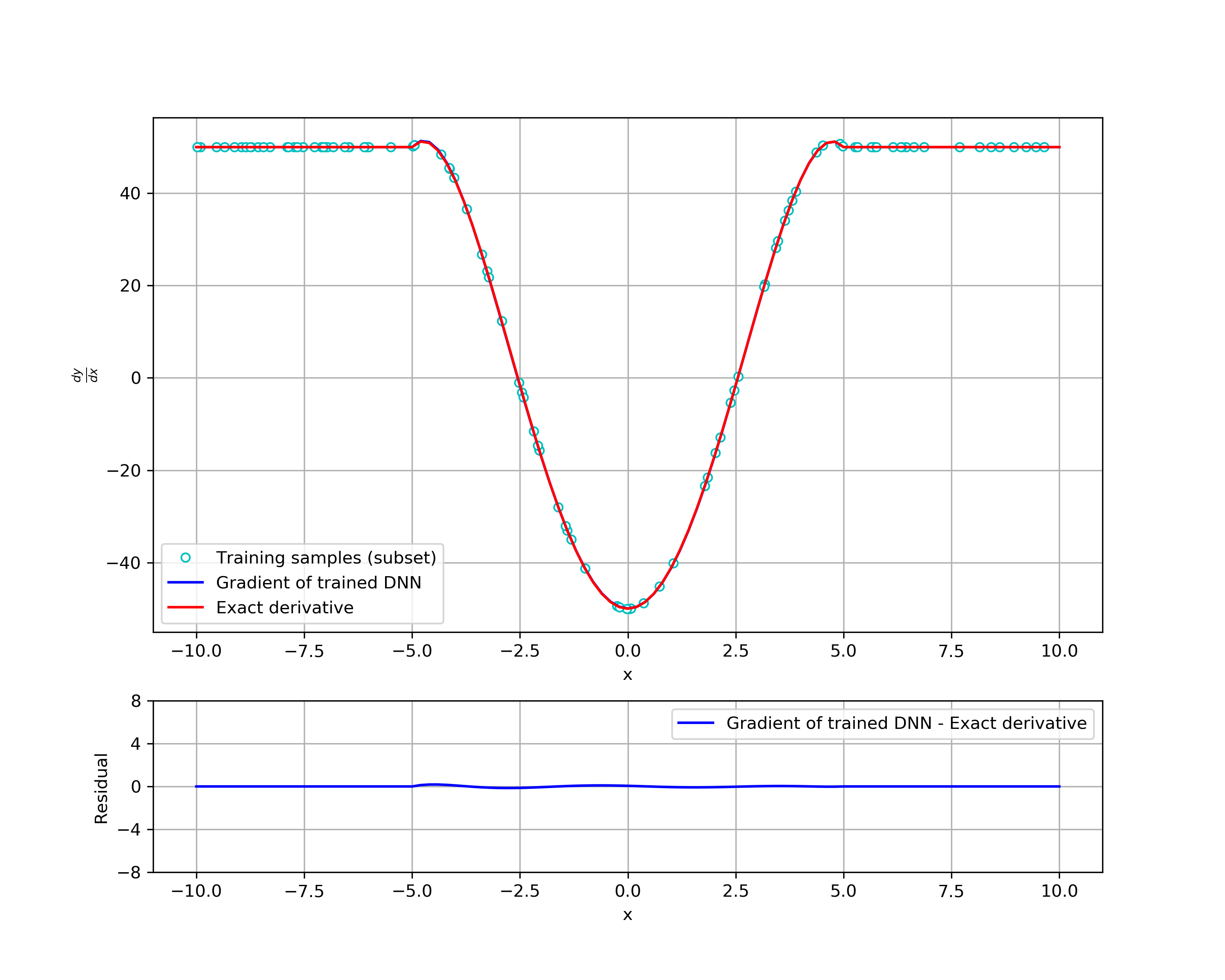} }}%
    \caption{First function approximation:  Differential Machine Learning with asymptotic treatment (fixed parameters) where results from DNN are in solid blue, 
    true values are in solid red, and the blue scatter points represent a subset of samples used in training chosen randomly for visualization purposes only.}%
    \label{fig:results_dml_fixedasymp}%
\end{figure}

\begin{figure}%
    \centering
    \subfloat[\centering Difference to true value]{{\includegraphics[width=7cm]{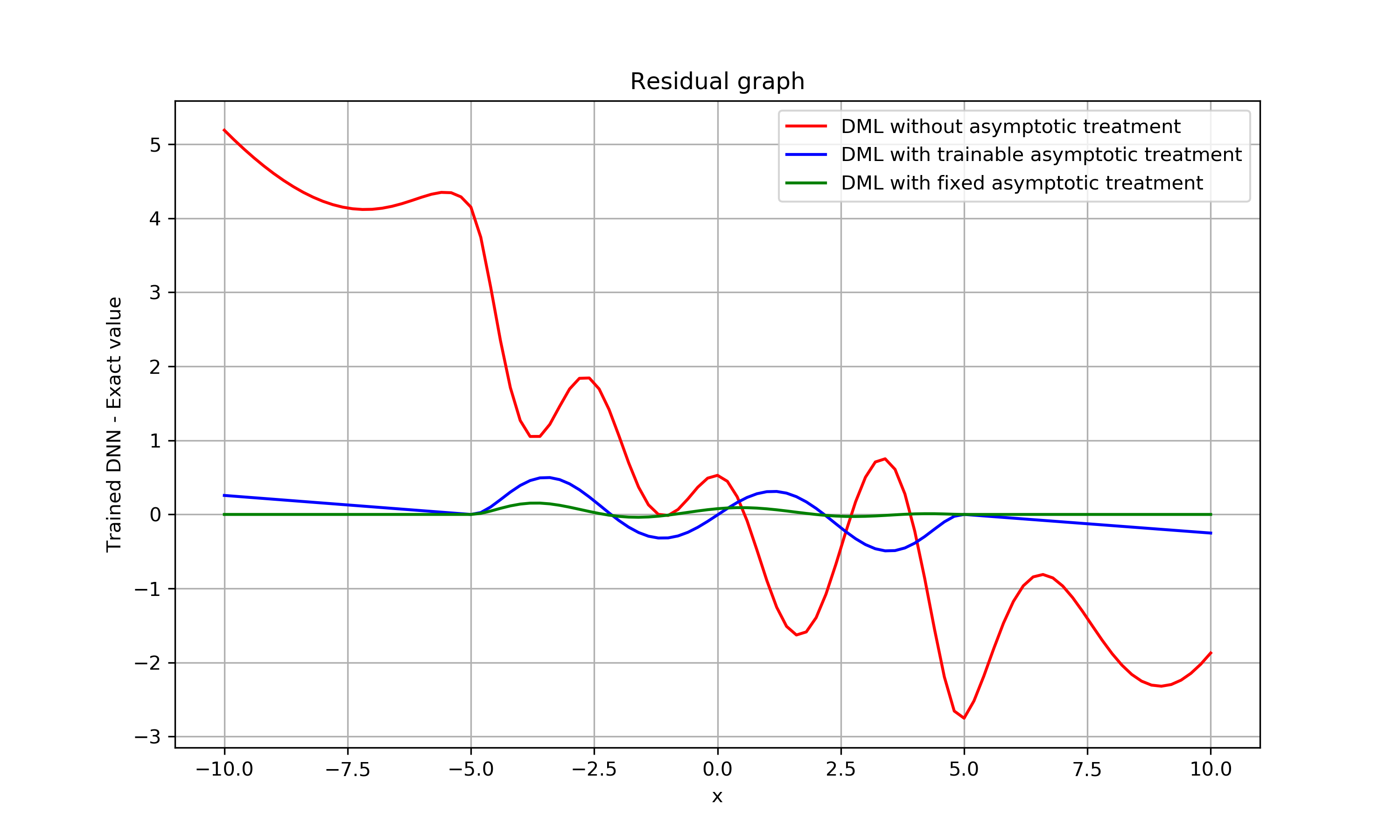} }}%
    \qquad
    \subfloat[\centering Difference to true derivative]{{\includegraphics[width=7cm]{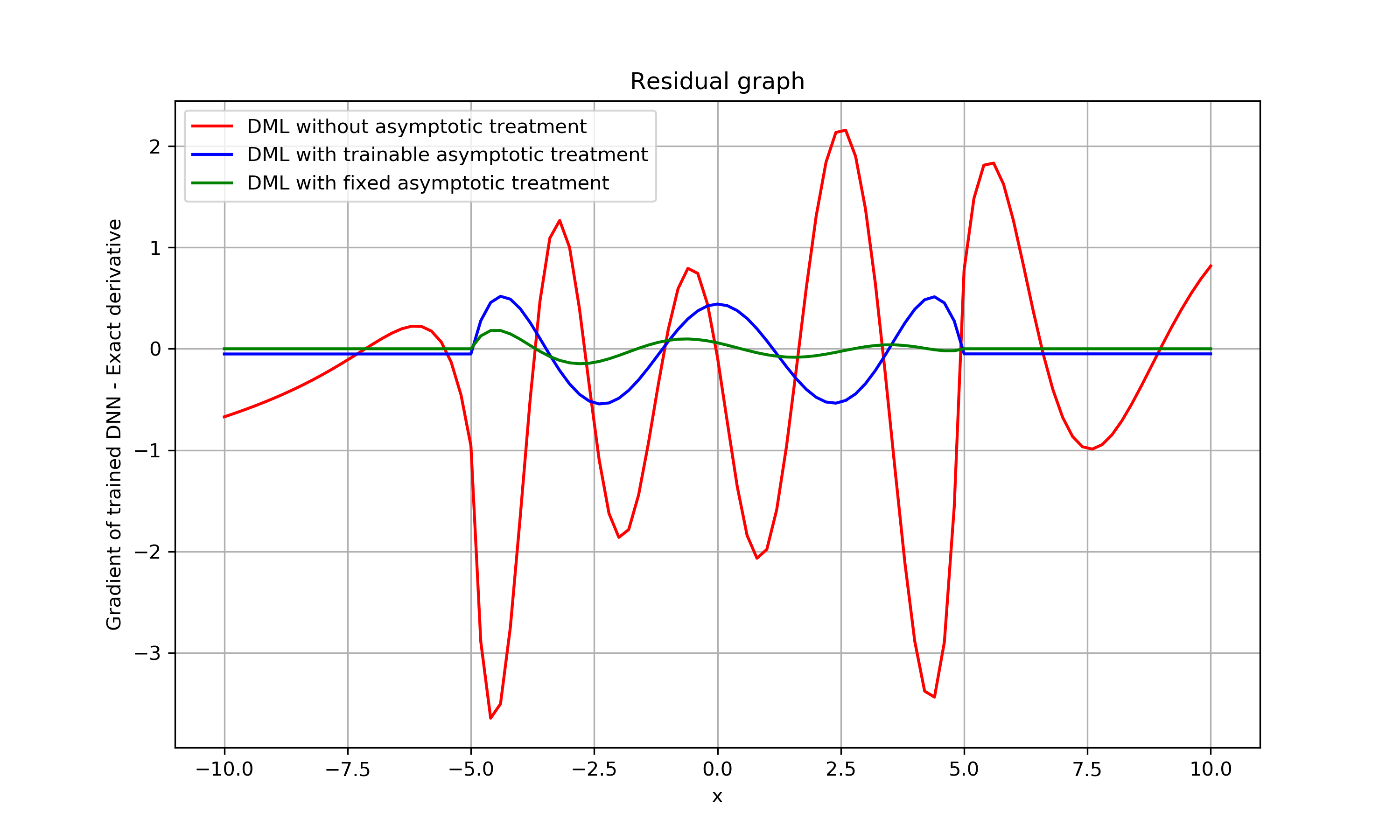} }}%
    \qquad
     \subfloat[\centering Training loss graph (DML loss)]{{\includegraphics[width=7cm]{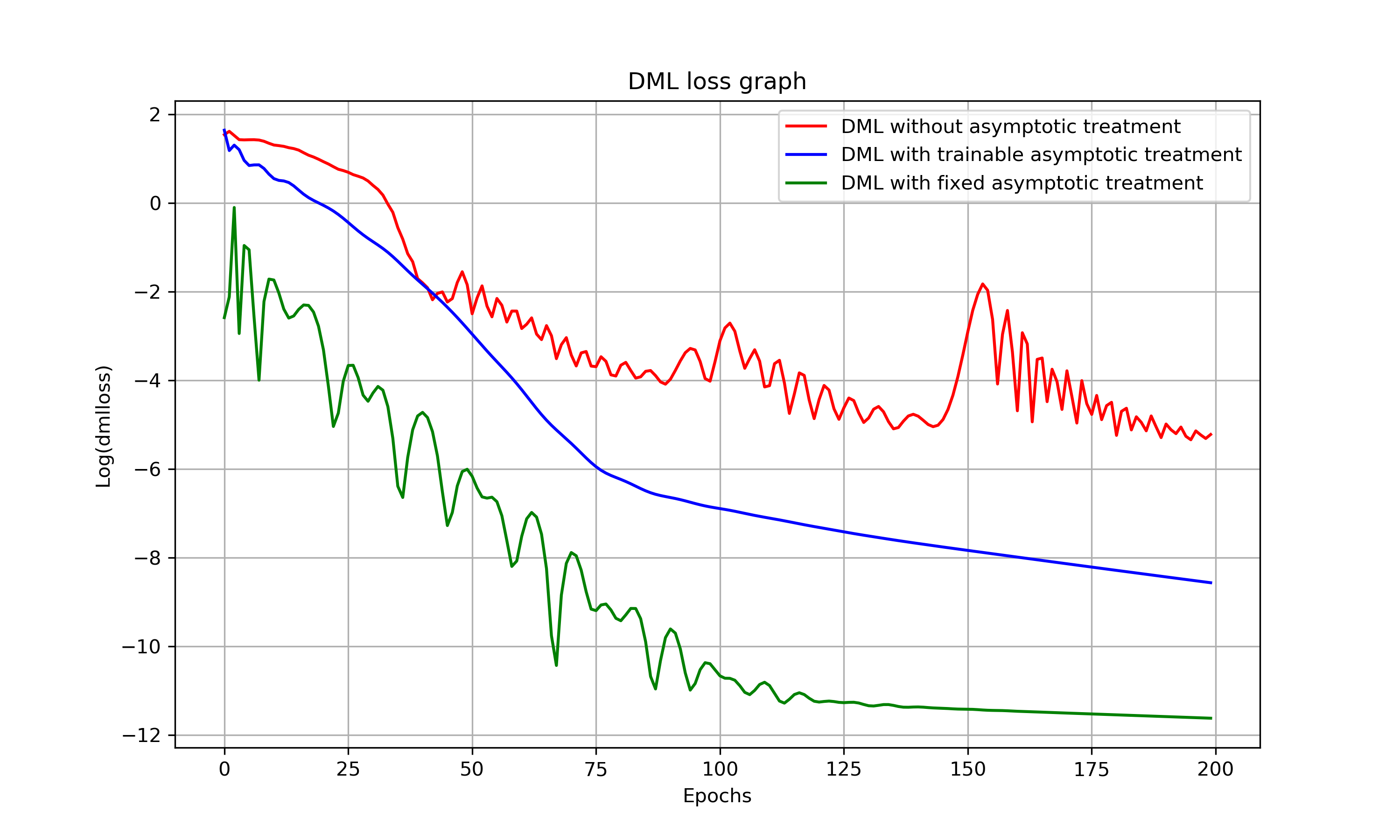} }}%
    \caption{First function approximation: Difference and loss graphs for DML without and with asymptotic treatment}%
    \label{fig:lossgraphs}%
\end{figure}

\begin{figure}%
    \centering
    \subfloat[\centering Difference to true value]{{\includegraphics[width=7cm]{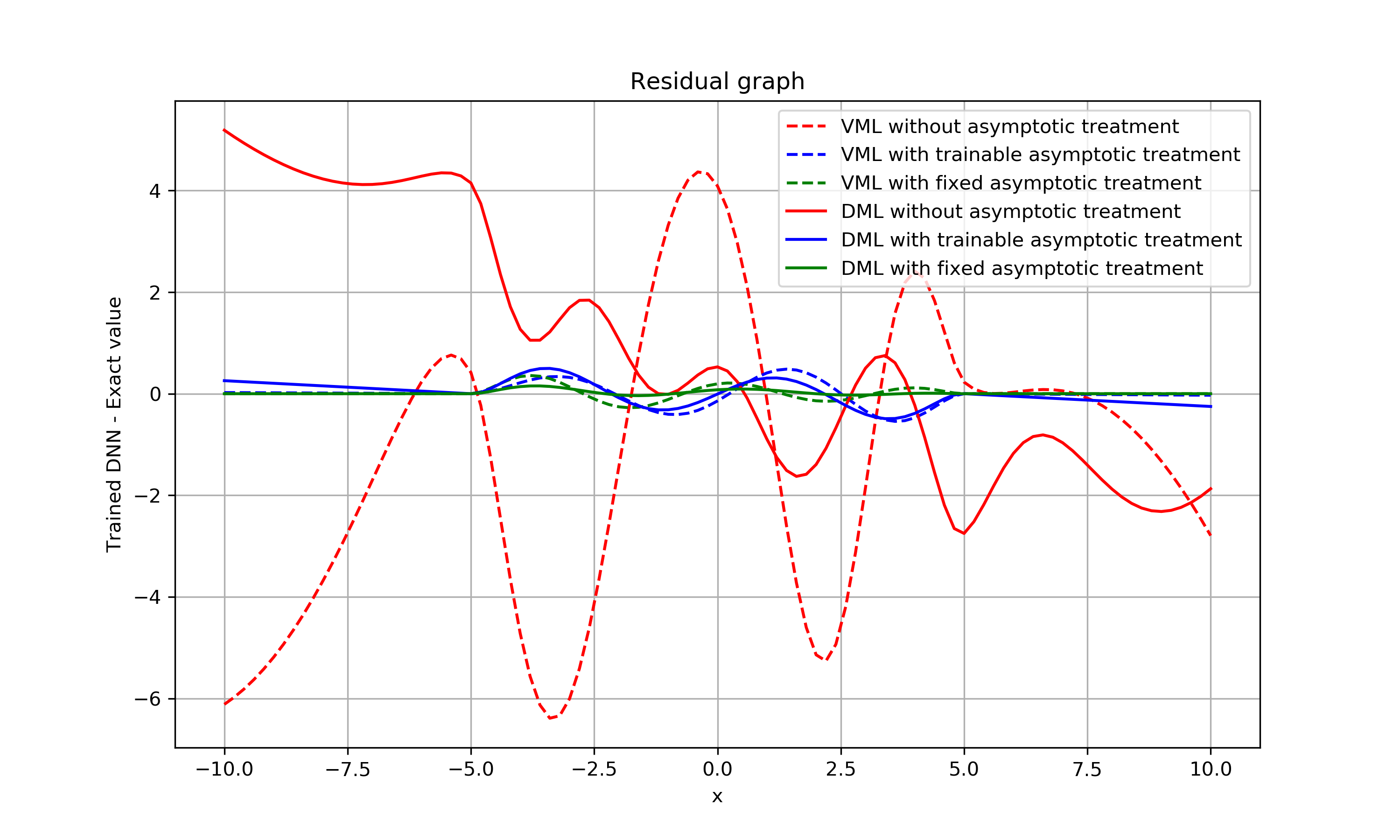} }}%
    \qquad
    \subfloat[\centering Difference to true derivative]{{\includegraphics[width=7cm]{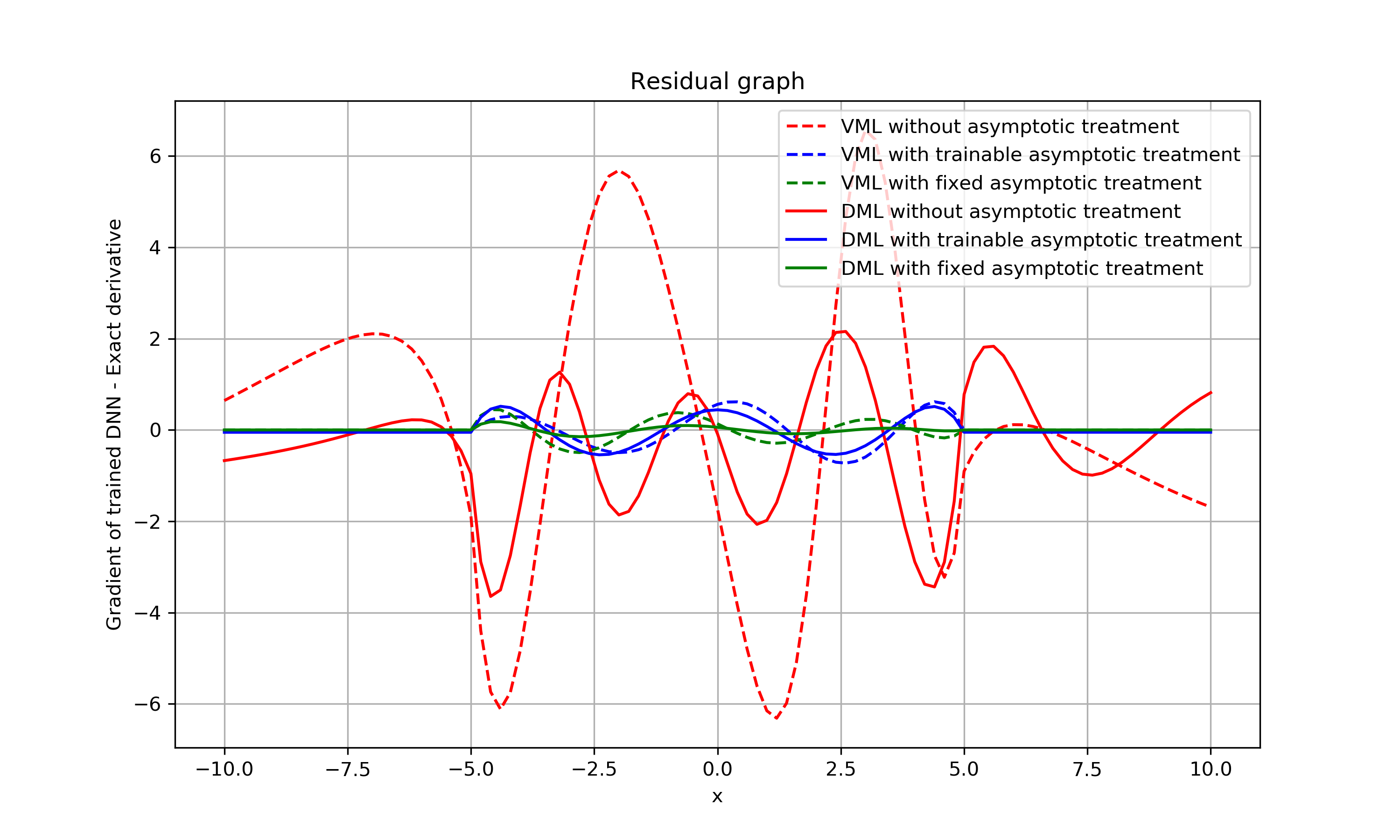} }}%
    \qquad
     \subfloat[\centering Training loss graph (VML loss)]{{\includegraphics[width=7cm]{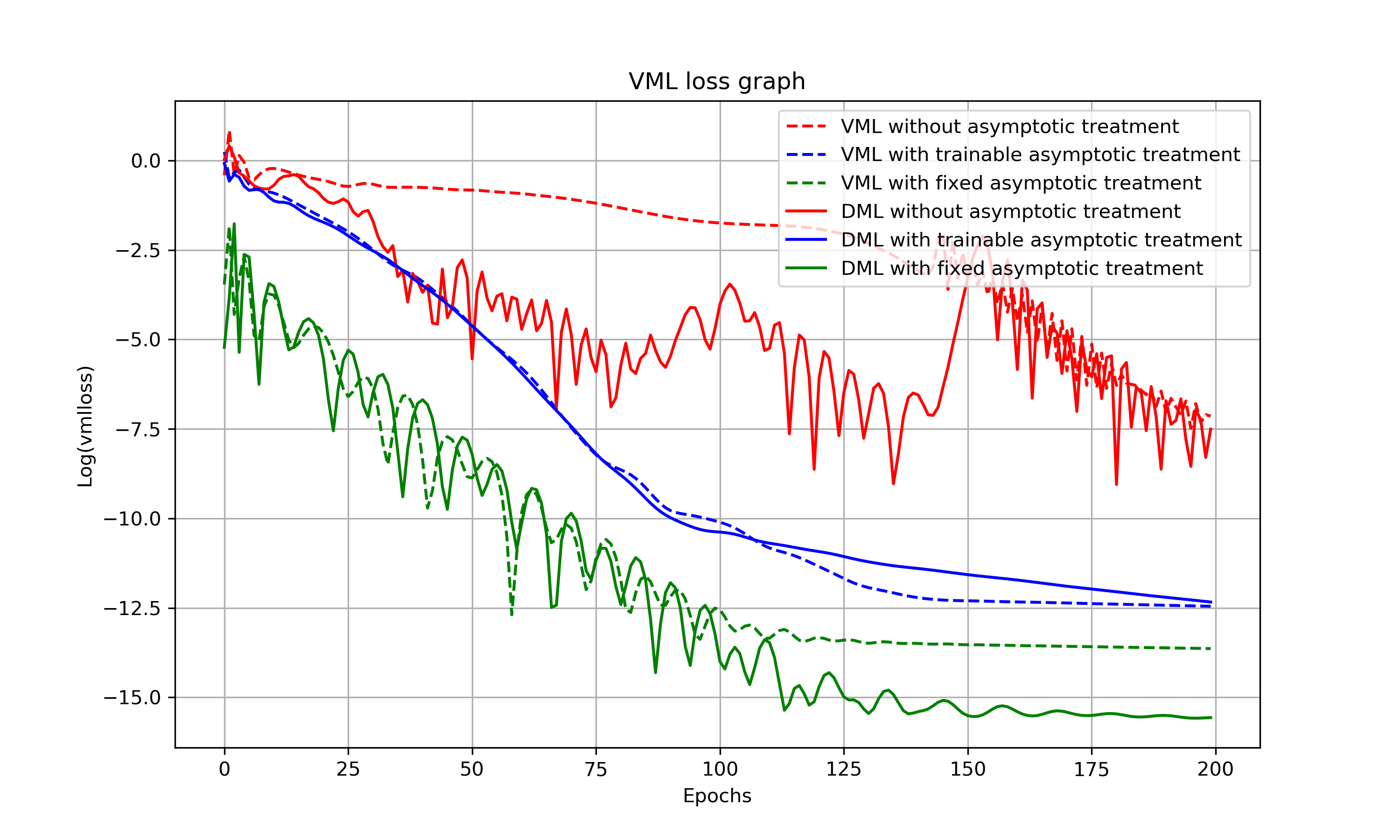} }}%
     \subfloat[\centering Training loss graph (DML loss)]{{\includegraphics[width=7cm]{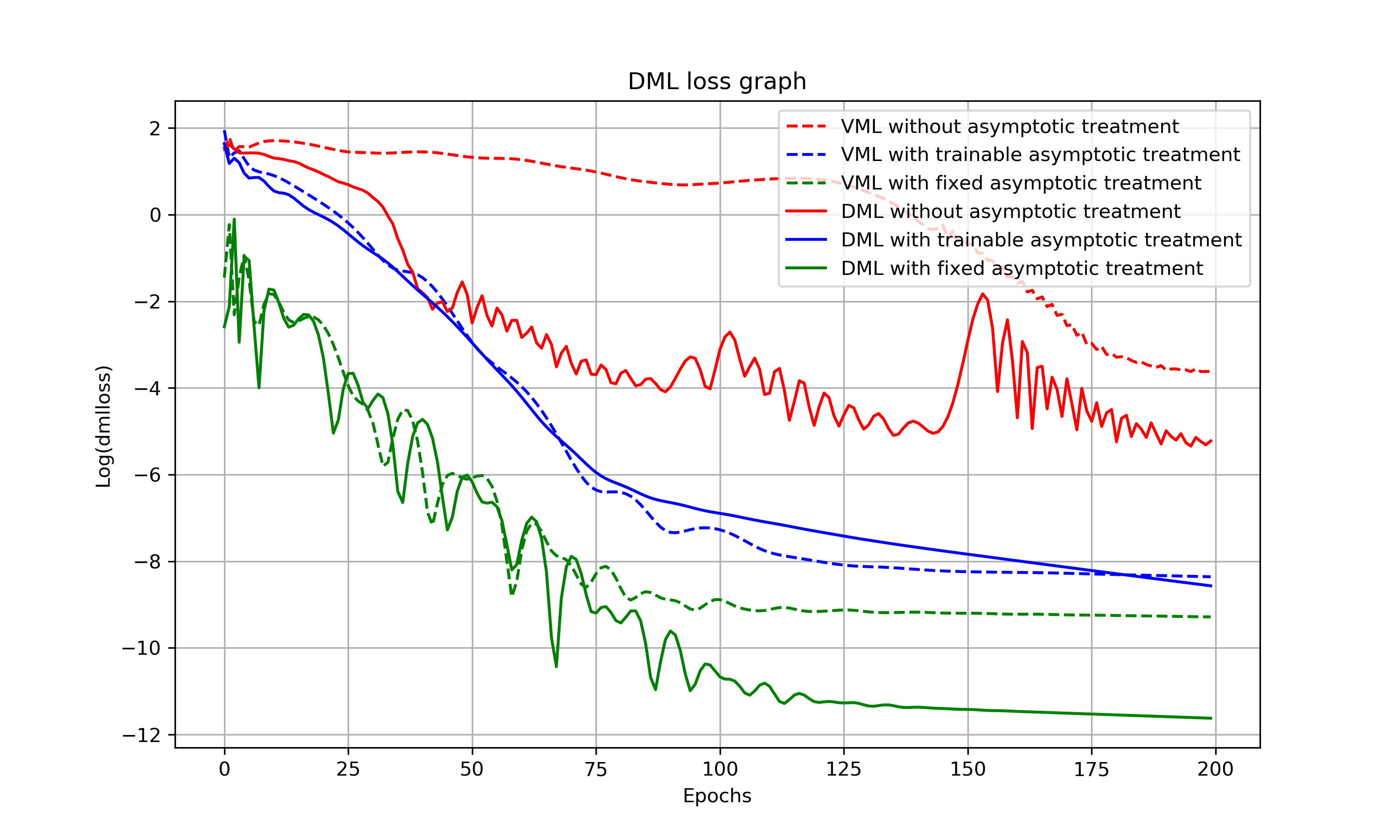} }}%
    \caption{First function approximation: Difference and loss graphs for DML and VML (without and with asymptotic treatment)}%
    \label{fig:lossgraphs_all}%
\end{figure}

We evaluate the performance of the trained DNNs from different methods on a test dataset. 
Fig.~\ref{fig:results_vml} shows the results of the trained deep neural network using VML without asymptotic treatment,
 Fig.~\ref{fig:results_vml_trainasymp} shows the results of the trained deep neural network using VML with asymptotic treatment (trainable parameters) 
 and Fig.~\ref{fig:results_vml_fixedasymp} shows the results of the trained deep neural network using VML with asymptotic treatment (fixed parameters).
 When looking at the difference to the true values, it is evident that VML in conjunction with asymptotic treatment leads to faster and accurate convergence, compared to 
 VML without asymptotic treatment. This conclusion can also be obtained on examining the corresponding difference and loss graphs as shown in Fig.~\ref{fig:lossgraphs_vml}. 
 Among asymptotic methods, we observe that fixed asymptotic parameters lead to better results than trainable asymptotic parameters as is natural to expect. 
 Comparing DML without and with asymptotic treatment, we are led to the same conclusion. 
 Fig.~\ref{fig:results_dml} shows the results of the trained deep neural network using DML without asymptotic treatment, 
 Fig.~\ref{fig:results_dml_trainasymp} shows the results of the trained deep neural network using DML with asymptotic treatment (trainable parameters), 
 Fig.~\ref{fig:results_dml_fixedasymp} shows the results of the trained deep neural network using DML with asymptotic treatment (fixed parameters) 
 and Fig.~\ref{fig:lossgraphs} shows the corresponding difference and loss graphs. We also note that DML with asymptotic treatment performs better 
 than just DML without any asymptotic treatment, demonstrating the effectiveness of our methodology. The scales for the graphs are kept fixed across different methods to facilitate easier comparison.
Fig.~\ref{fig:lossgraphs_all} shows the comparison of DML and VML in difference and loss graphs.

\clearpage

\section{Learning Black-Scholes formula with function approximation}
\label{sec:bsfuncapprox}

In this section, we apply our methodology in function approximation to approximate the Black-Scholes function. 
The price of a vanilla option on a non-dividend paying stock $V(t) = e^{-r(T-t)}\mathbb E_t[(\phi (S(T) - K))^+]$ with Black-Scholes model is
\begin{equation}
\begin{split}
&V(t;S(t),K,r,\sigma,T) =\phi( N(\phi d_+) \cdot S(t) - N(\phi d_-) \cdot K \cdot e^{-r(T-t)} )\\
&d_+ = \frac{1}{\sigma\sqrt{T-t}}(ln\frac{S(t)}{K} + (r+\frac{\sigma^2}{2})(T-t))\\
&d_- = d_+ - \sigma \sqrt{T-t}
\end{split}
\end{equation}
\label{eq:2.4}
where $\phi$ is +1 (-1) for call (put), $S$ is the (stochastic) price of equity, $r$ is the (here constant) short rate, 
$K$ is the strike, and $T$ is the maturity. 

We test on a call option with intermediate time $t=1$ year, maturity $T = 2$ years, volatility $\sigma$ = 0.1, short rate $r$ = 0$\%$ and 
strike $K$ = 10. We choose to vary $X=S(t)$ and denote the function to be approximated $Black-Scholes(x)$.\\

\begin{figure}%
    \centering
    \subfloat[\centering Black-Scholes and asymptotic and nonasymptotic components]{{\includegraphics[width=7cm]{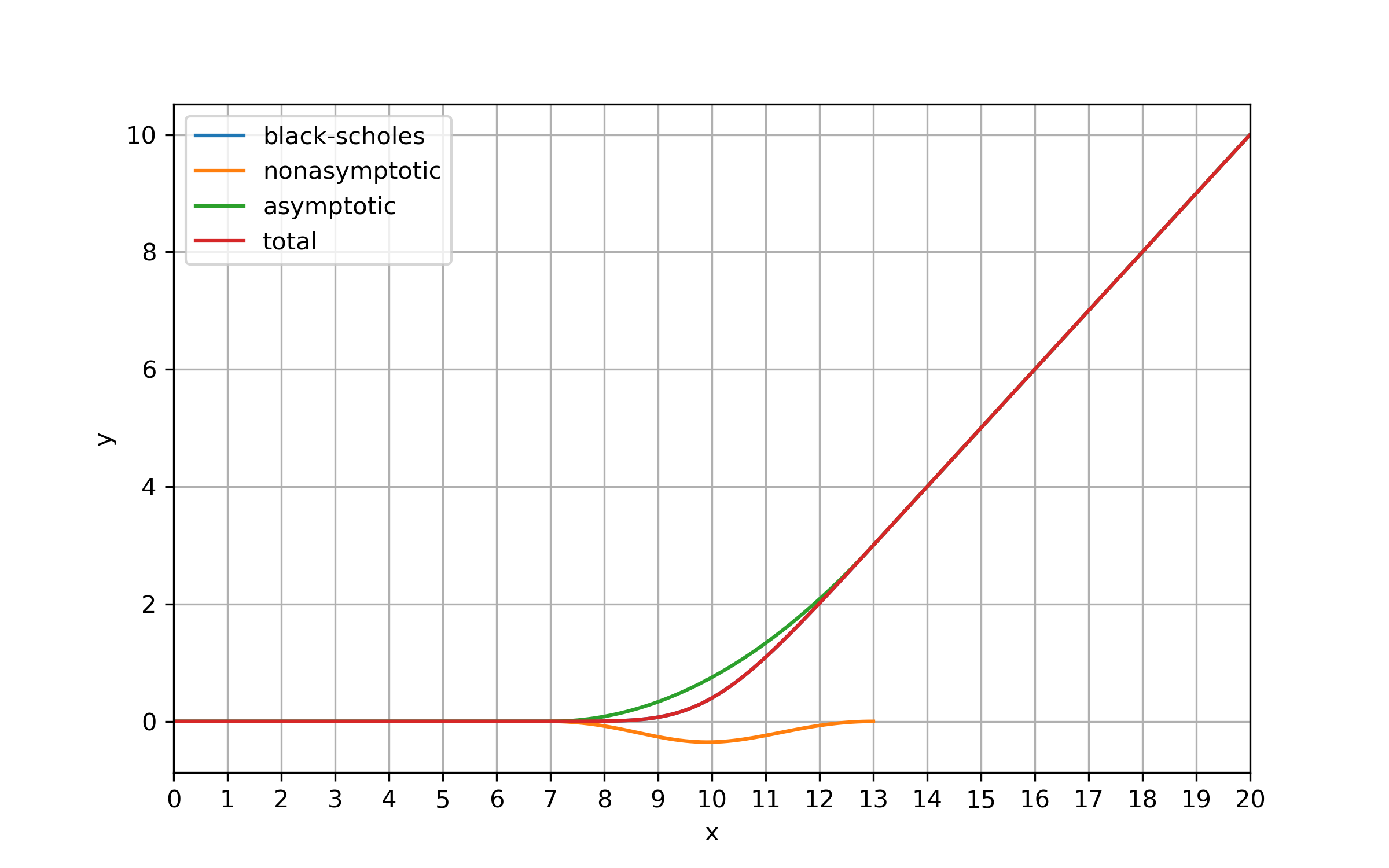} }}%
    \qquad
    \subfloat[\centering Derivative of Black-Scholes along with the derivative of associated components]{{\includegraphics[width=7cm]{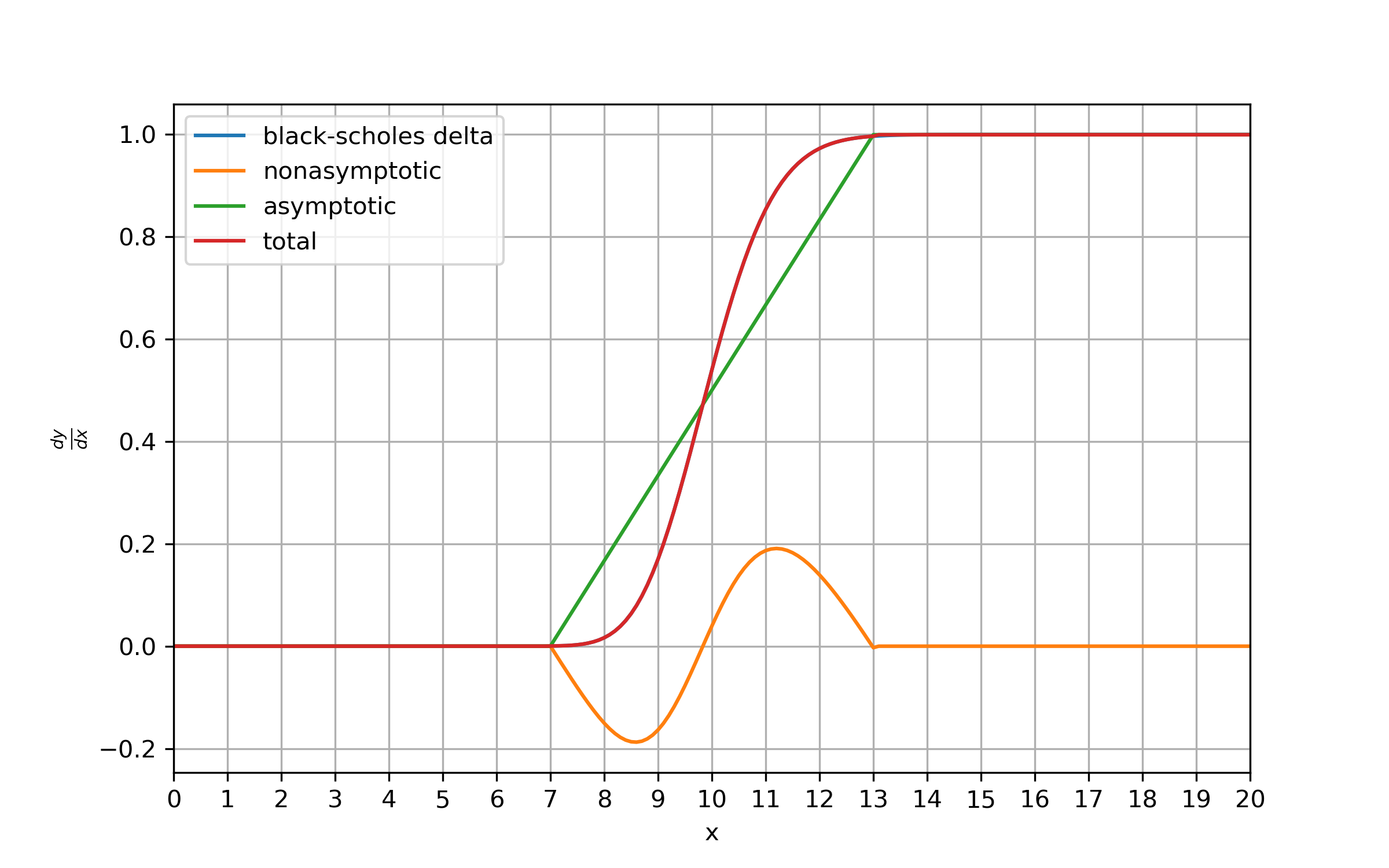} }}%
    \caption{Black-Scholes function approximation: Function and its asymptotic and nonasymptotic components - values and derivatives.}%
    \label{fig:func_bs_total}%
\end{figure}

The Black-Scholes function\footnote{Or, more accurately, its representation within some nonasymptotic 
domain together with asymptotic forms.} along with its asymptotic and non-asymptotic components 
is shown in Fig.~\ref{fig:func_bs_total}. Nominal parameters for $LL$ and $UL$ are chosen to be 7 and 13 respectively. 
The asymptotic parameters ($LS$,$LI$) and ($US$,$UI$) are either kept trainable or estimated via least squares.  
We generate the input data for training using 50,000 randomly selected $x$ from a uniform distribution 
between 0 and 20 along with associated true function and derivatives values $y$ and $\frac{dy}{dx}$. 
The hyperparameters for the neural network used for training on normalized inputs are: 
input dimension = 1, output dimension = 1, hidden layers with nodes = [20, 20], activation function = softplus,
 optimizer = Adam, epochs = 200, and learning rate = 0.05. \\

\begin{figure}%
    \centering
    \subfloat[\centering Trained DNN along with true value]{{\includegraphics[width=7cm]{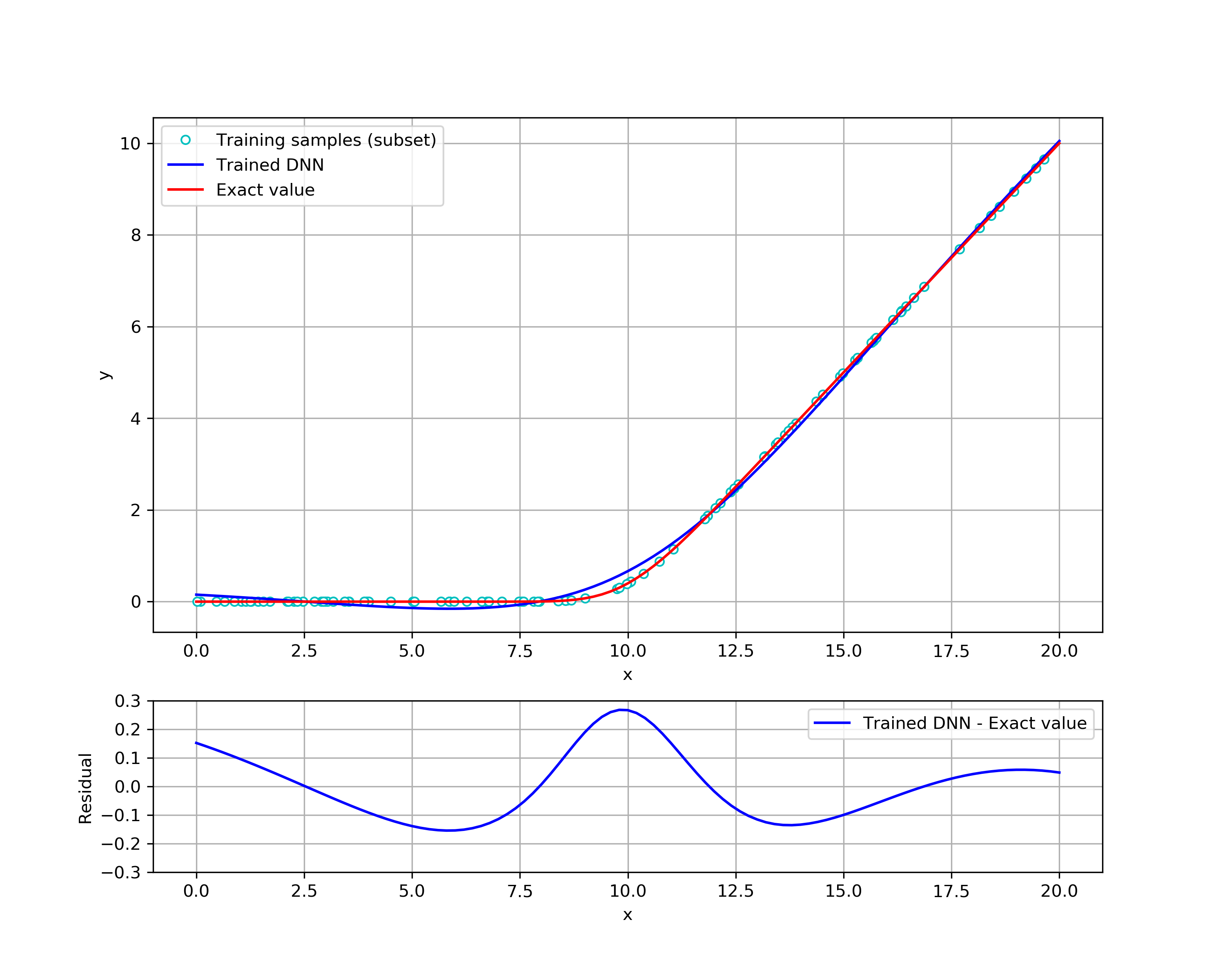} }}%
    \qquad
    \subfloat[\centering Gradient of trained DNN along with true derivative]{{\includegraphics[width=7cm]{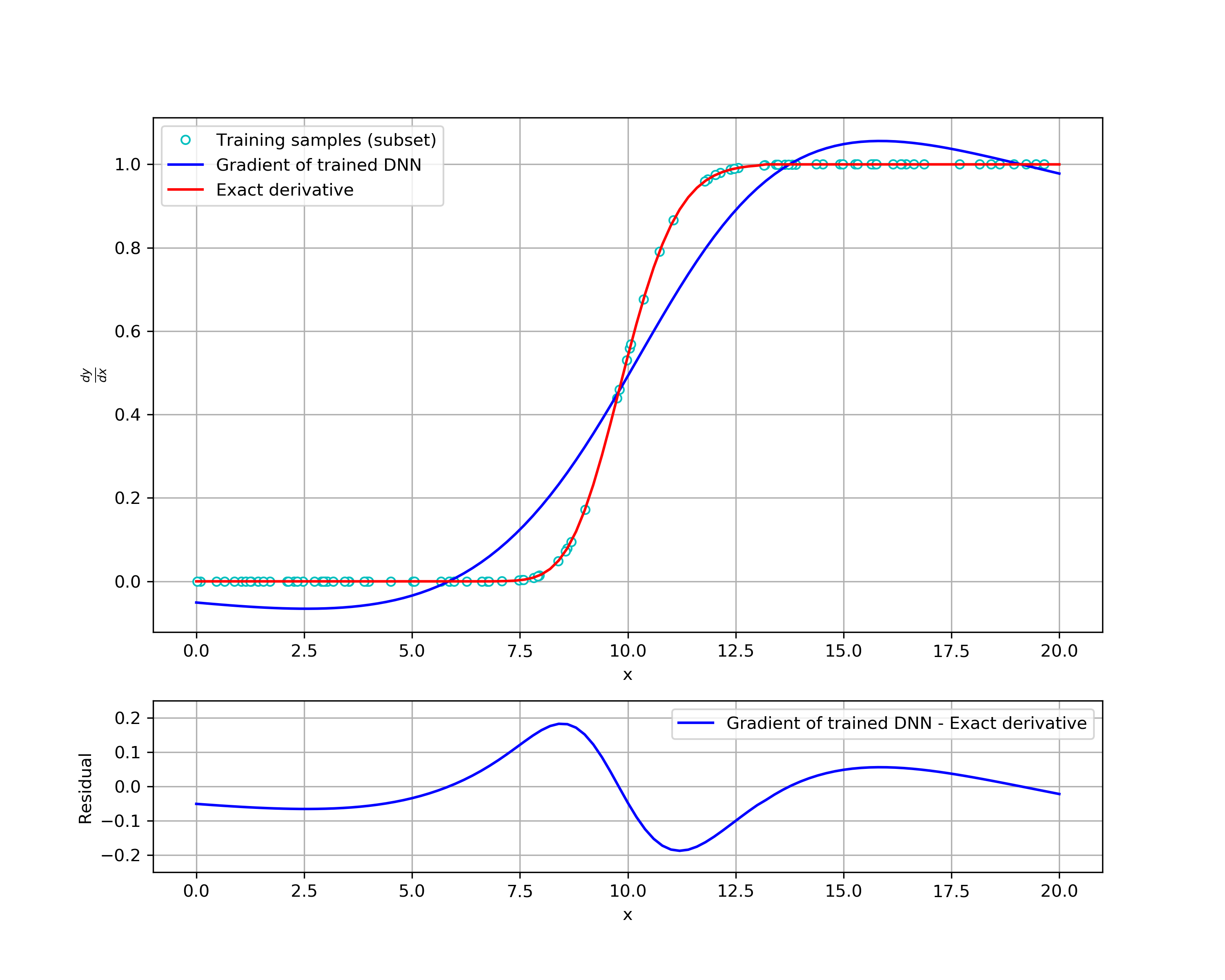} }}%
    \caption{Black-Scholes function approximation:  Vanilla Machine Learning without asymptotic treatment where results from DNN are in solid blue, 
    true values are in solid red, and the blue scatter points represent a subset of samples used in training chosen randomly for visualization purposes only.}%
    \label{fig:results_bs_vml}%
\end{figure}

\begin{figure}%
    \centering
    \subfloat[\centering Trained DNN along with true value]{{\includegraphics[width=7cm]{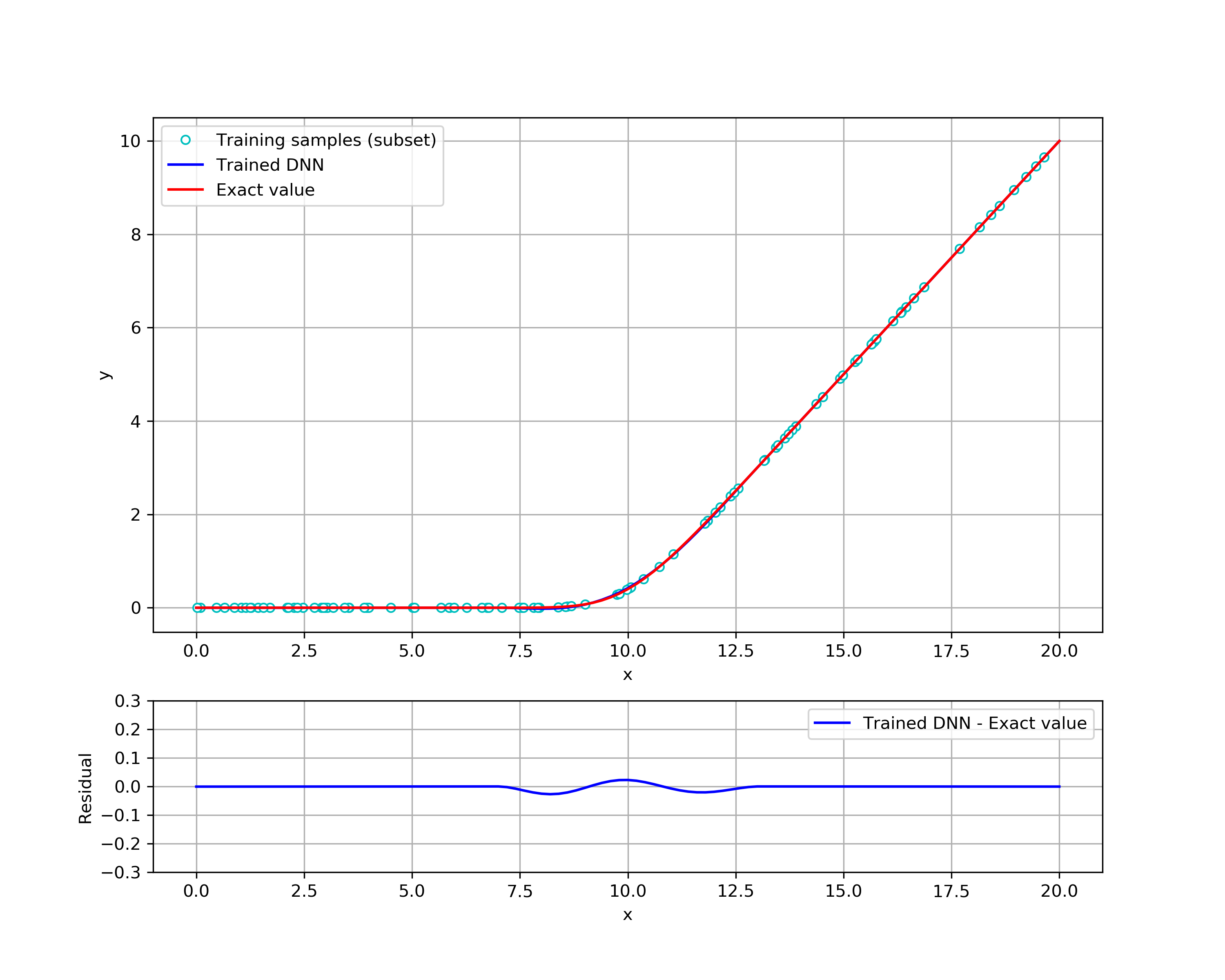} }}%
    \qquad
    \subfloat[\centering Gradient of trained DNN along with true derivative]{{\includegraphics[width=7cm]{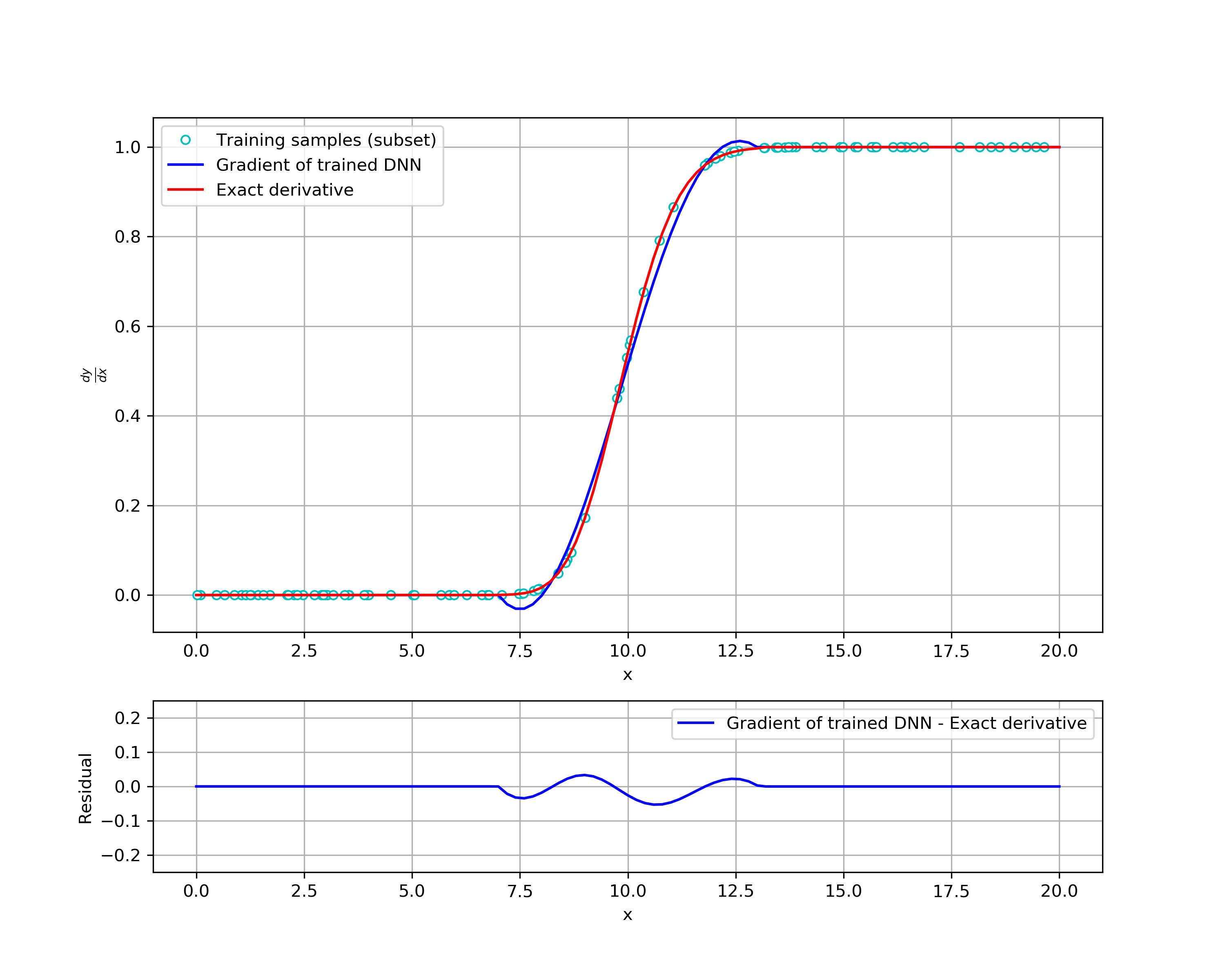} }}%
    \caption{Black-Scholes function approximation:  Vanilla Machine Learning with asymptotic treatment (trainable parameters) where results from DNN are in solid blue, 
    true values are in solid red, and the blue scatter points represent a subset of samples used in training chosen randomly for visualization purposes only.}%
    \label{fig:results_bs_vml_trainasymp}%
\end{figure}

\begin{figure}%
    \centering
    \subfloat[\centering Trained DNN along with true value]{{\includegraphics[width=7cm]{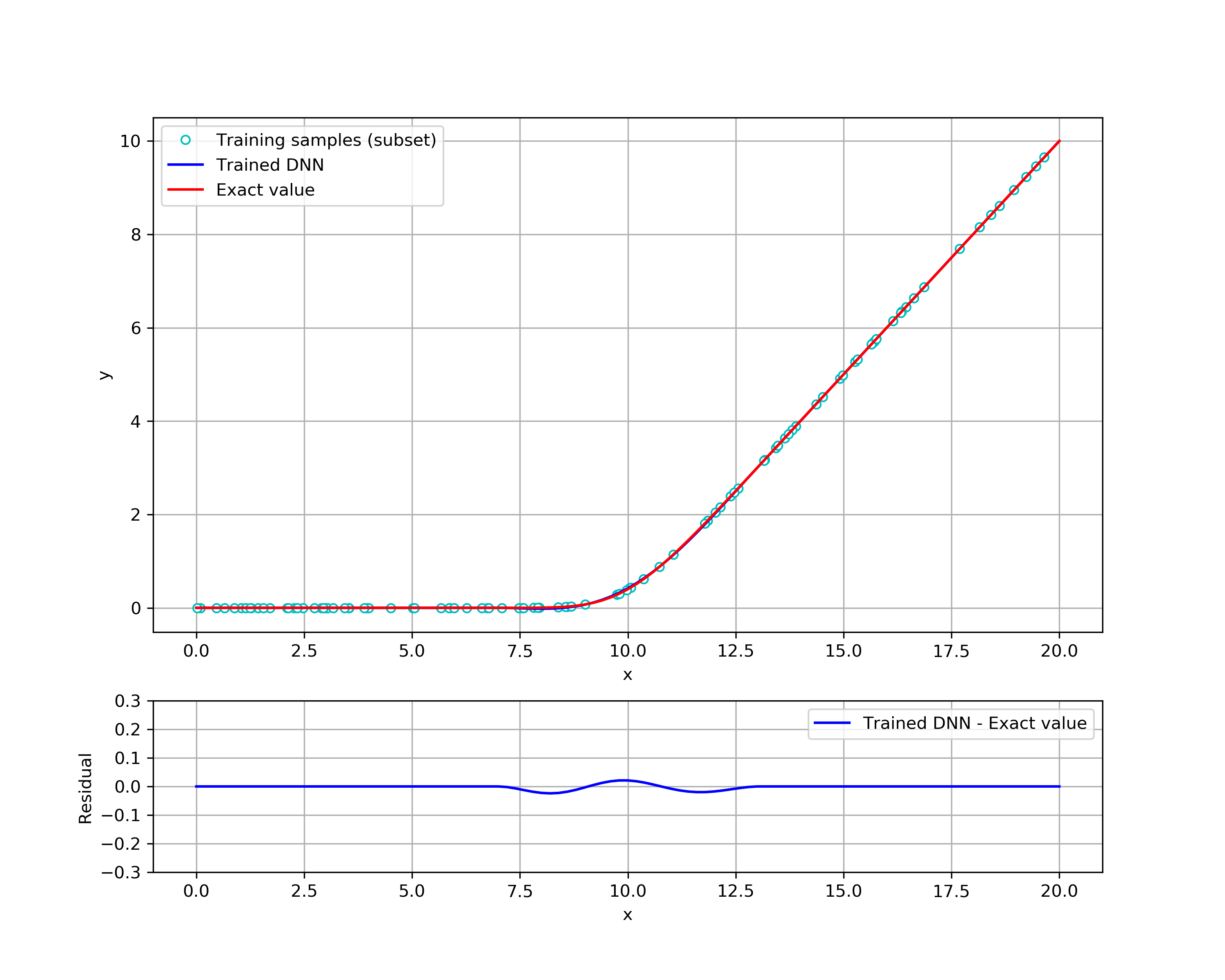} }}%
    \qquad
    \subfloat[\centering Gradient of trained DNN along with true derivative]{{\includegraphics[width=7cm]{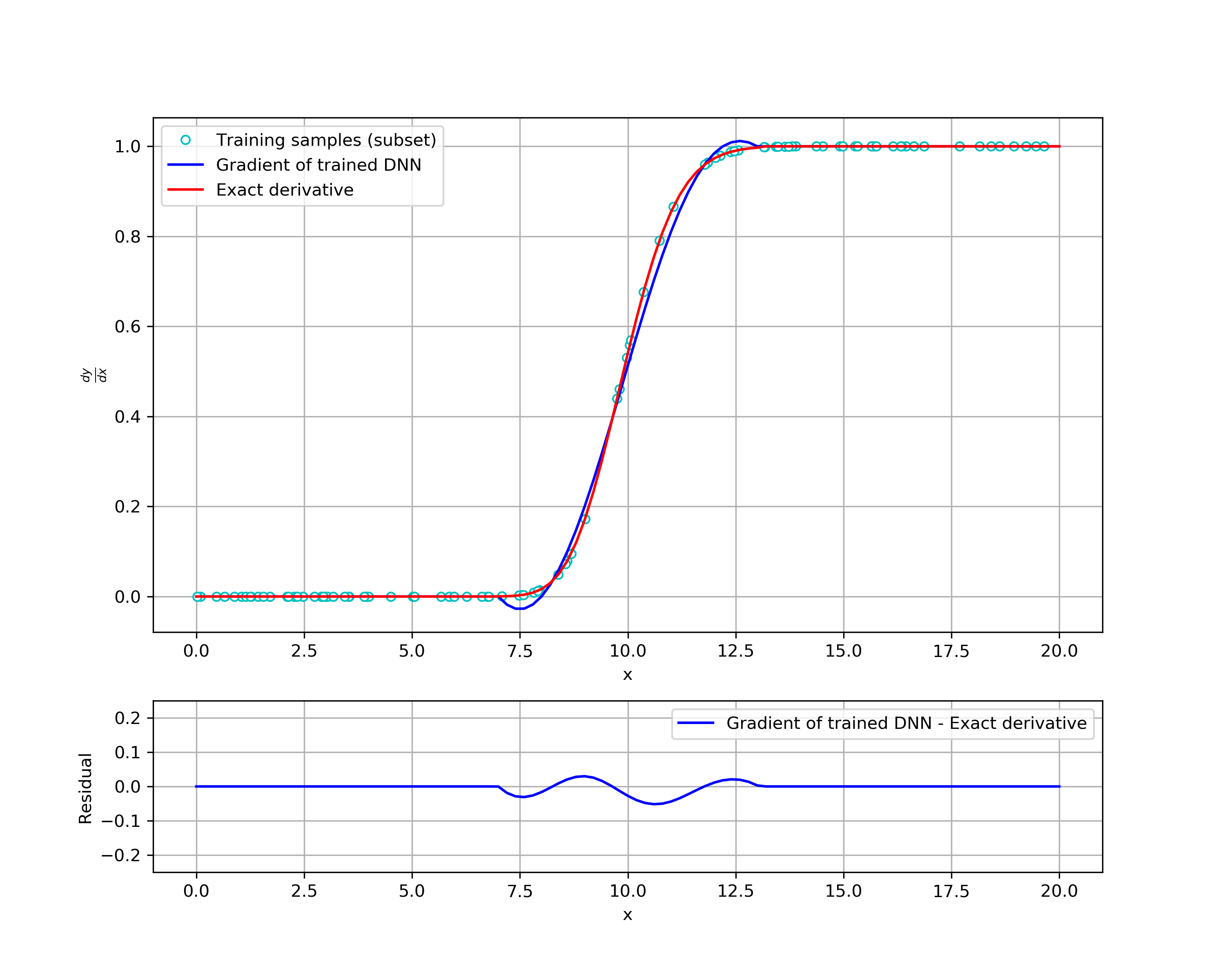} }}%
    \caption{Black-Scholes function approximation:  Vanilla Machine Learning with asymptotic treatment (fixed parameters) where results from DNN are in solid blue, 
    true values are in solid red, and the blue scatter points represent a subset of samples used in training chosen randomly for visualization purposes only.}%
    \label{fig:results_bs_vml_fixedasymp}%
\end{figure}

\begin{figure}%
    \centering
    \subfloat[\centering  Difference to true value]{{\includegraphics[width=7cm]{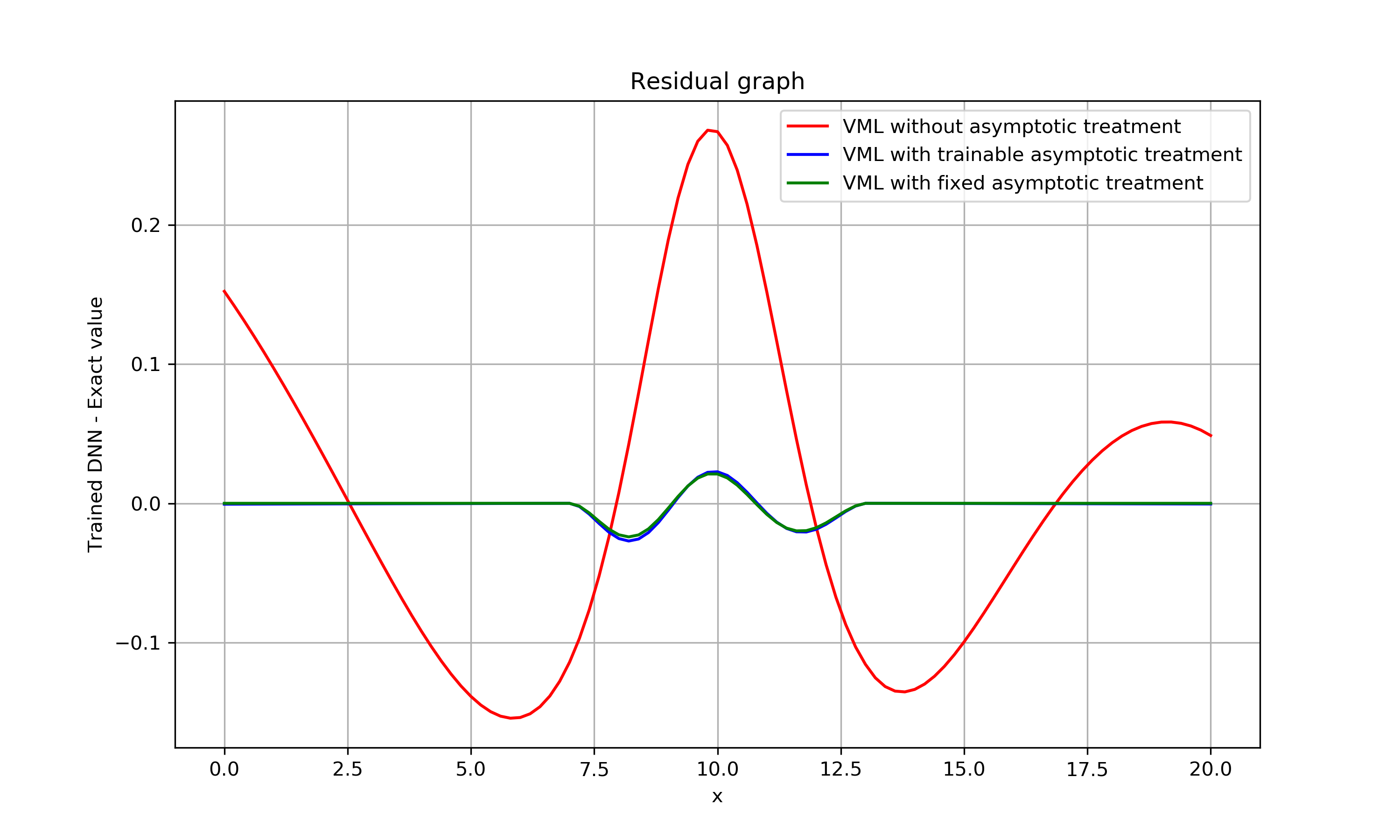} }}%
    \qquad
    \subfloat[\centering  Difference to true derivative]{{\includegraphics[width=7cm]{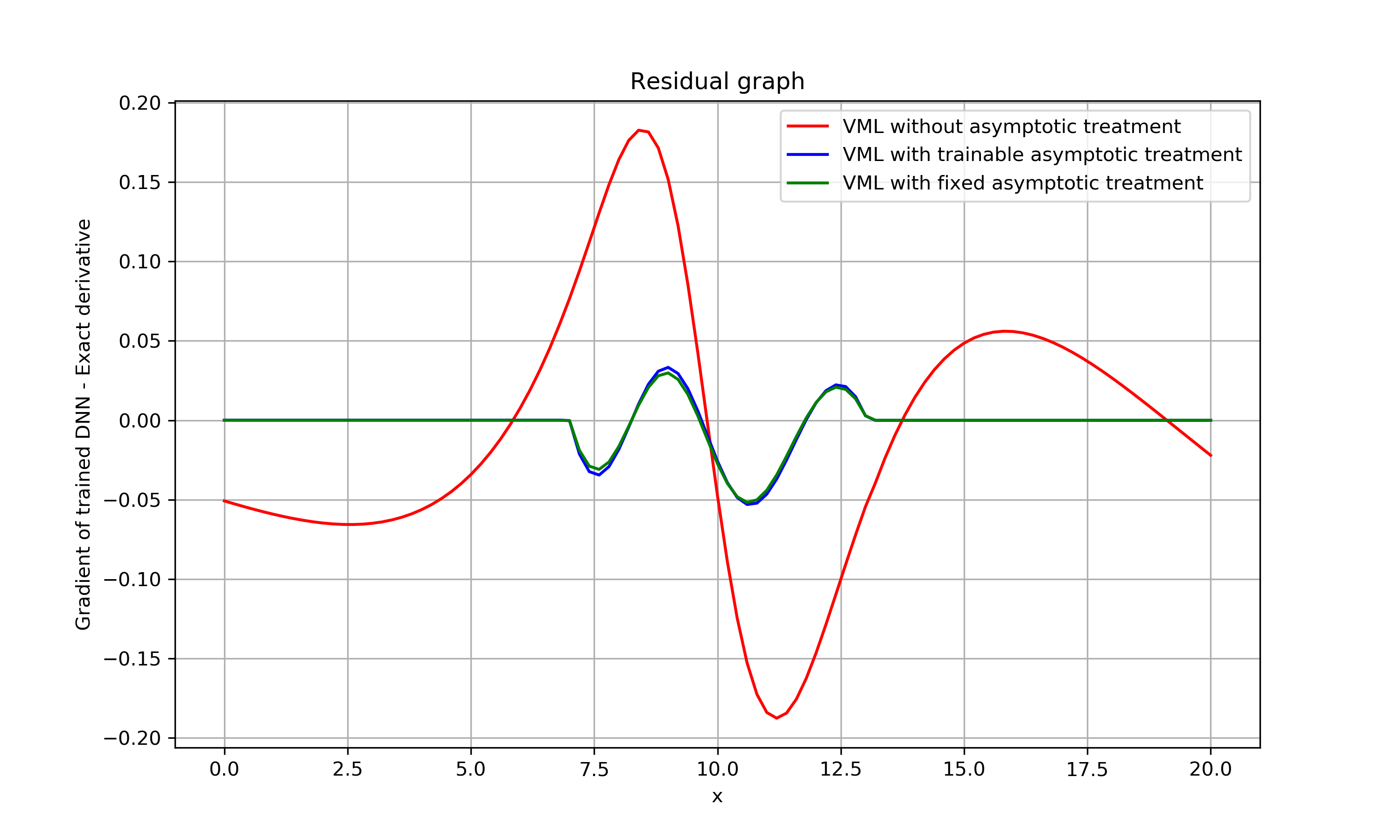} }}%
    \qquad
     \subfloat[\centering Training loss graph (VML loss)]{{\includegraphics[width=7cm]{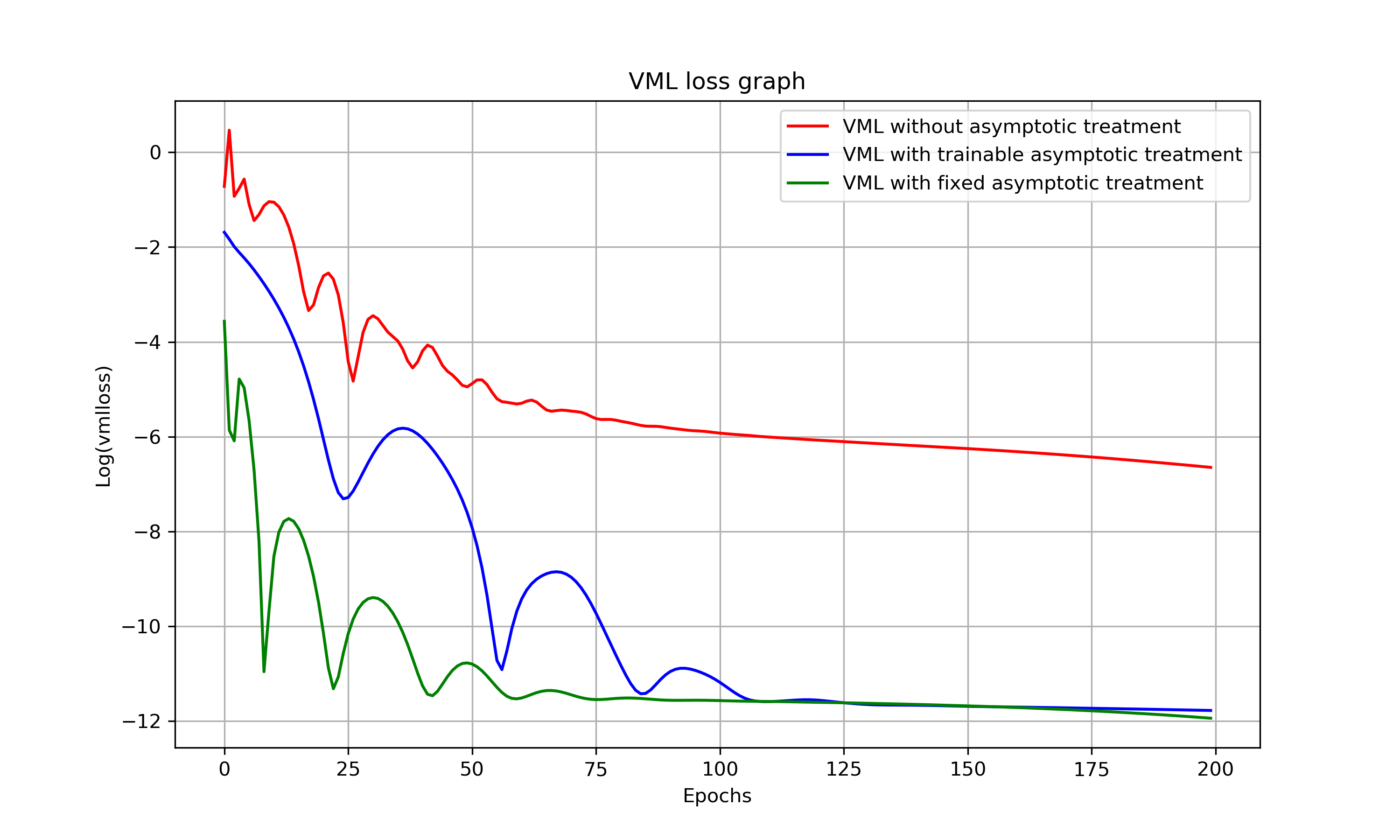} }}%
    \caption{Black-Scholes function approximation: Difference and loss graphs for VML without and with asymptotic treatment}%
    \label{fig:lossgraphs_bs_vml}%
\end{figure}

\begin{figure}%
    \centering
    \subfloat[\centering Trained DNN along with true value]{{\includegraphics[width=7cm]{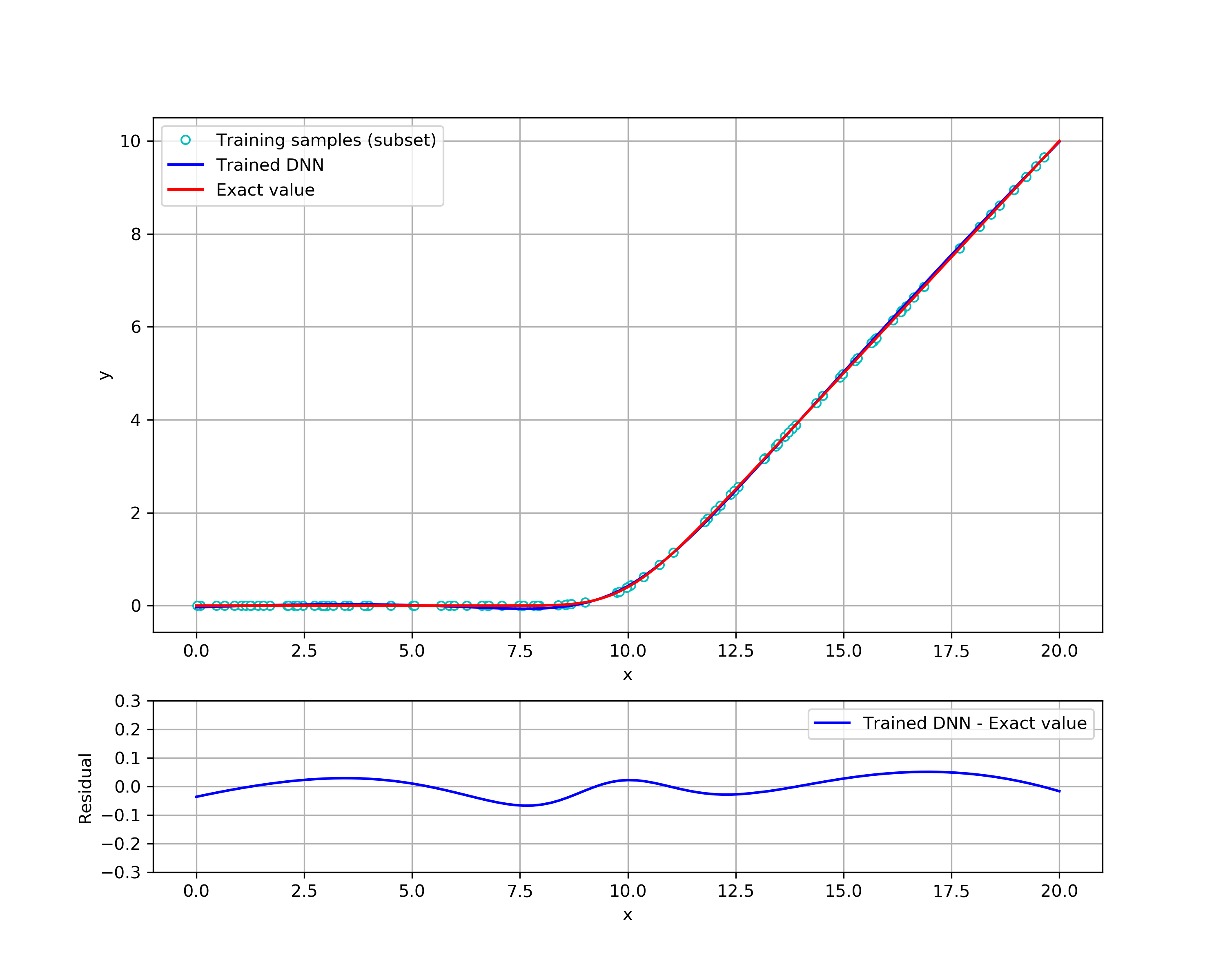} }}%
    \qquad
    \subfloat[\centering Gradient of trained DNN along with true derivative]{{\includegraphics[width=7cm]{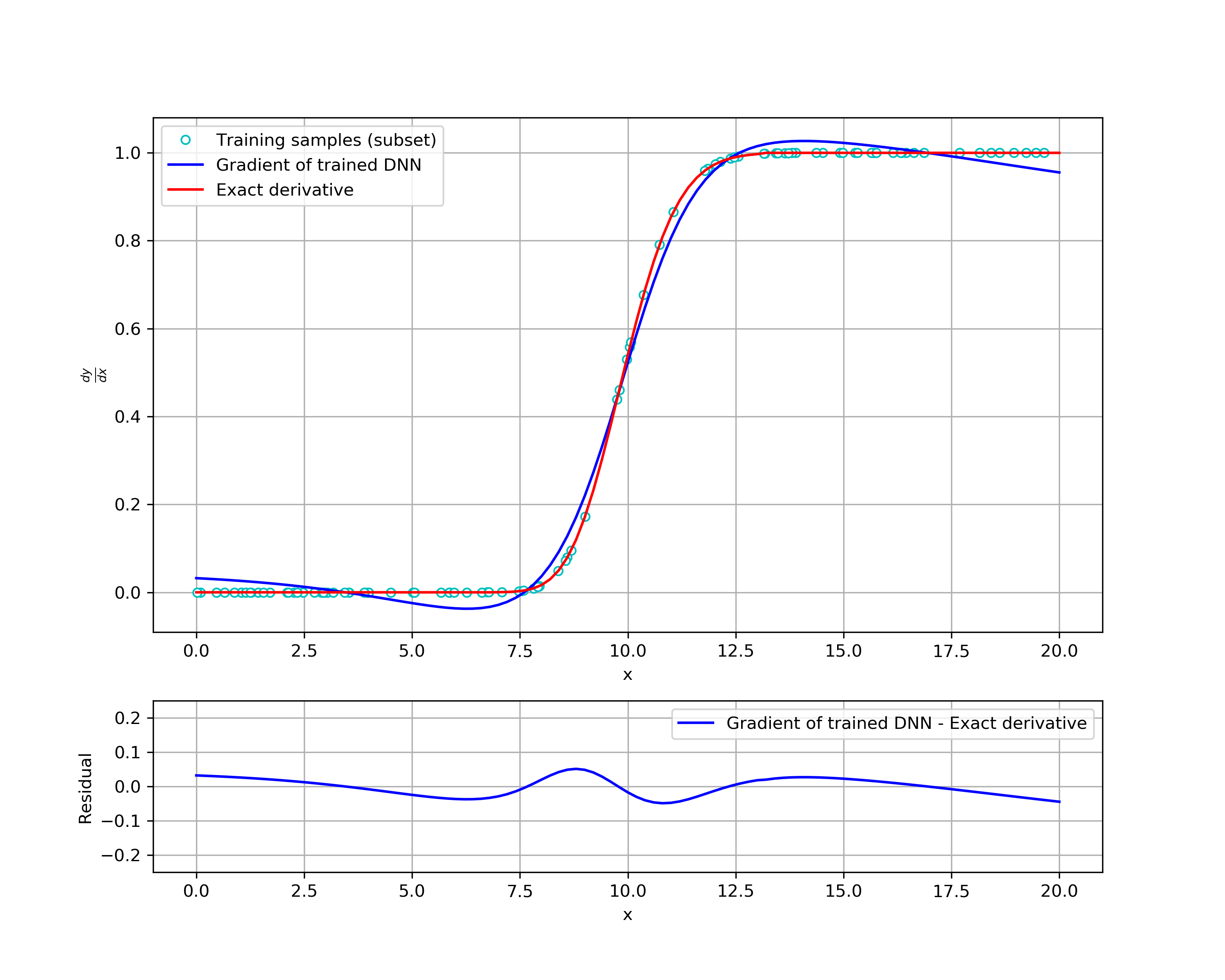} }}%
    \caption{Black-Scholes function approximation:  Differential Machine Learning without asymptotic treatment where results from DNN are in solid blue, 
    true values are in solid red, and the blue scatter points represent a subset of samples used in training chosen randomly for visualization purposes only.}%
    \label{fig:results_bs_dml}%
\end{figure}

\begin{figure}%
    \centering
    \subfloat[\centering Trained DNN along with true value]{{\includegraphics[width=7cm]{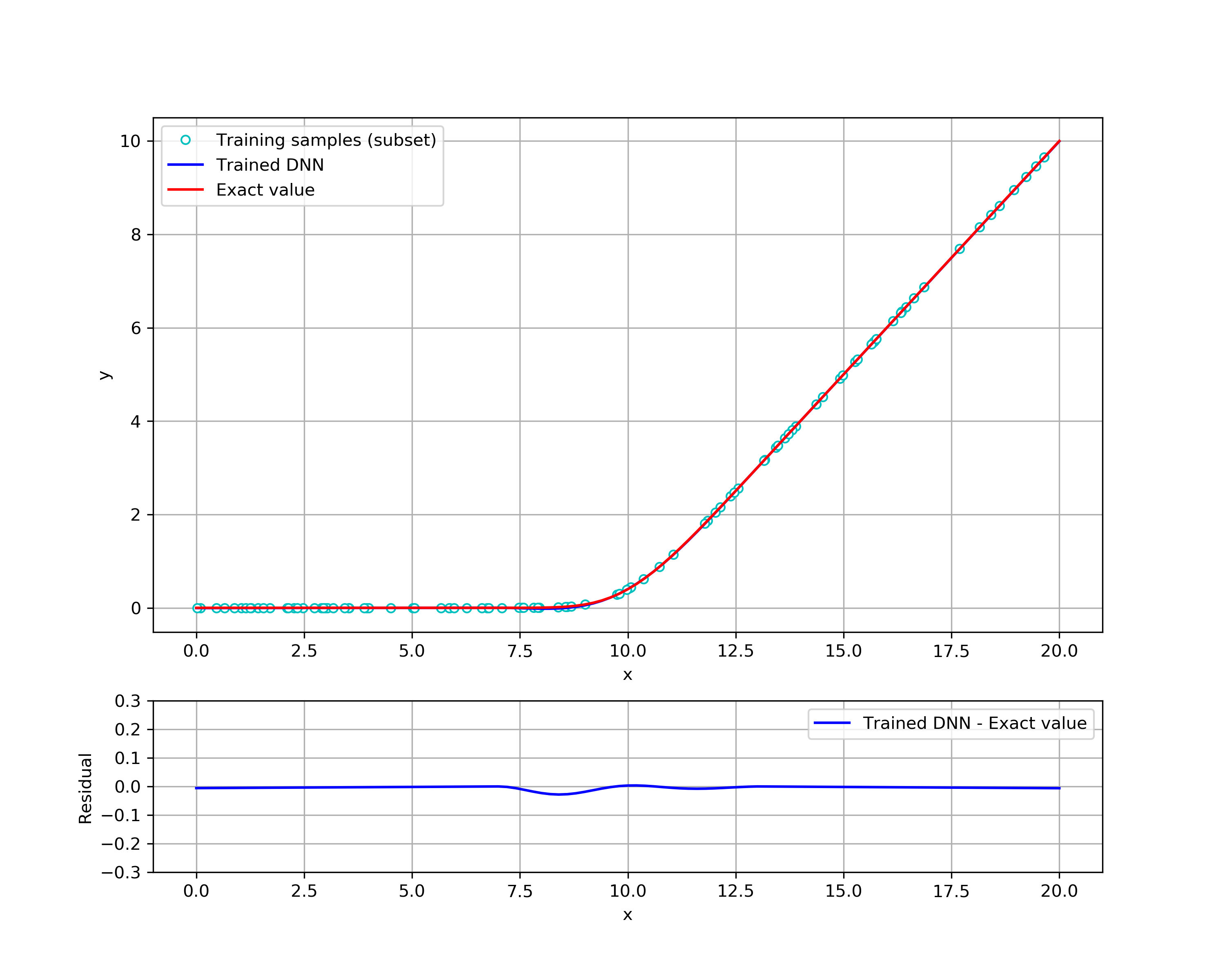} }}%
    \qquad
    \subfloat[\centering Gradient of trained DNN along with true derivative]{{\includegraphics[width=7cm]{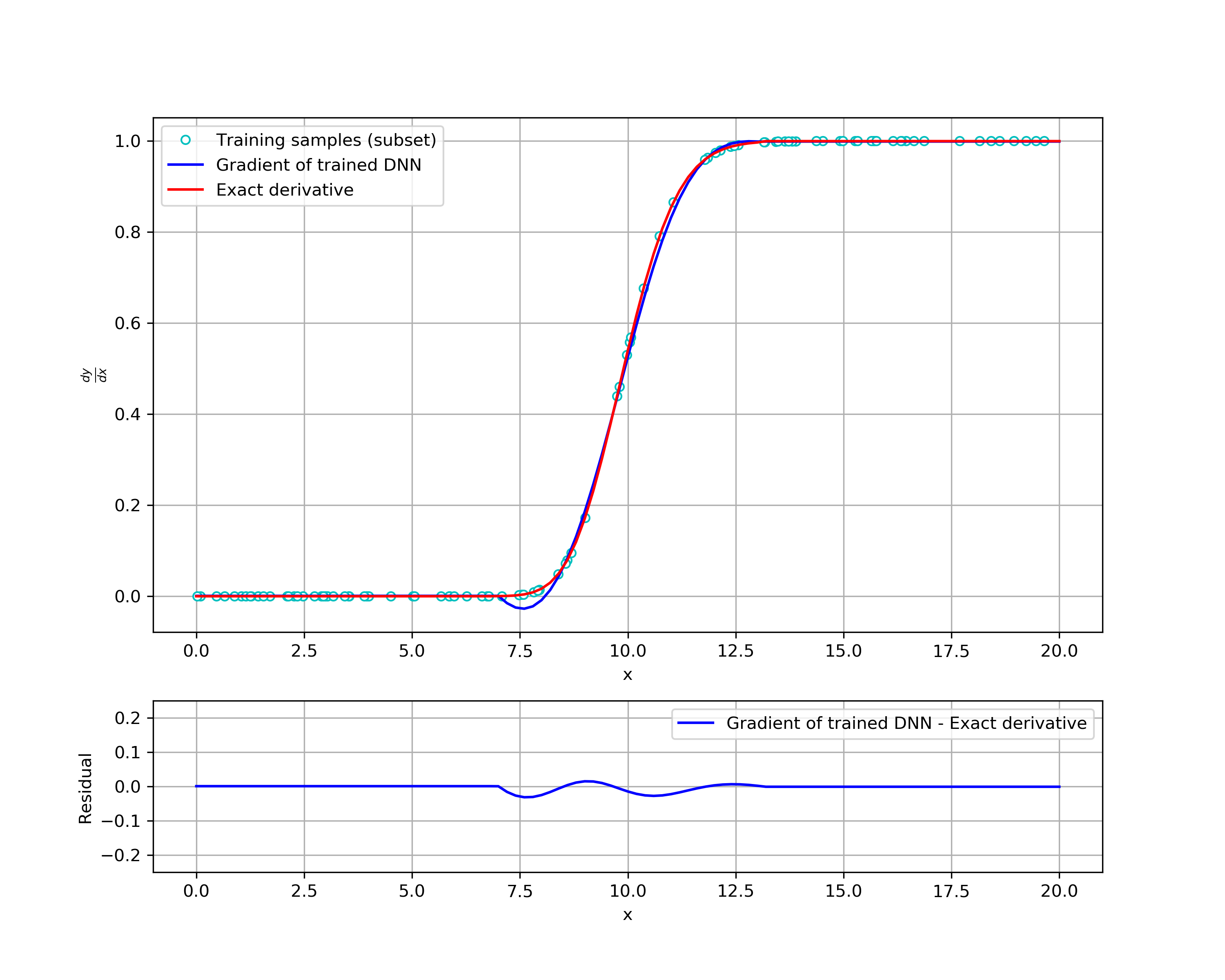} }}%
    \caption{Black-Scholes function approximation:  Differential Machine Learning with asymptotic treatment (trainable parameters) where results from DNN are in solid blue, 
    true values are in solid red, and the blue scatter points represent a subset of samples used in training chosen randomly for visualization purposes only.}%
    \label{fig:results_bs_dml_trainasymp}%
\end{figure}

\begin{figure}%
    \centering
    \subfloat[\centering Trained DNN along with true value]{{\includegraphics[width=7cm]{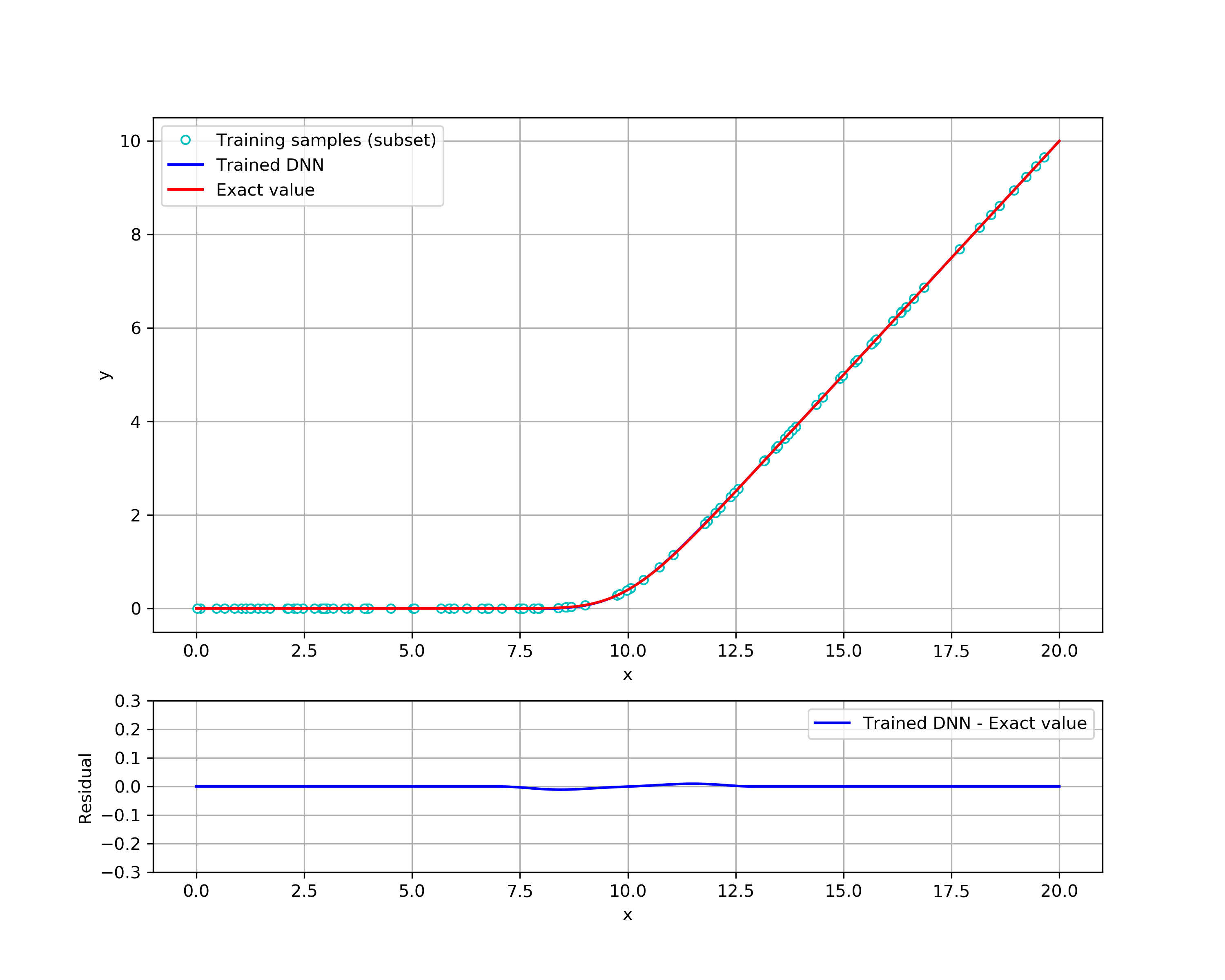} }}%
    \qquad
    \subfloat[\centering Gradient of trained DNN along with true derivative]{{\includegraphics[width=7cm]{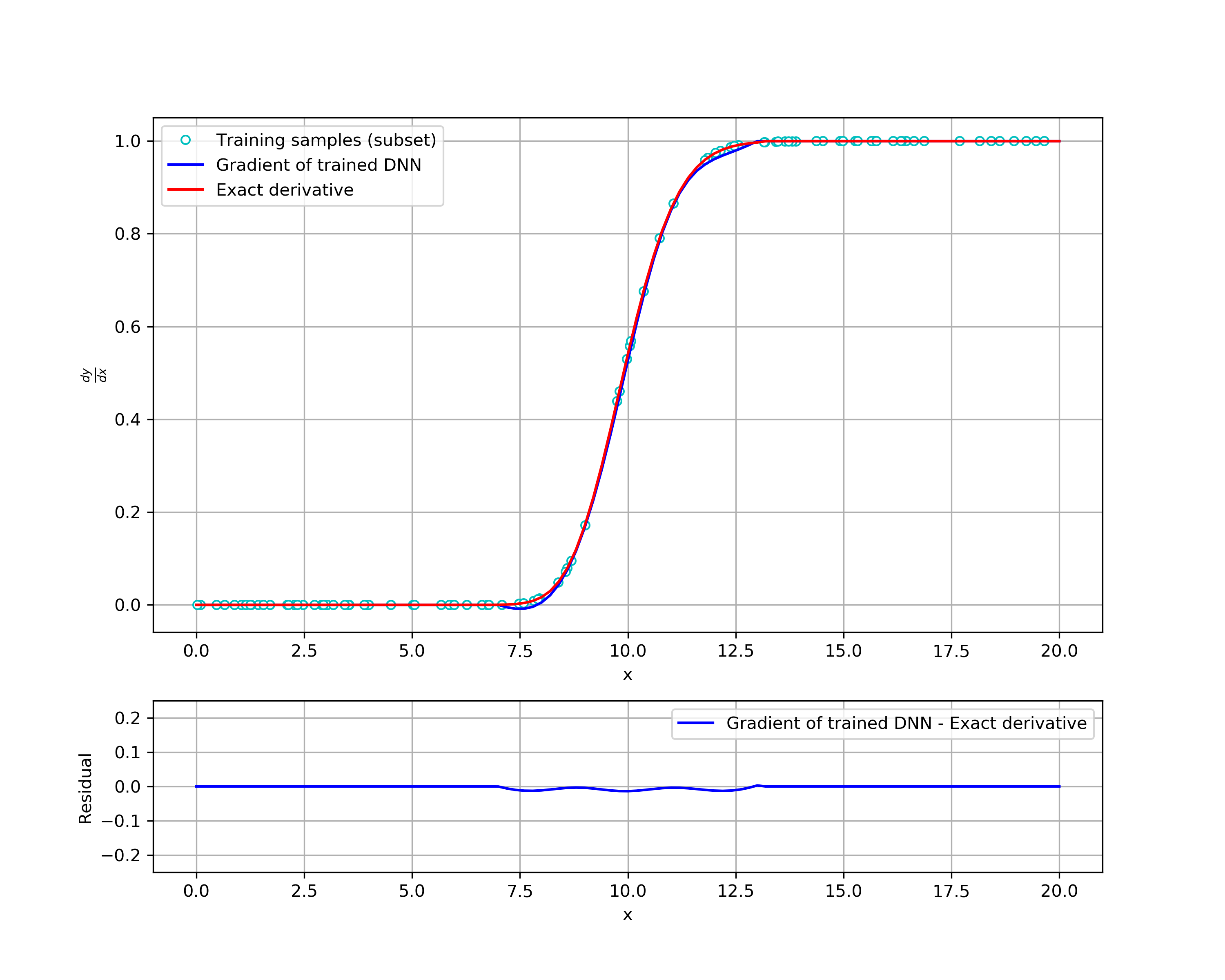} }}%
    \caption{Black-Scholes function approximation:  Differential Machine Learning with asymptotic treatment (fixed parameters) where results from DNN are in solid blue, 
    true values are in solid red, and the blue scatter points represent a subset of samples used in training chosen randomly for visualization purposes only.}%
    \label{fig:results_bs_dml_fixedasymp}%
\end{figure}

\begin{figure}%
    \centering
    \subfloat[\centering  Difference to true value]{{\includegraphics[width=7cm]{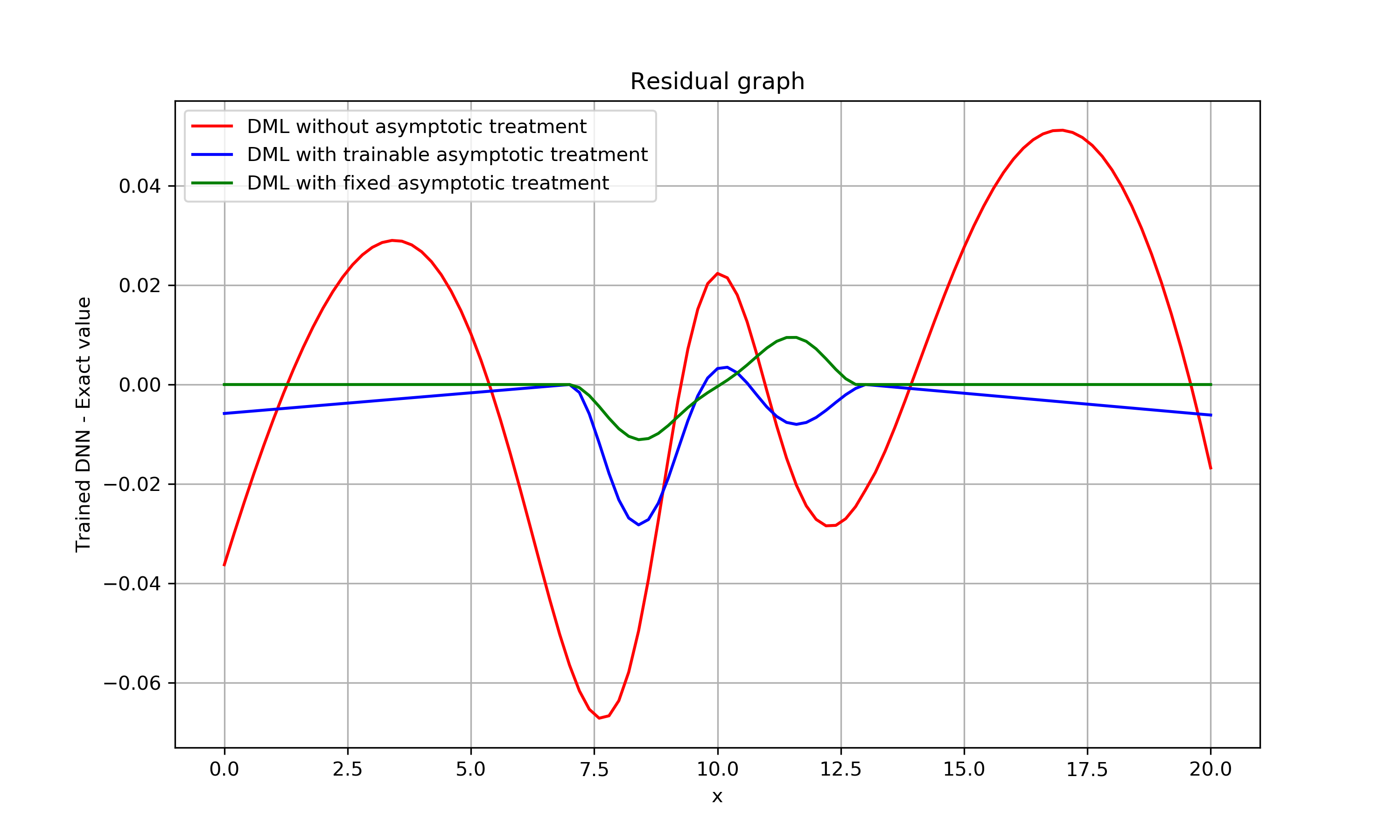} }}%
    \qquad
    \subfloat[\centering  Difference to true derivative]{{\includegraphics[width=7cm]{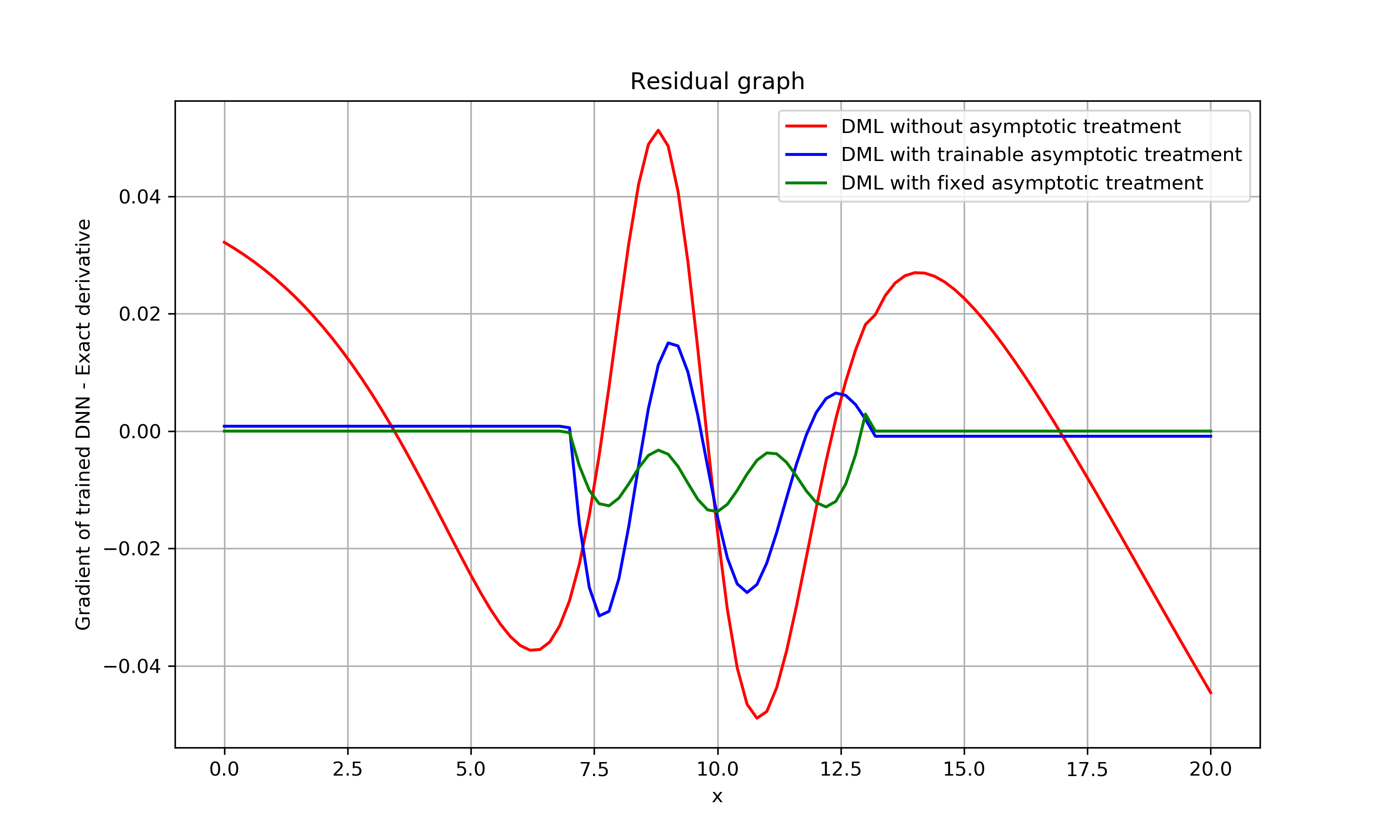} }}%
    \qquad
     \subfloat[\centering Training loss graph (DML loss)]{{\includegraphics[width=7cm]{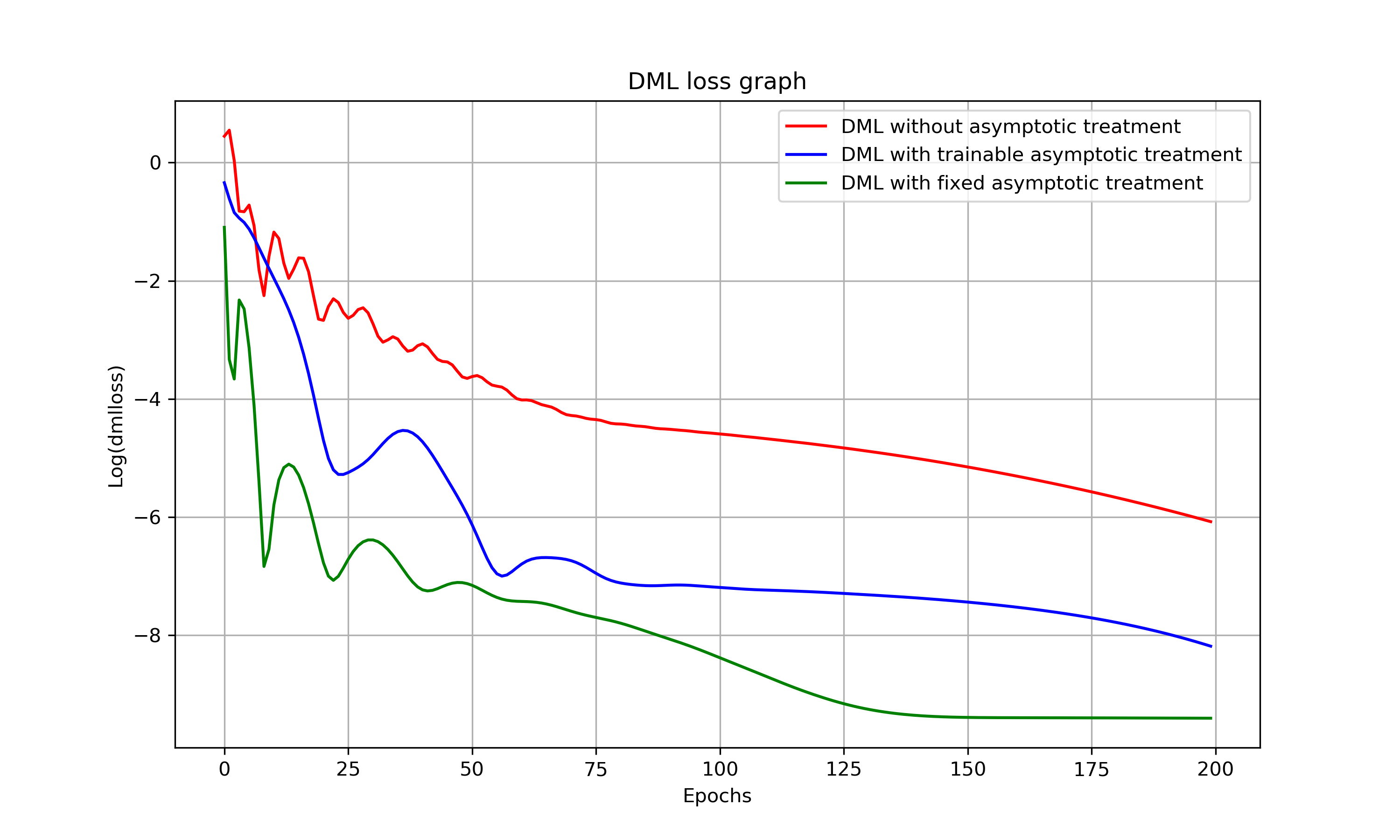} }}%
    \caption{Black-Scholes function approximation: Difference and loss graphs for DML without and with asymptotic treatment}%
    \label{fig:lossgraphs_bs}%
\end{figure}

\begin{figure}%
    \centering
    \subfloat[\centering Difference to true value]{{\includegraphics[width=7cm]{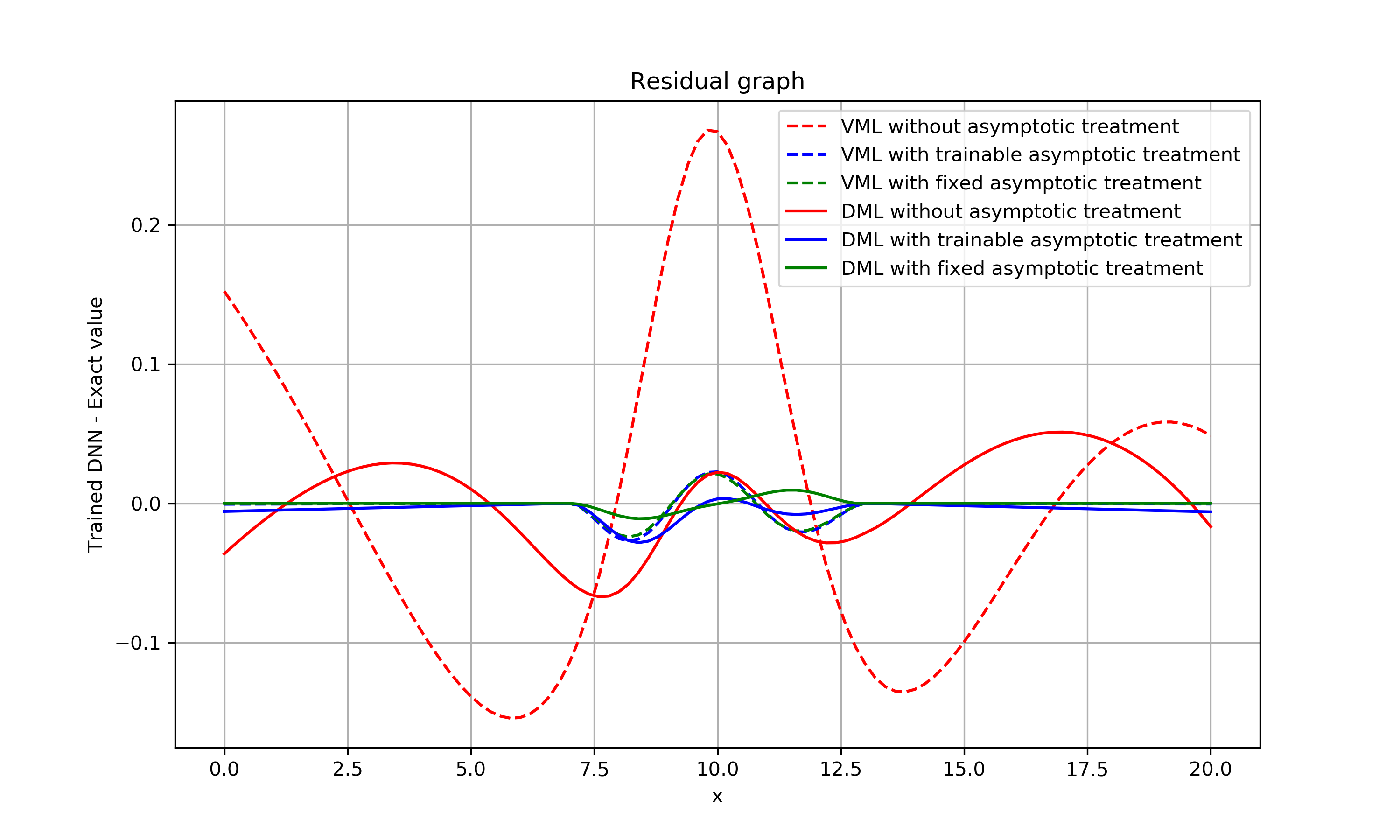} }}%
    \qquad
    \subfloat[\centering Difference to true derivative]{{\includegraphics[width=7cm]{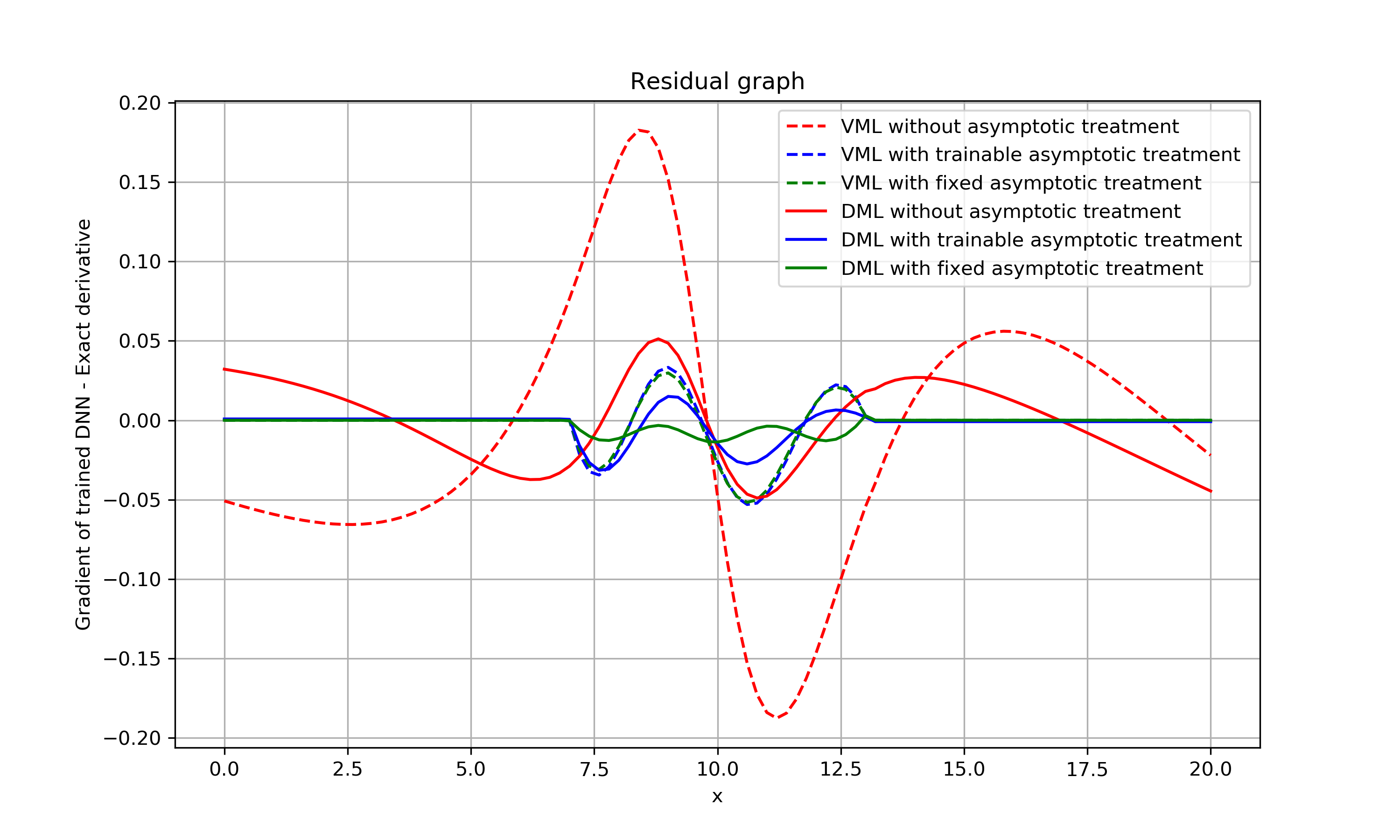} }}%
    \qquad
     \subfloat[\centering Training loss graph (VML loss)]{{\includegraphics[width=7cm]{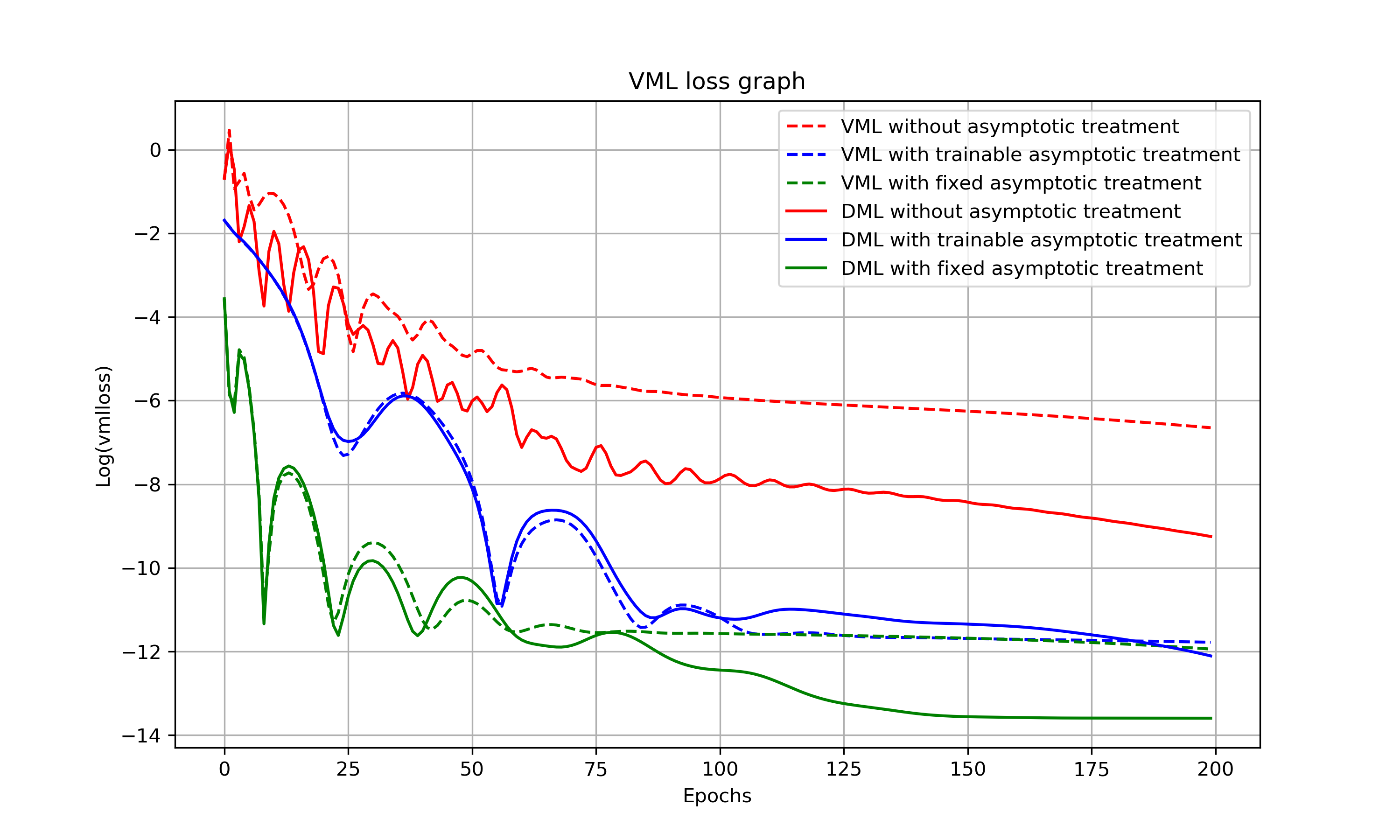} }}%
    \qquad
     \subfloat[\centering Training loss graph (DML loss)]{{\includegraphics[width=7cm]{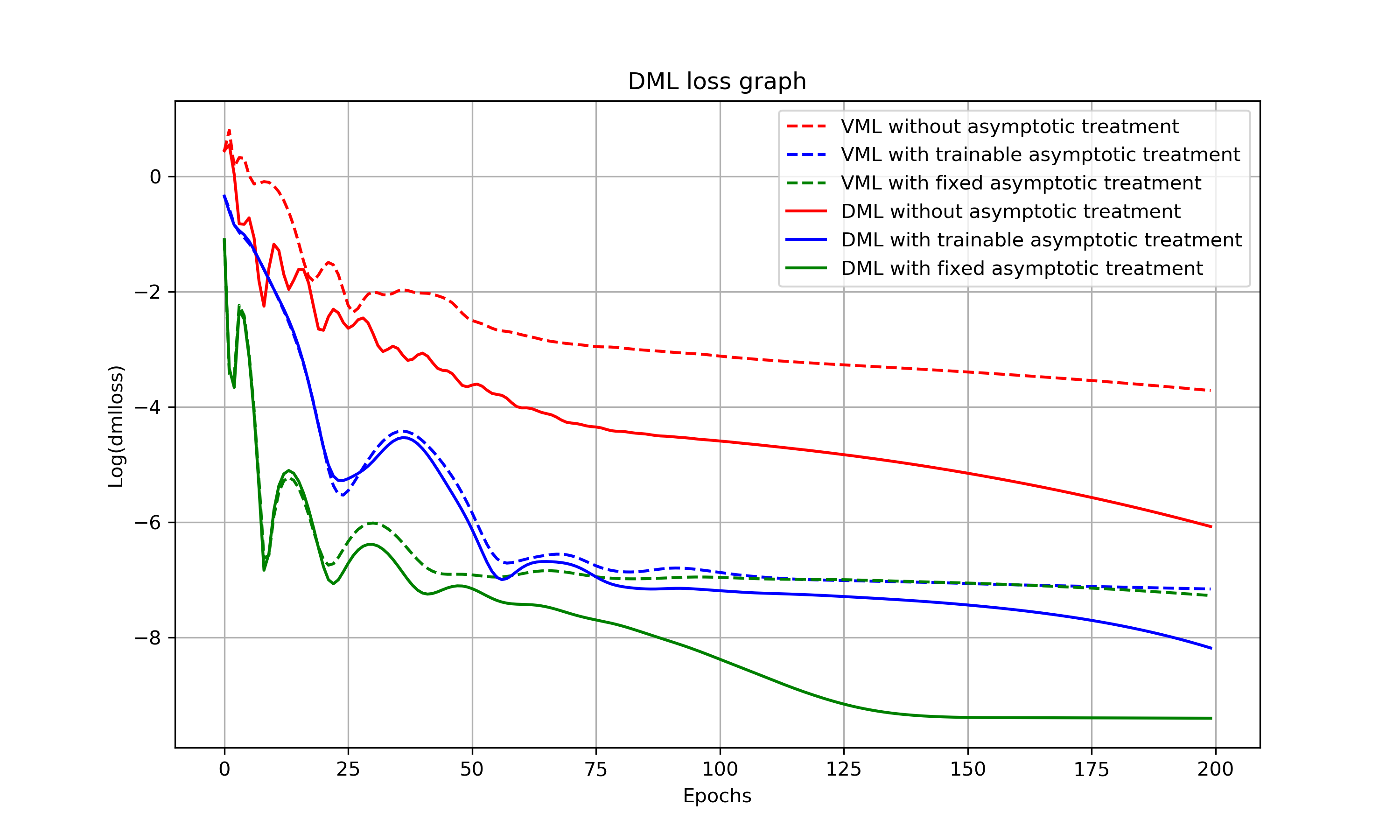} }}%
    \caption{Black-Scholes function approximation: Difference and loss graphs for DML and VML (without and with asymptotic treatment)}%
    \label{fig:lossgraphs_bs_all}%
\end{figure}

Proceeding similarly as in the previous section, we evaluate the performance of the trained DNNs from different methods on a 
test dataset. Fig.~\ref{fig:results_bs_vml} shows the results of the trained deep neural network using VML without asymptotic treatment,
 Fig.~\ref{fig:results_bs_vml_trainasymp} shows the results of the trained deep neural network using VML with asymptotic treatment
  (trainable parameters) and Fig.~\ref{fig:results_bs_vml_fixedasymp} shows the results of the trained deep neural network using VML 
  with asymptotic treatment (fixed parameters). Inspecting the difference as well as the loss graphs in Fig.~\ref{fig:lossgraphs_bs_vml}, 
  it is again evident that using VML in conjunction with asymptotic treatment leads to faster and accurate convergence as compared to VML 
  without any asymptotic treatment.  Fig.~\ref{fig:results_bs_dml} shows the results of the trained deep neural network using DML without asymptotic treatment,
 Fig.~\ref{fig:results_bs_dml_trainasymp} shows the results of the trained deep neural network using DML with asymptotic treatment
  (trainable parameters) and Fig.~\ref{fig:results_bs_dml_fixedasymp} shows the results of the trained deep neural network using DML 
  with asymptotic treatment (fixed parameters). Observing the difference as well as the loss graphs in Fig.~\ref{fig:lossgraphs_bs}, 
  we come to the same conclusion that DML with asymptotic treatment performs better than just 
  DML without any asymptotic treatment, again demonstrating the effectiveness of our methodology. 
  The scales for the graphs are kept fixed across different methods to facilitate easier comparison. 
  Fig.~\ref{fig:lossgraphs_bs_all} shows the comparison of DML and VML using difference and loss graphs.

\clearpage

\begin{figure}%
    \centering
    \subfloat[\centering  Difference to true value]{{\includegraphics[width=7cm]{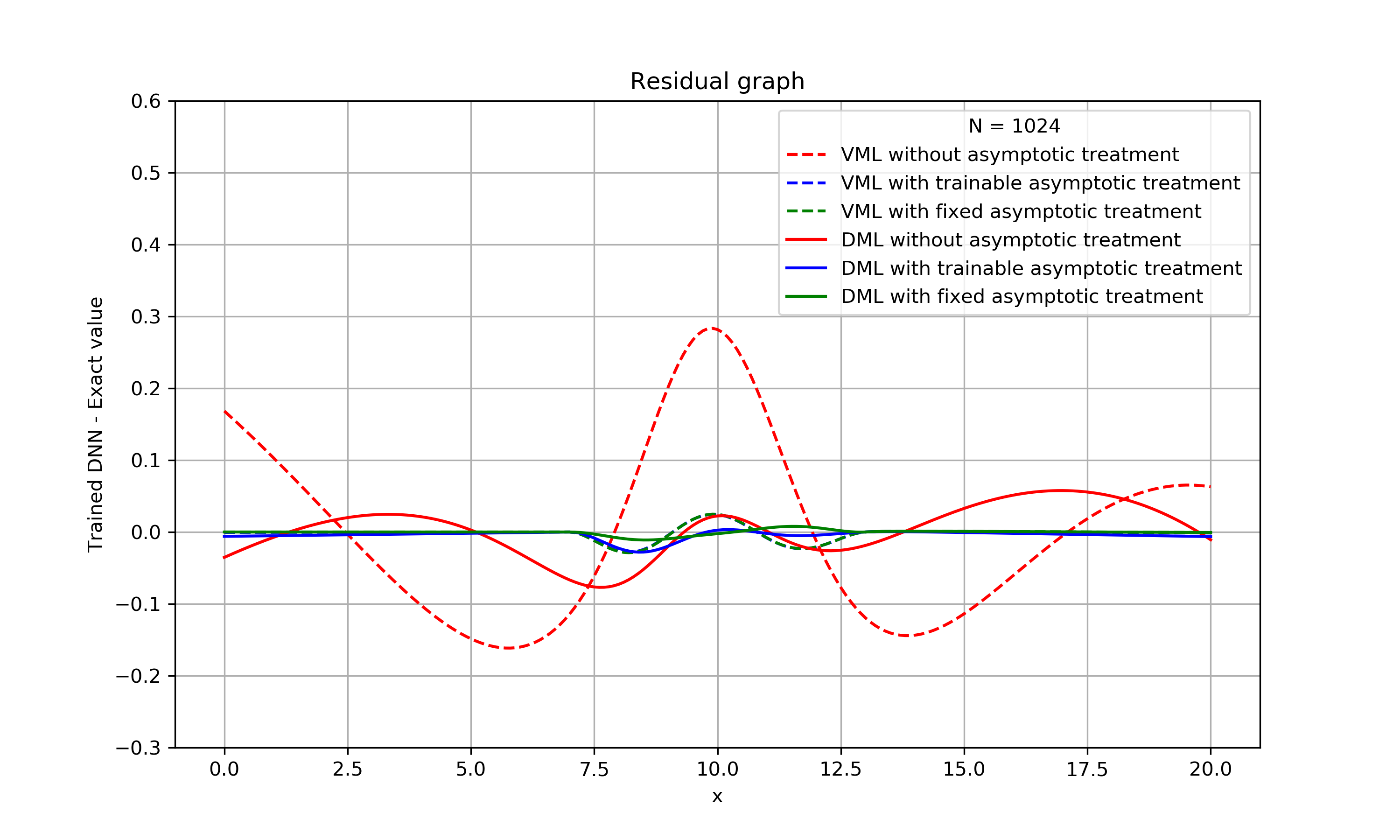} }}%
    \qquad
    \subfloat[\centering  Difference to true derivative]{{\includegraphics[width=7cm]{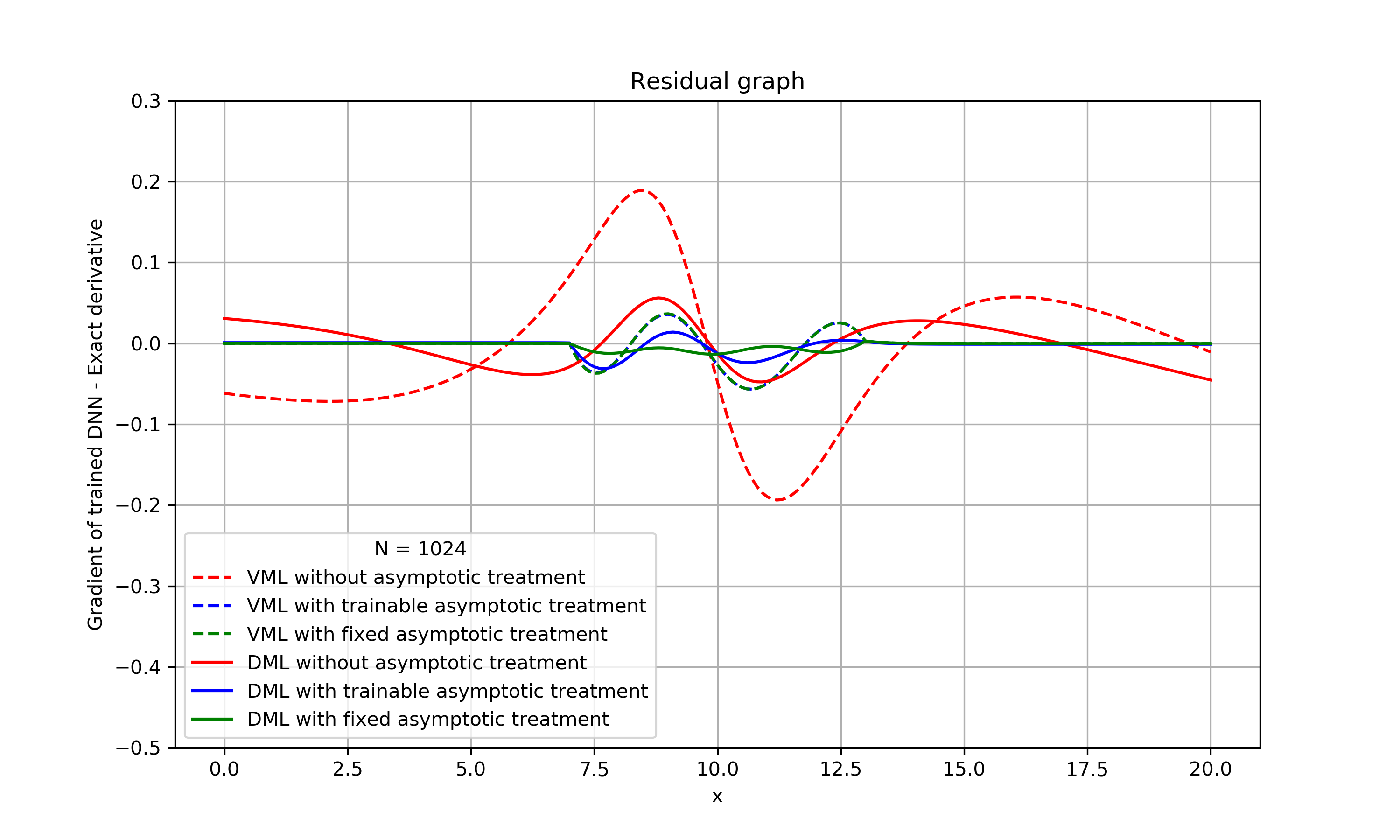} }}%
    \caption{Black-Scholes function approximation: Difference graphs for DML and VML (without and with asymptotic treatment) for sample size = $2^{10}$}%
    \label{fig:residual_regression_bs_1024_ex2}%
\end{figure}

\begin{figure}%
    \centering
    \subfloat[\centering  Difference to true value]{{\includegraphics[width=7cm]{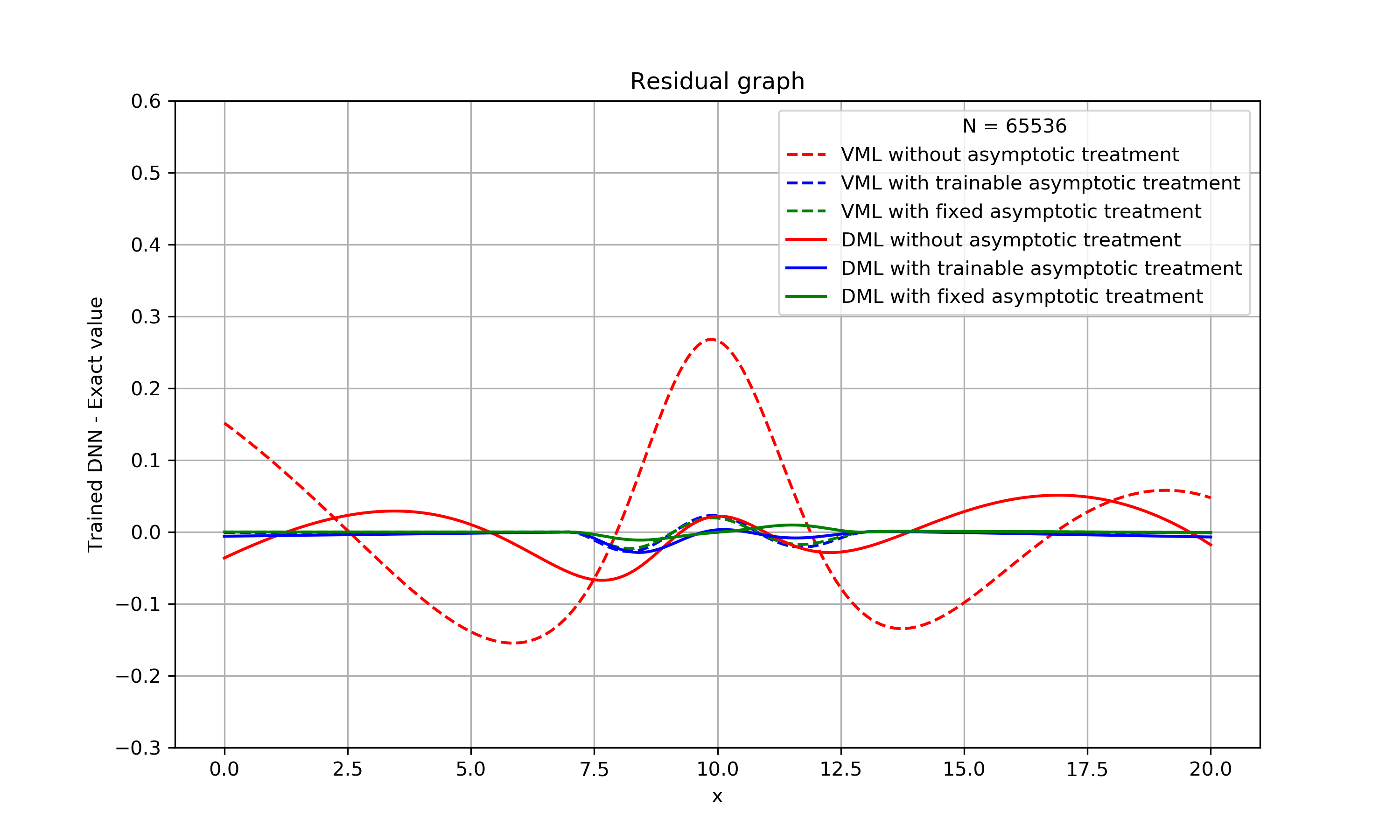} }}%
    \qquad
    \subfloat[\centering  Difference to true derivative]{{\includegraphics[width=7cm]{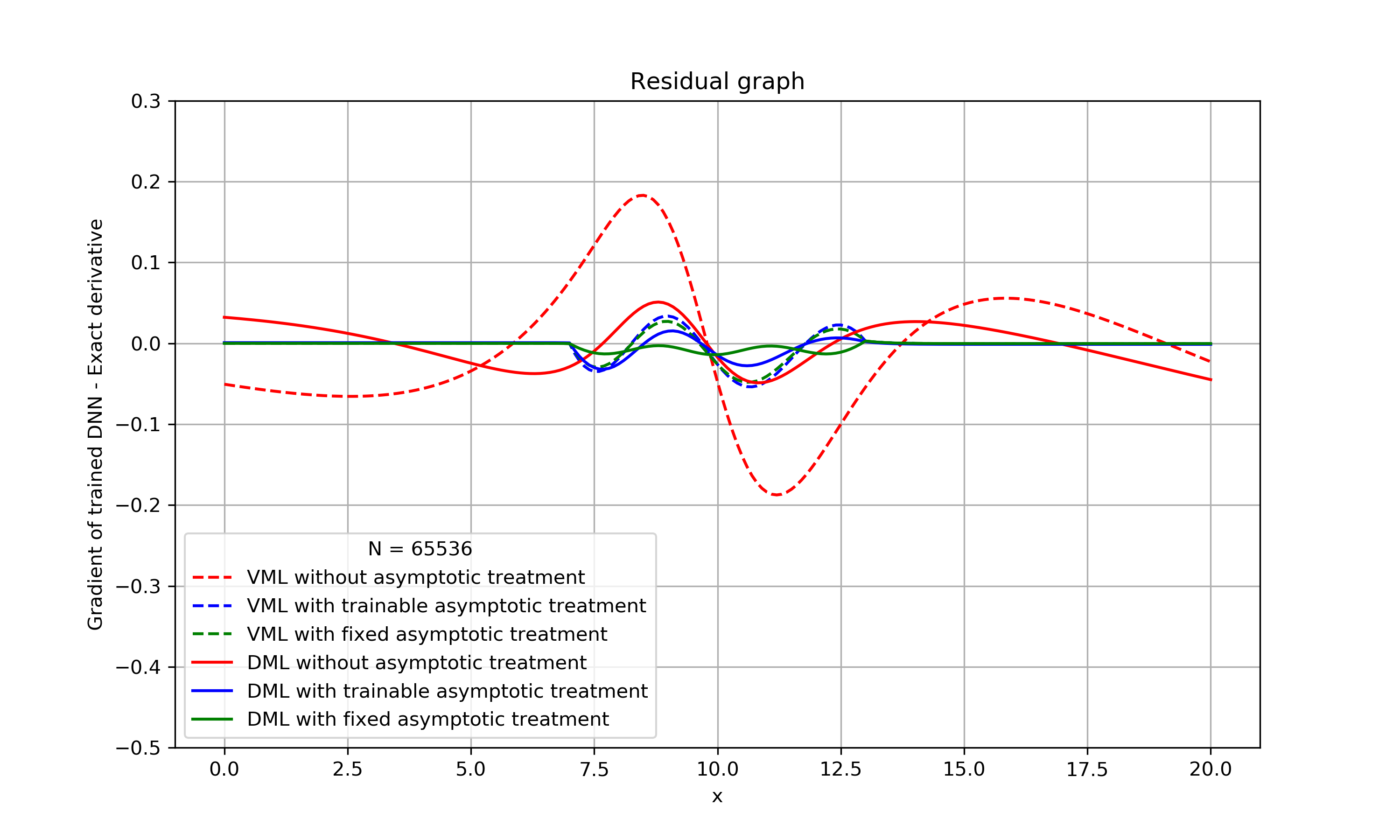} }}%
    \caption{Black-Scholes function approximation: Difference graphs for DML and VML (without and with asymptotic treatment) for sample size = $2^{16}$}%
    \label{fig:residual_regression_bs_65536_ex2}%
\end{figure}

We also performed tests on effects of sample size on results using VML and DML (without and with asymptotic treatment). 
Fig.~\ref{fig:residual_regression_bs_1024_ex2} and Fig.~\ref{fig:residual_regression_bs_65536_ex2} shows the results for a 
sample size of $2^{10}$ and $2^{16}$ respectively. 
It can be observed that the impact of sample size is minimal on the results of function approximation. There is some 
very slight improvement with increase in sample size, somewhat more pronounced in the loss function values. 

Lastly, we also performed tests with non-zero interest rates $r$ and observed very similar results. Adding $r$ as a drift and in 
the discounting will change the shape and the coefficients slightly but otherwise does not seem to impact the results and we thus 
omit the plots.

\clearpage
\section{Learning Black-Scholes formula through regression from simulated discounted payoff samples}
\label{sec:bsfunccondex}

Now that we have established the effectiveness of asymptotic treatment for approximating functions including Black-Scholes with faster and accurate convergence,
 we move on to the next natural step of learning the conditional expectation $\mathbb E[Y|X]$ given samples of initial or intermediate states $x$, 
 values $y$, and sample-wise derivatives $\frac{dy}{dx}$. 
 In the particular case of call option considered in the previous section, we use the spot price as conditioning $X$ 
 and the discounted samples of the payoff as $Y$. The conditional expectation function that we are after will
 represent the net present value (NPV) and will be approximated by the output of our neural network respective 
 function.  
 We simulate 10,000 stock price trajectories - here only prices $S_t$ and $S_T$ at intermediate time and maturity - under the Black-Scholes model, 
 with log-Forward Euler (which leads to exact simulation) with one time-step from $t$ to $T$, and record the 
 discounted call option payoff $\tilde{C}_T=e^{-(T-t)} (S_T-K)^+$.  
 We use $X=S_t$ and $Y=\tilde{C}_T$.   
In this test, we use the same setting as in the last section: intermediate time $t=1$ year, 
maturity $T = 2$ years, volatility $\sigma = 0.1$, short rate $r = 0\%$ and strike $K = 10$. 
We follow the same procedure as in the preceding section for training the neural network on the 
simulated dataset and evaluate their performance on a test set. \\

\begin{figure}%
    \centering
    \subfloat[\centering Trained DNN along with true value]{{\includegraphics[width=7cm]{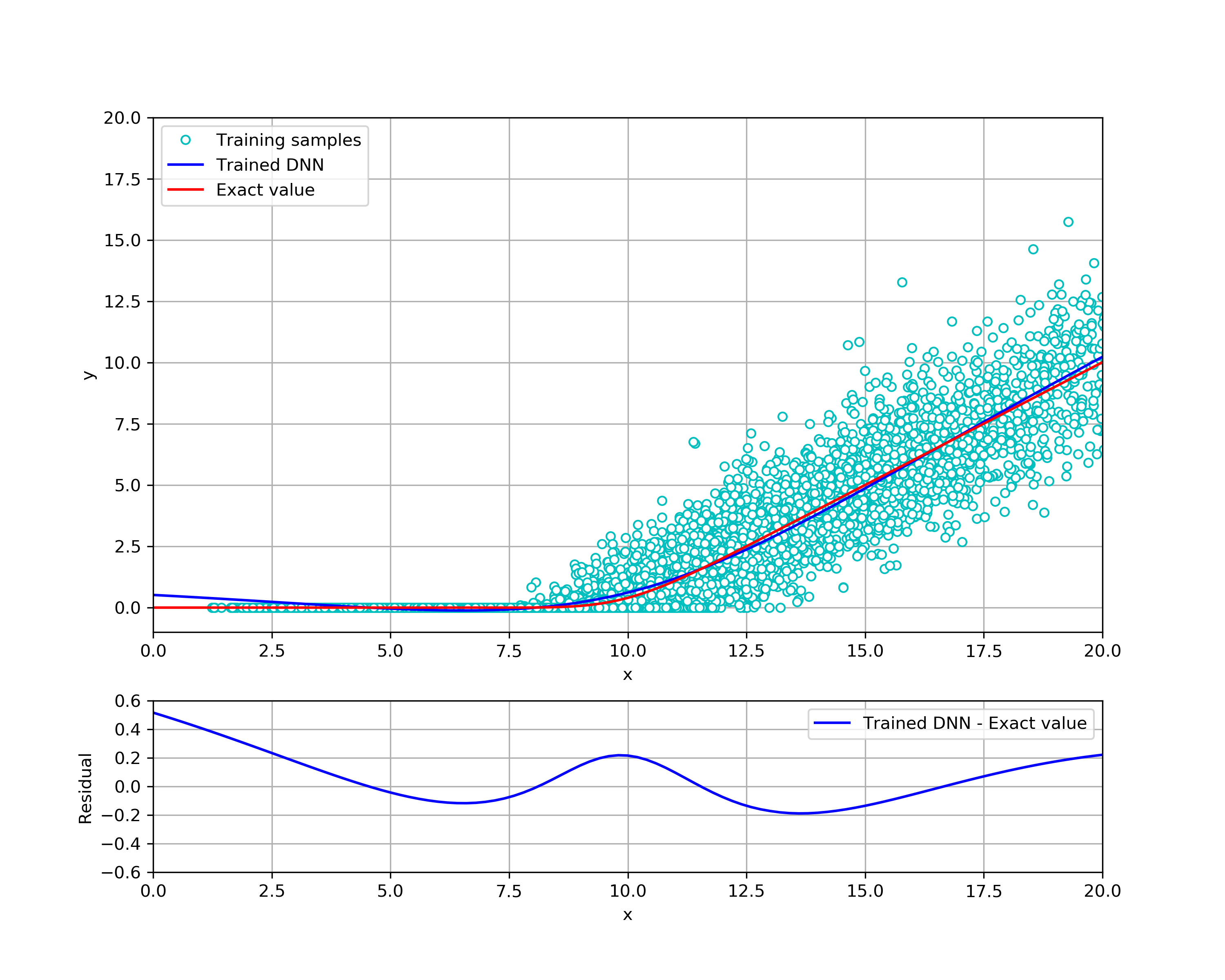} }}%
    \qquad
    \subfloat[\centering Gradient of trained DNN along with true derivative]{{\includegraphics[width=7cm]{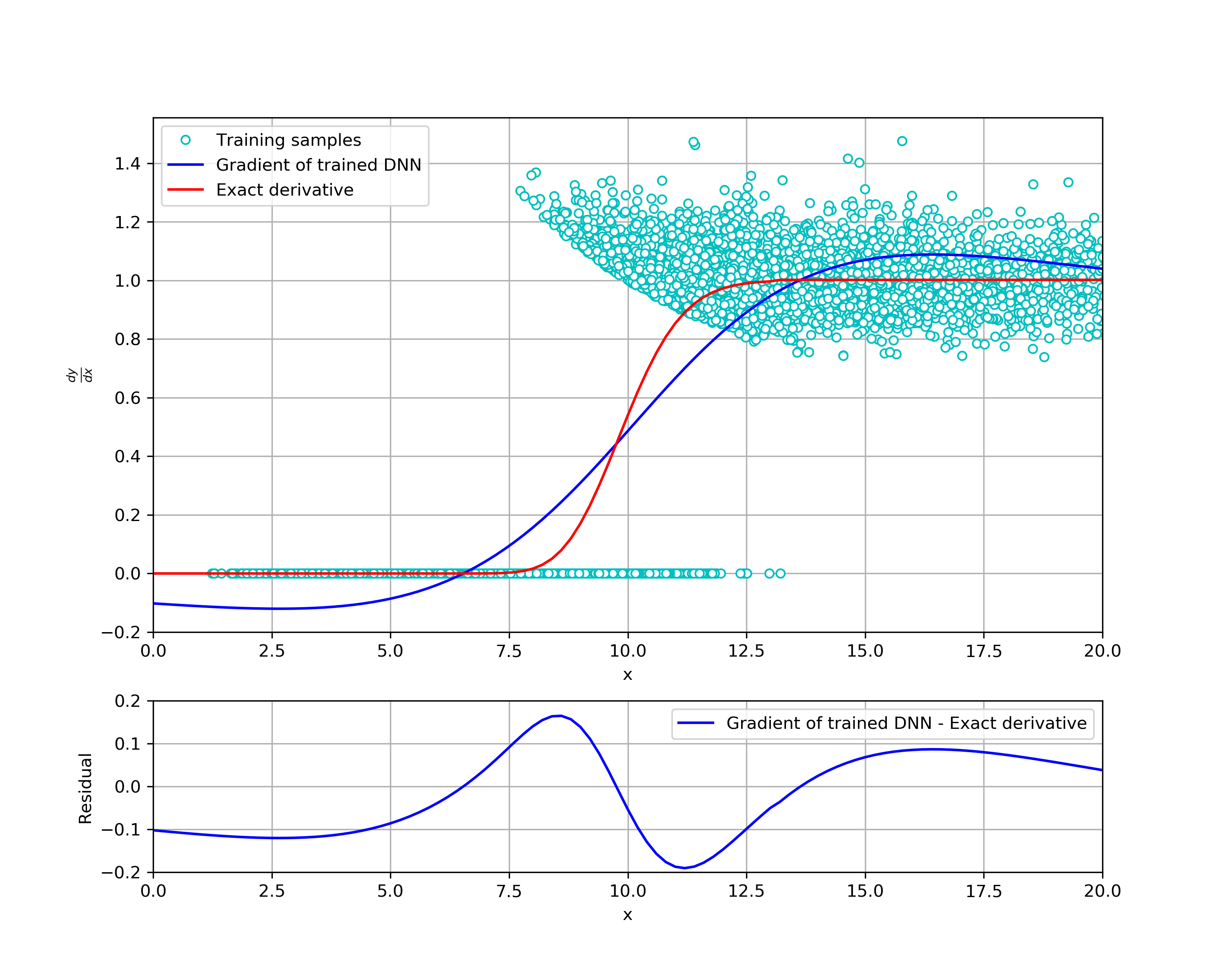} }}%
    \caption{Black-Scholes model regression: Vanilla Machine Learning without asymptotic treatment where the scatter points represent the samples used in training.}%
    \label{fig:results_regression_bs_vml}%
\end{figure}

\begin{figure}%
    \centering
    \subfloat[\centering Trained DNN along with true value]{{\includegraphics[width=7cm]{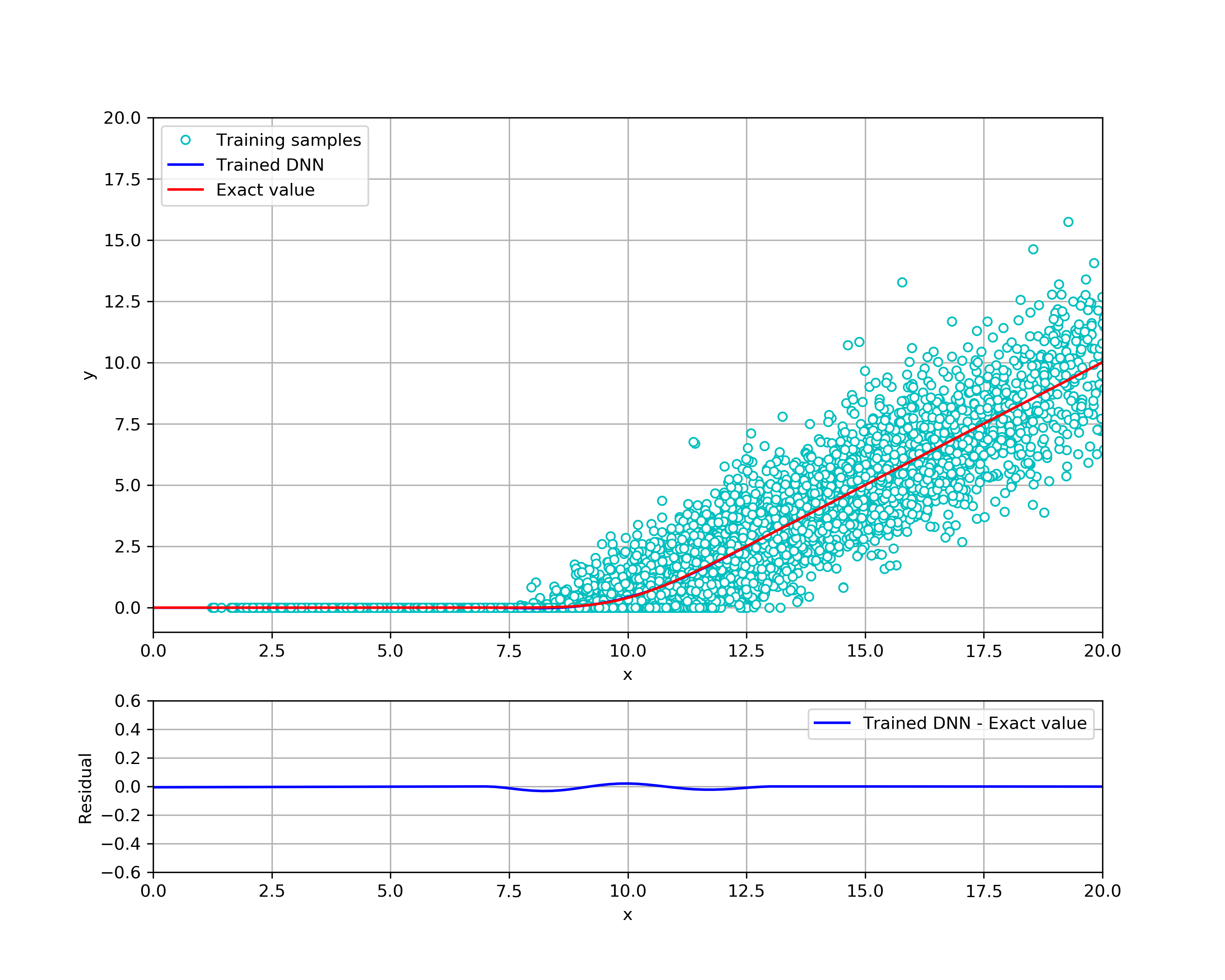} }}%
    \qquad
    \subfloat[\centering Gradient of trained DNN along with true derivative]{{\includegraphics[width=7cm]{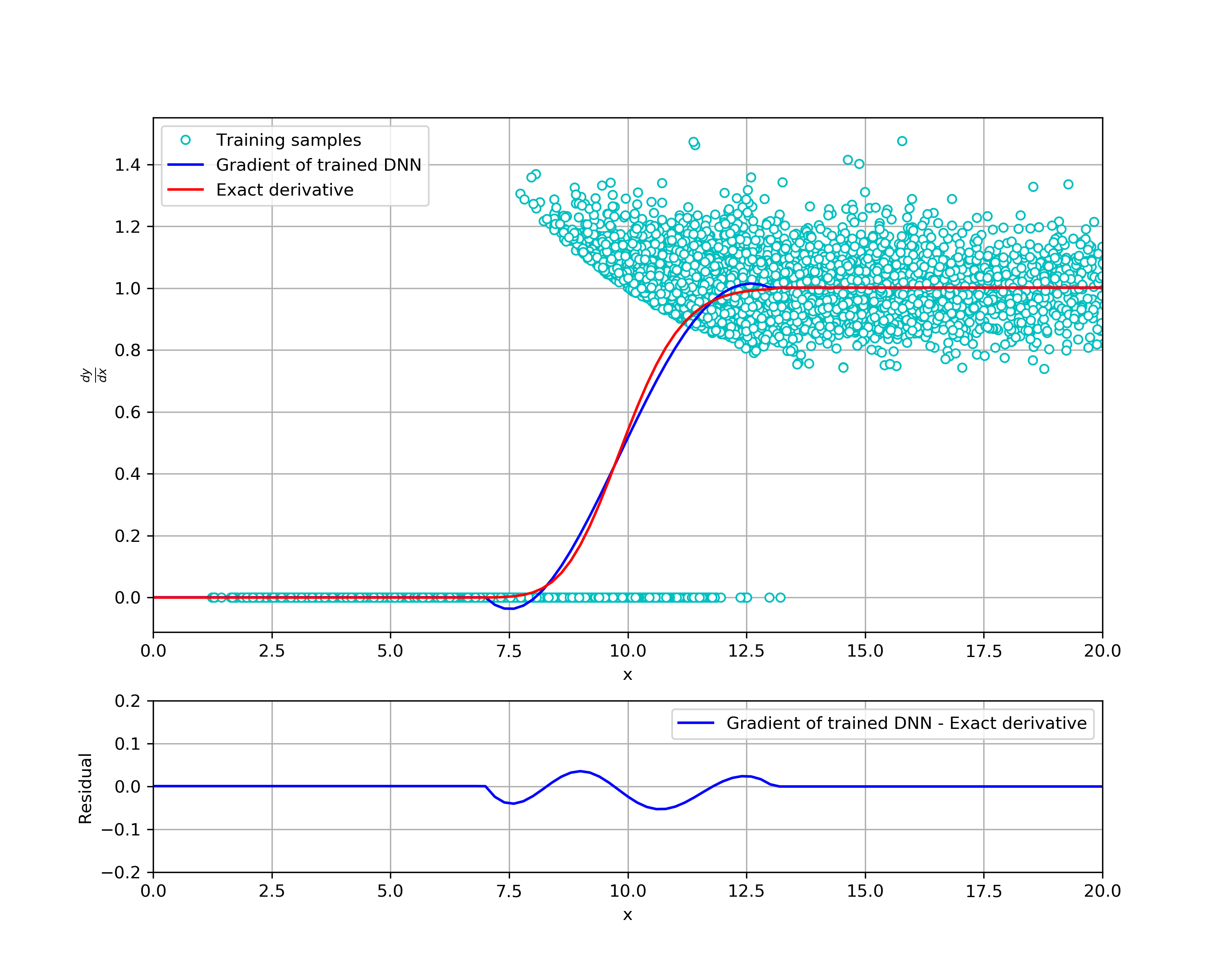} }}%
    \caption{Black-Scholes model regression:  Vanilla Machine Learning with asymptotic treatment (trainable parameters) where the scatter points represent the samples used in training.}%
    \label{fig:results_regression_bs_vml_trainasymp}%
\end{figure}

\begin{figure}%
    \centering
    \subfloat[\centering Trained DNN along with true value]{{\includegraphics[width=7cm]{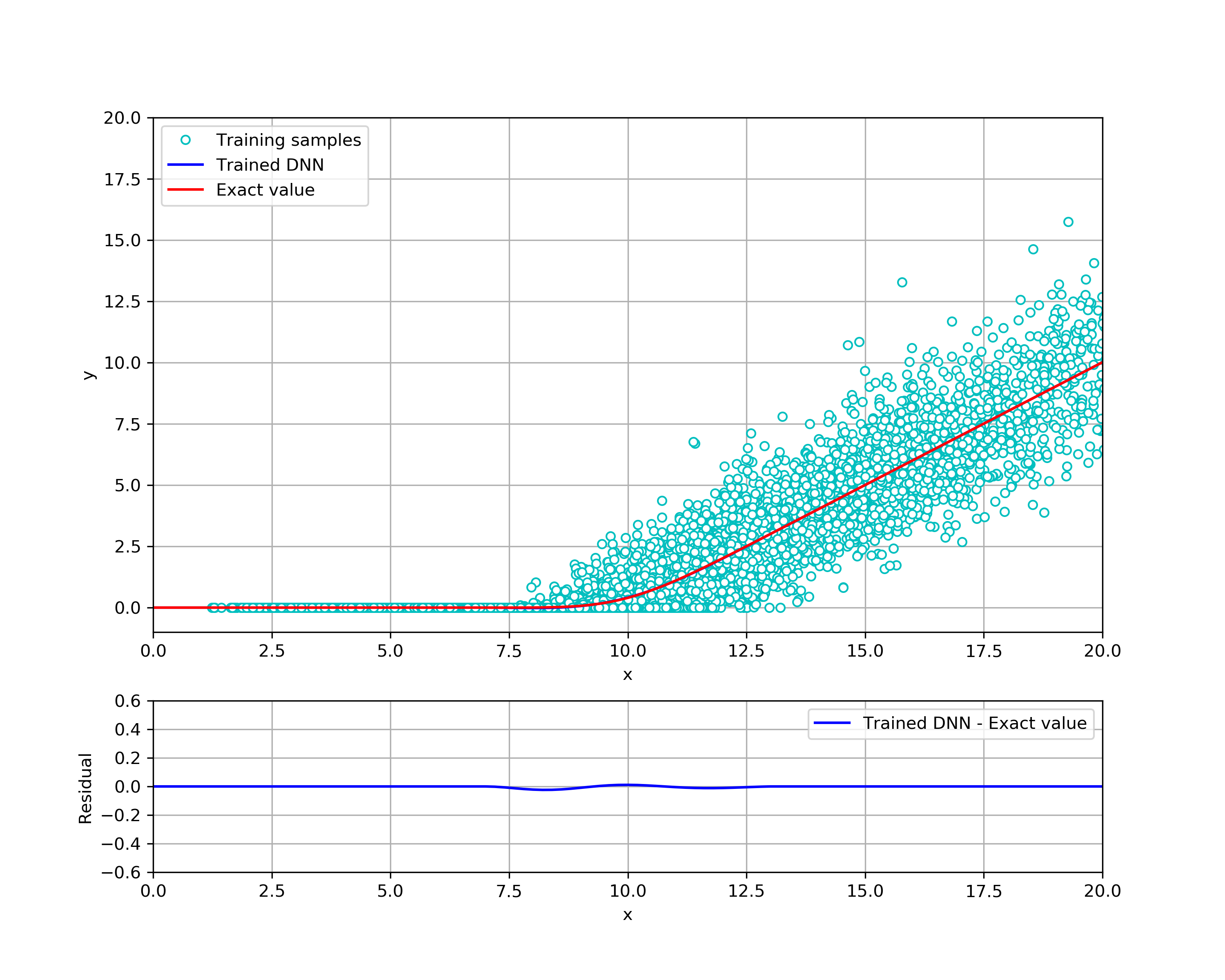} }}%
    \qquad
    \subfloat[\centering Gradient of trained DNN along with true derivative]{{\includegraphics[width=7cm]{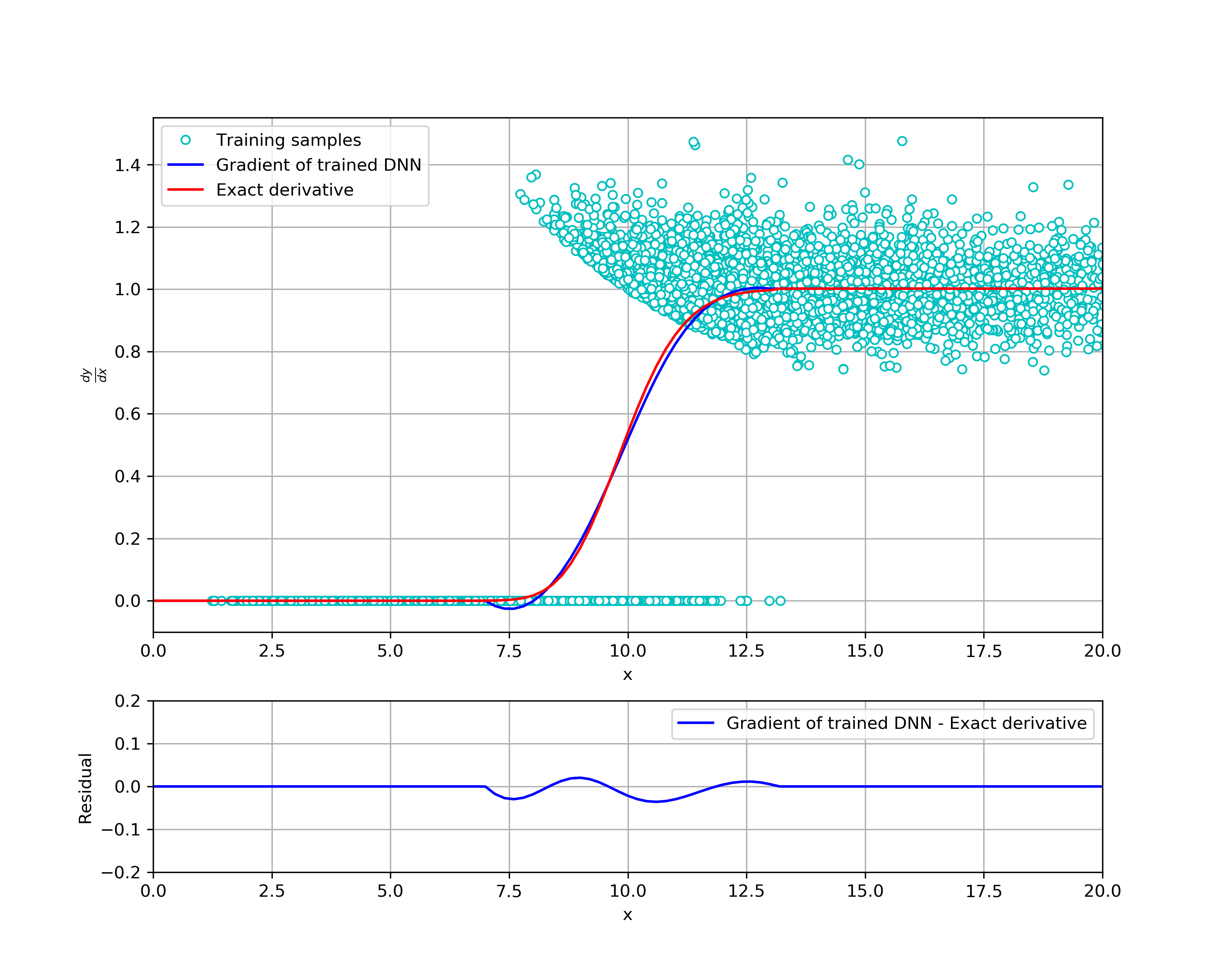} }}%
    \caption{Black-Scholes model regression: Vanilla Machine Learning with asymptotic treatment (fixed parameters) where the scatter points represent the samples used in training.}%
    \label{fig:results_regression_bs_vml_fixedasymp}%
\end{figure}

\begin{figure}%
    \centering
    \subfloat[\centering  Difference to true value]{{\includegraphics[width=7cm]{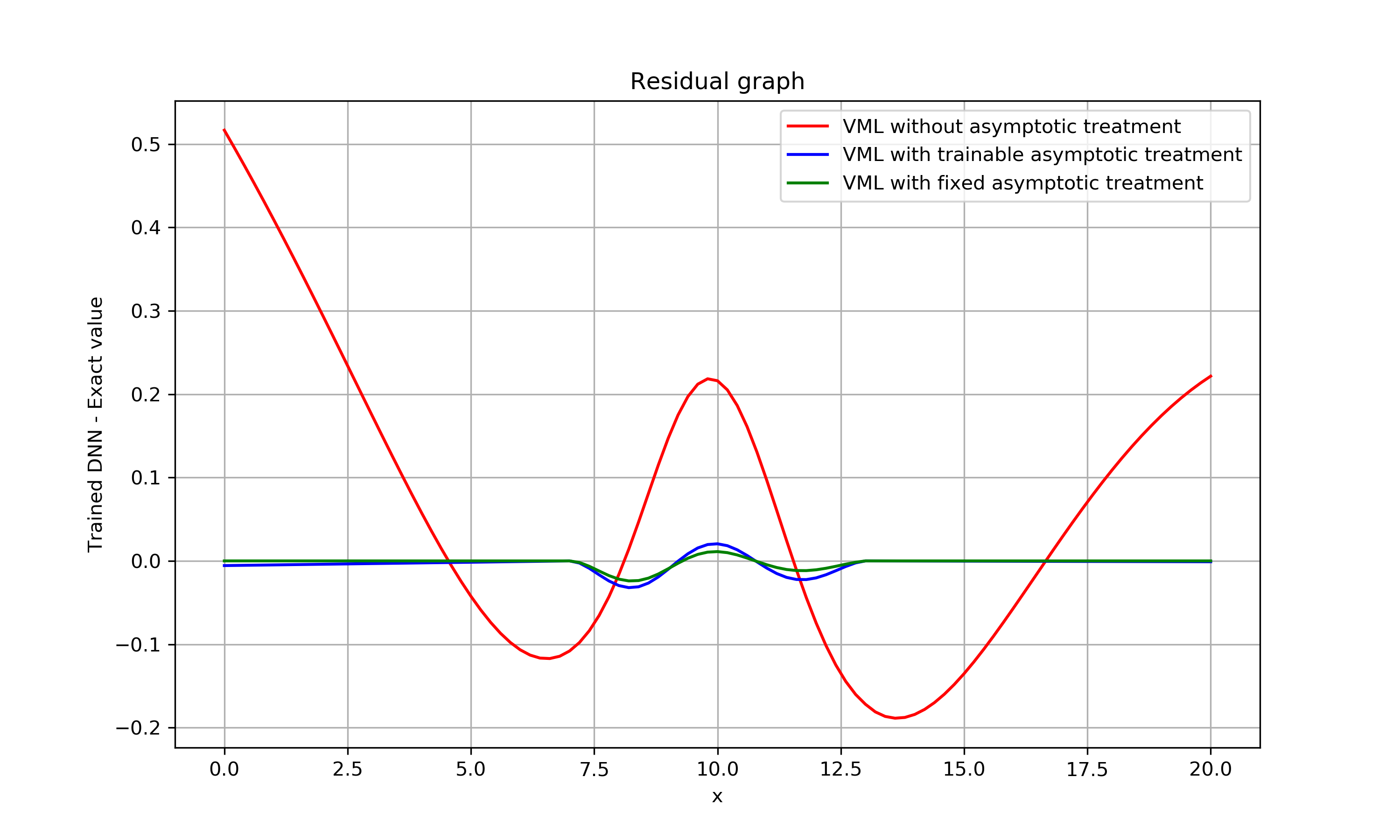} }}%
    \qquad
    \subfloat[\centering  Difference to true derivative]{{\includegraphics[width=7cm]{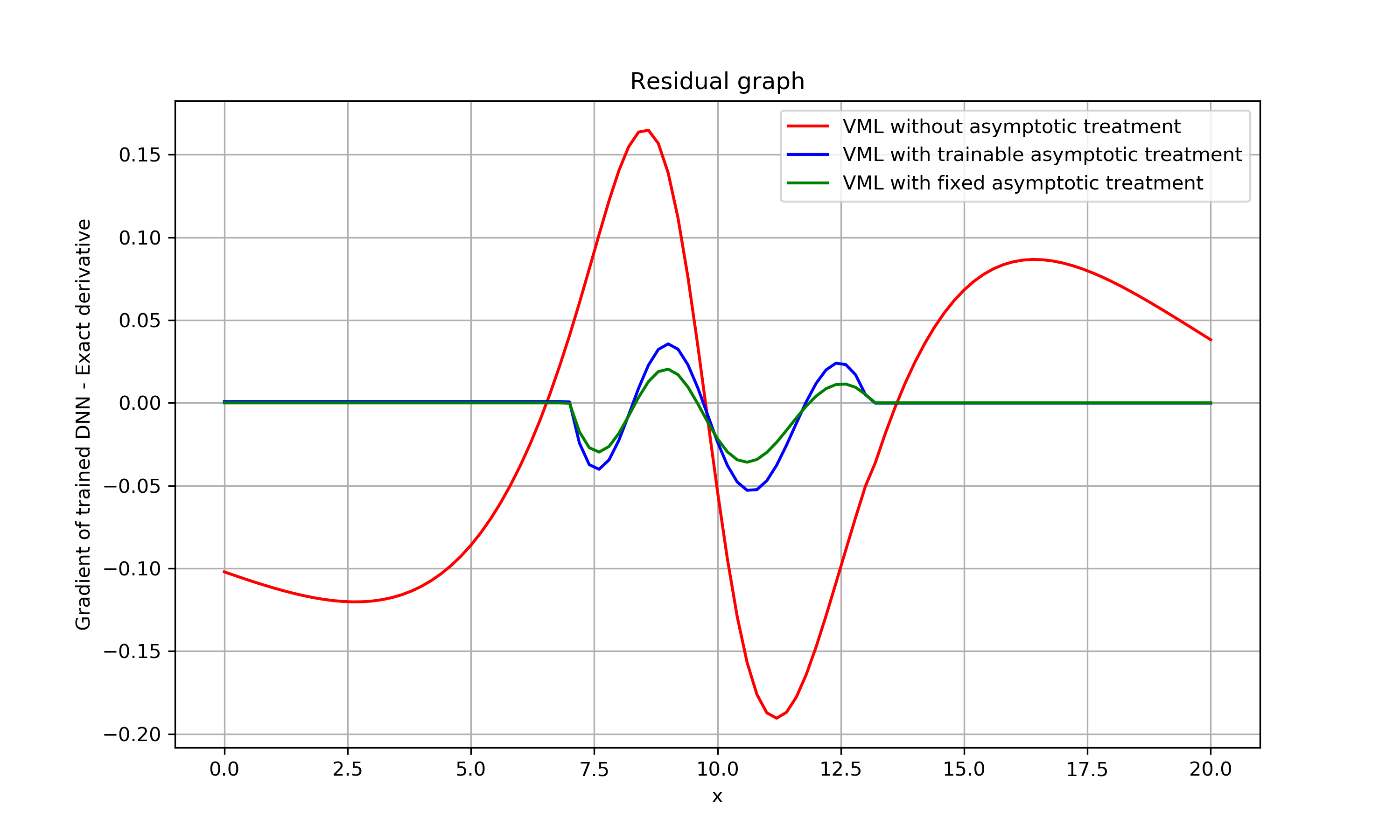} }}%
    \qquad
     \subfloat[\centering  Training loss graph (VML loss)]{{\includegraphics[width=7cm]{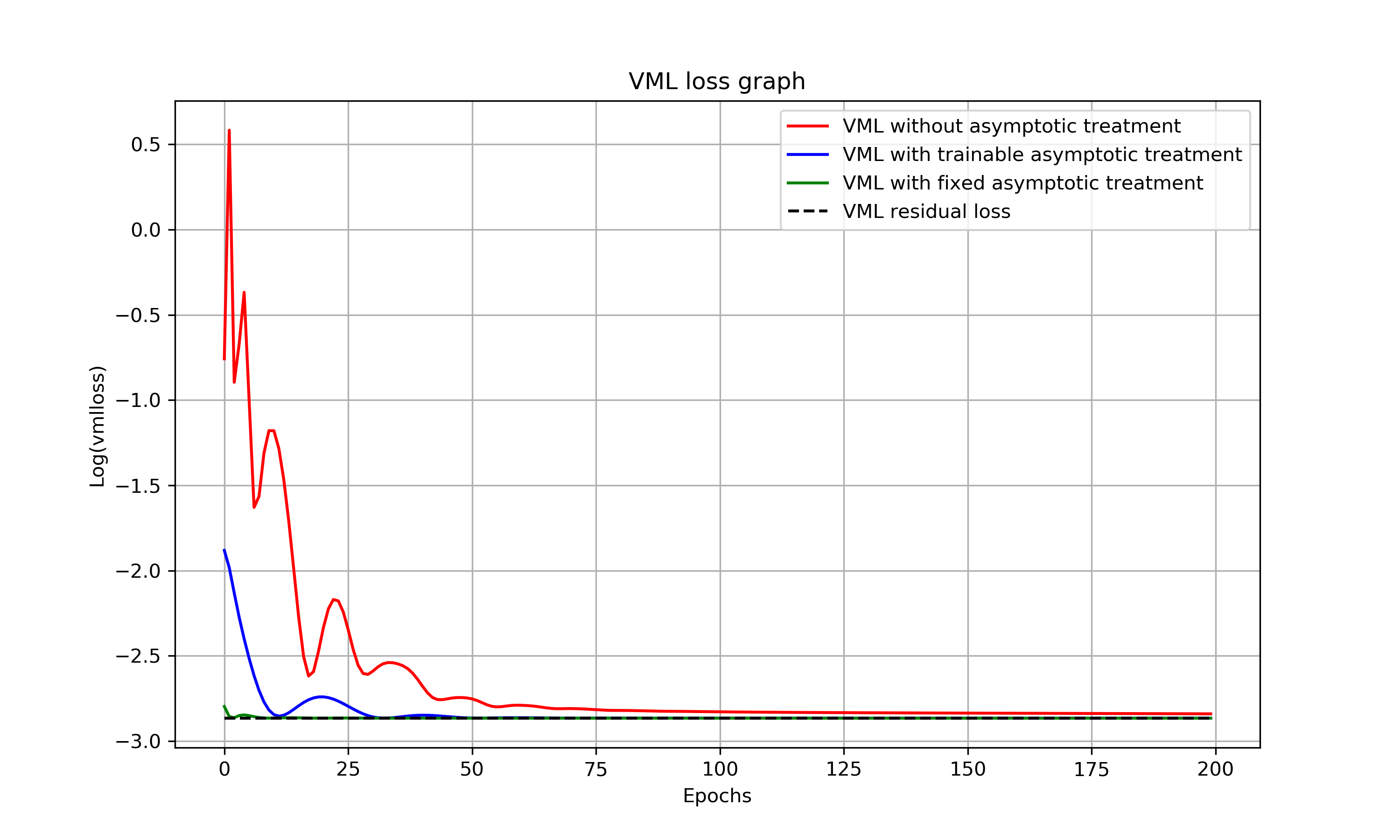} }}%
    \caption{Black-Scholes model regression: Difference and loss graphs for VML without and with asymptotic treatment}%
    \label{fig:lossgraphs_regression_bs_vml}%
\end{figure}

\begin{figure}%
    \centering
    \subfloat[\centering Trained DNN along with true value]{{\includegraphics[width=7cm]{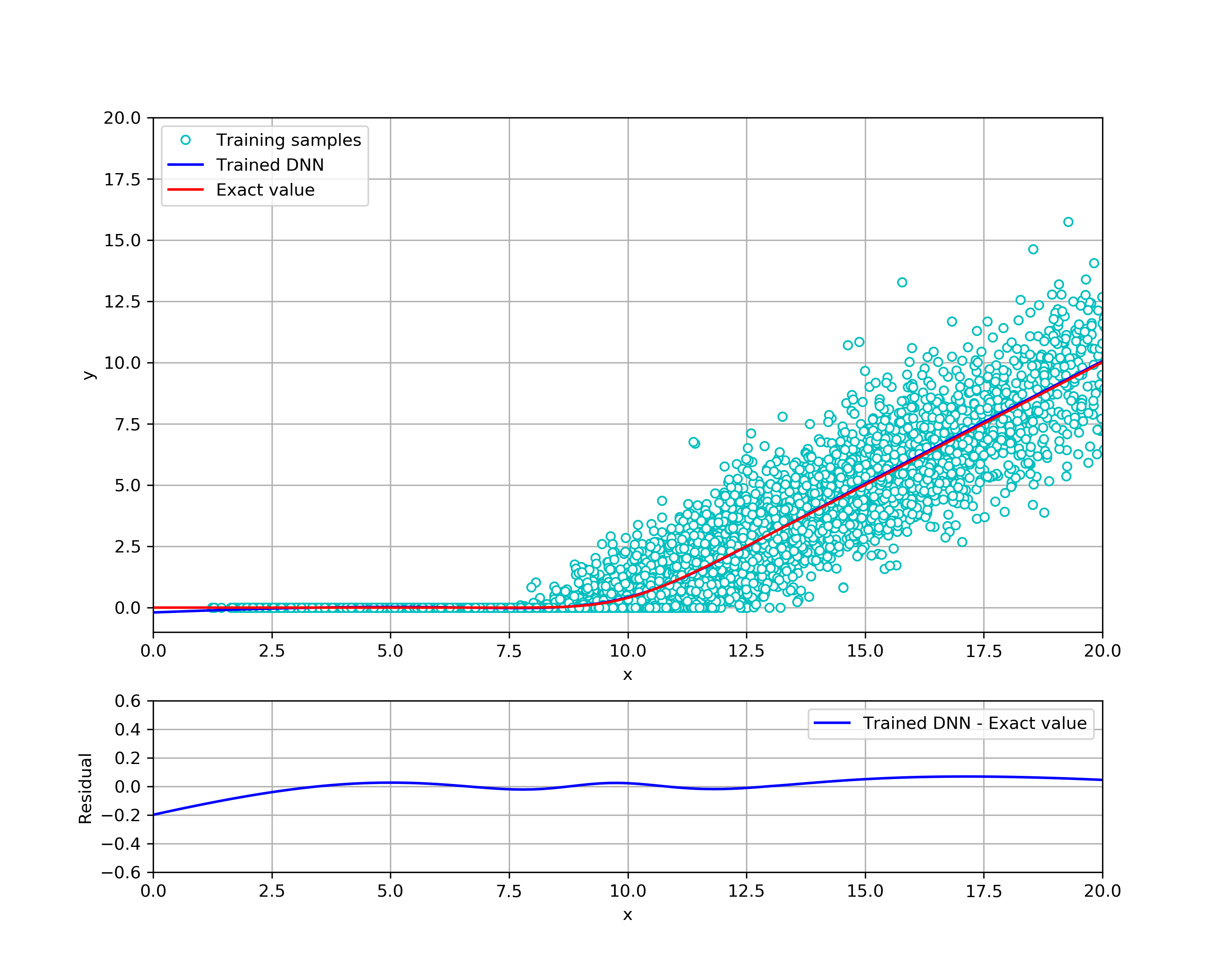} }}%
    \qquad
    \subfloat[\centering Gradient of trained DNN along with true derivative]{{\includegraphics[width=7cm]{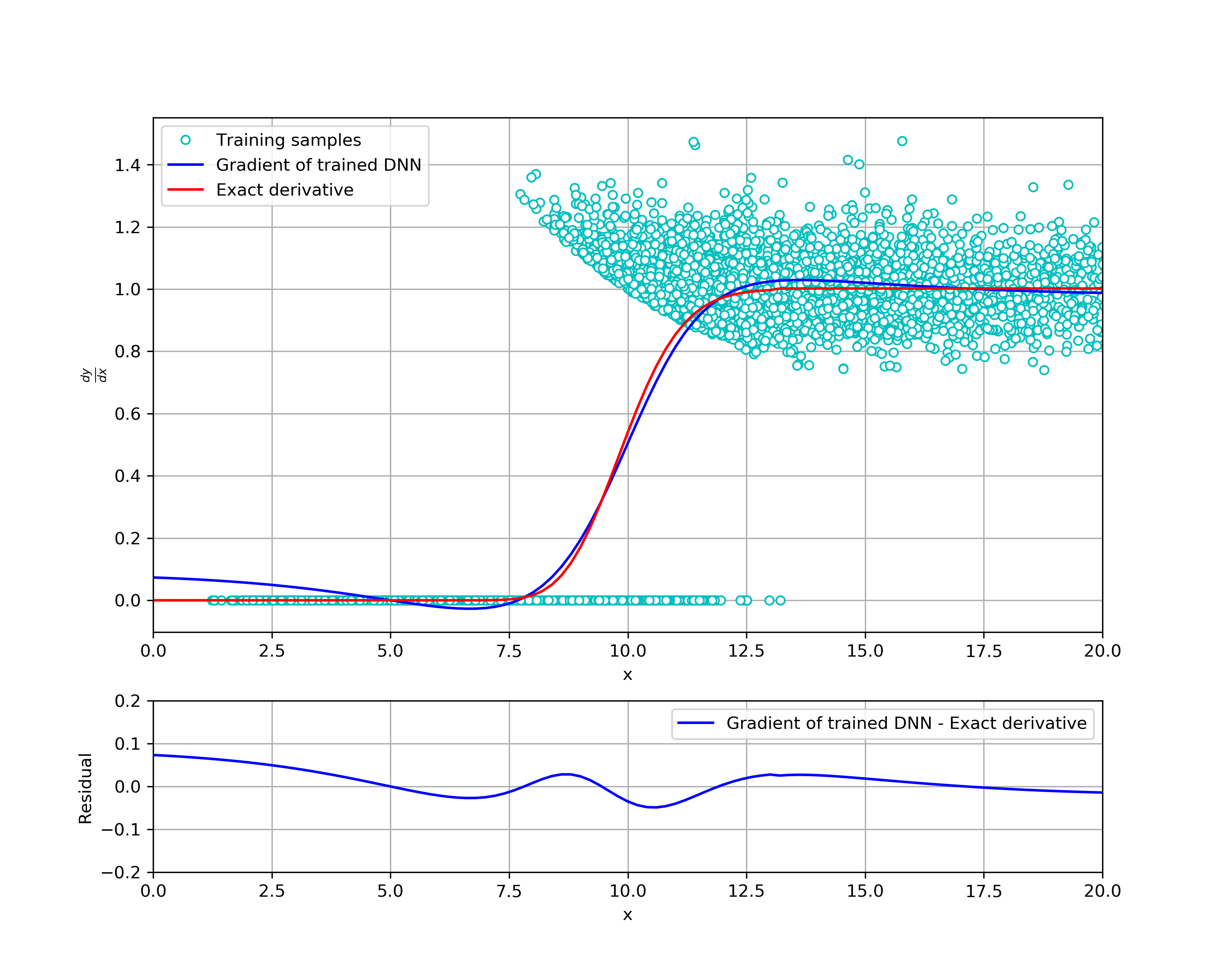} }}%
    \caption{Black-Scholes model regression:  Differential Machine Learning without asymptotic treatment where the scatter points represent the samples used in training.}%
    \label{fig:results_regression_bs_dml}%
\end{figure}

\begin{figure}%
    \centering
    \subfloat[\centering Trained DNN along with true value]{{\includegraphics[width=7cm]{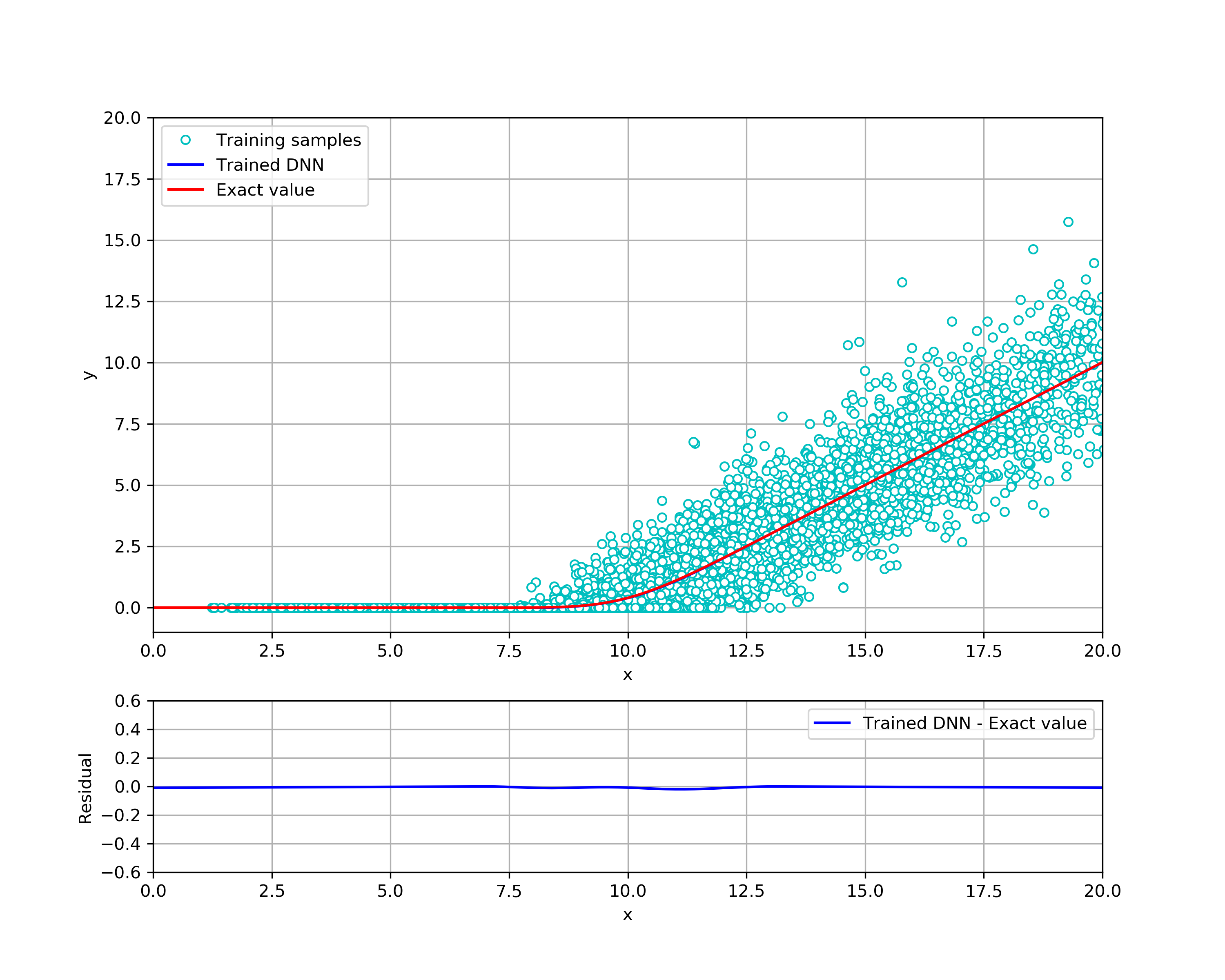} }}%
    \qquad
    \subfloat[\centering Gradient of trained DNN along with true derivative]{{\includegraphics[width=7cm]{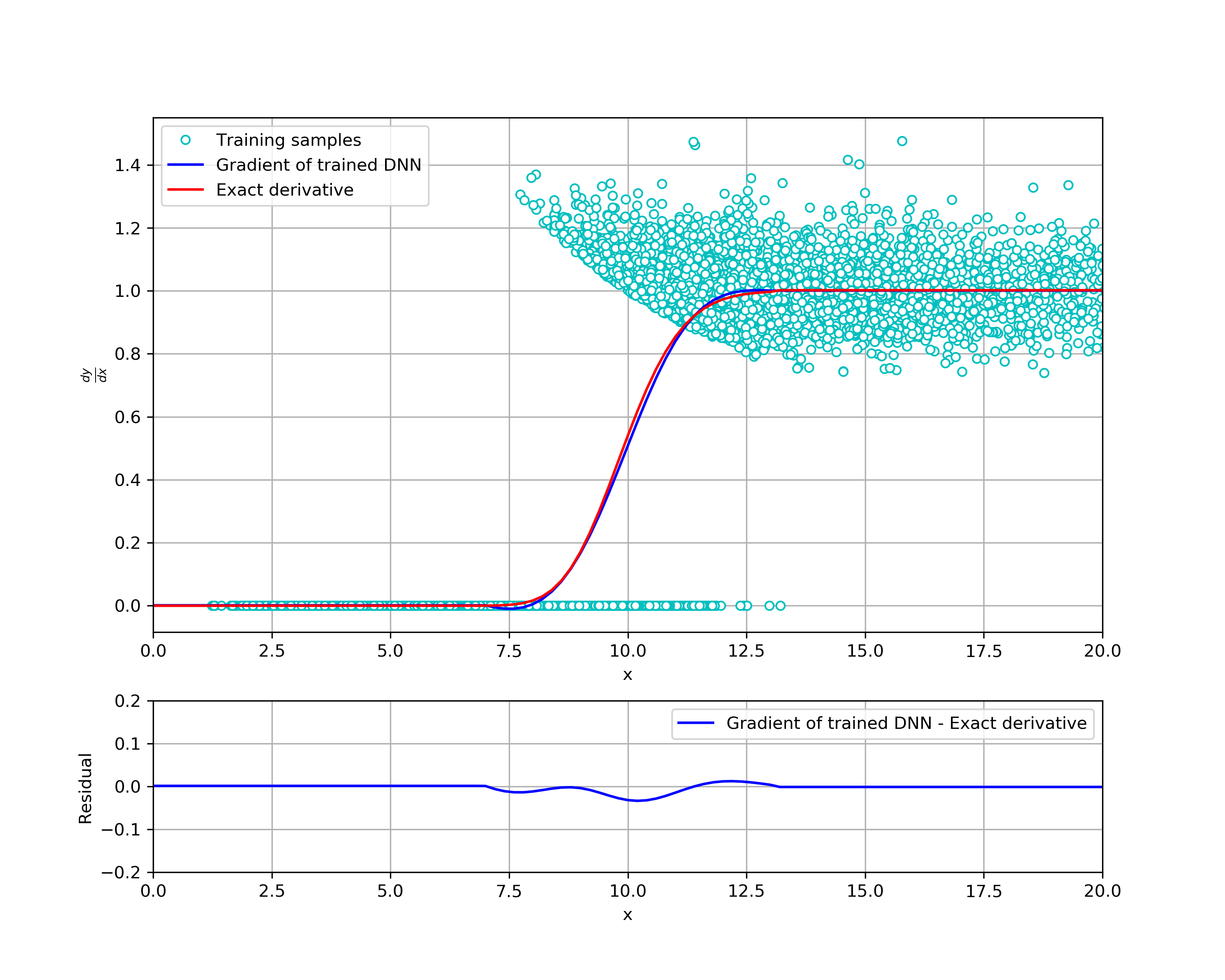} }}%
    \caption{Black-Scholes model regression: Differential Machine Learning with asymptotic treatment (trainable parameters) where the scatter points represent the samples used in training.}%
    \label{fig:results_regression_bs_dml_trainasymp}%
\end{figure}

\begin{figure}%
    \centering
    \subfloat[\centering Trained DNN along with true value]{{\includegraphics[width=7cm]{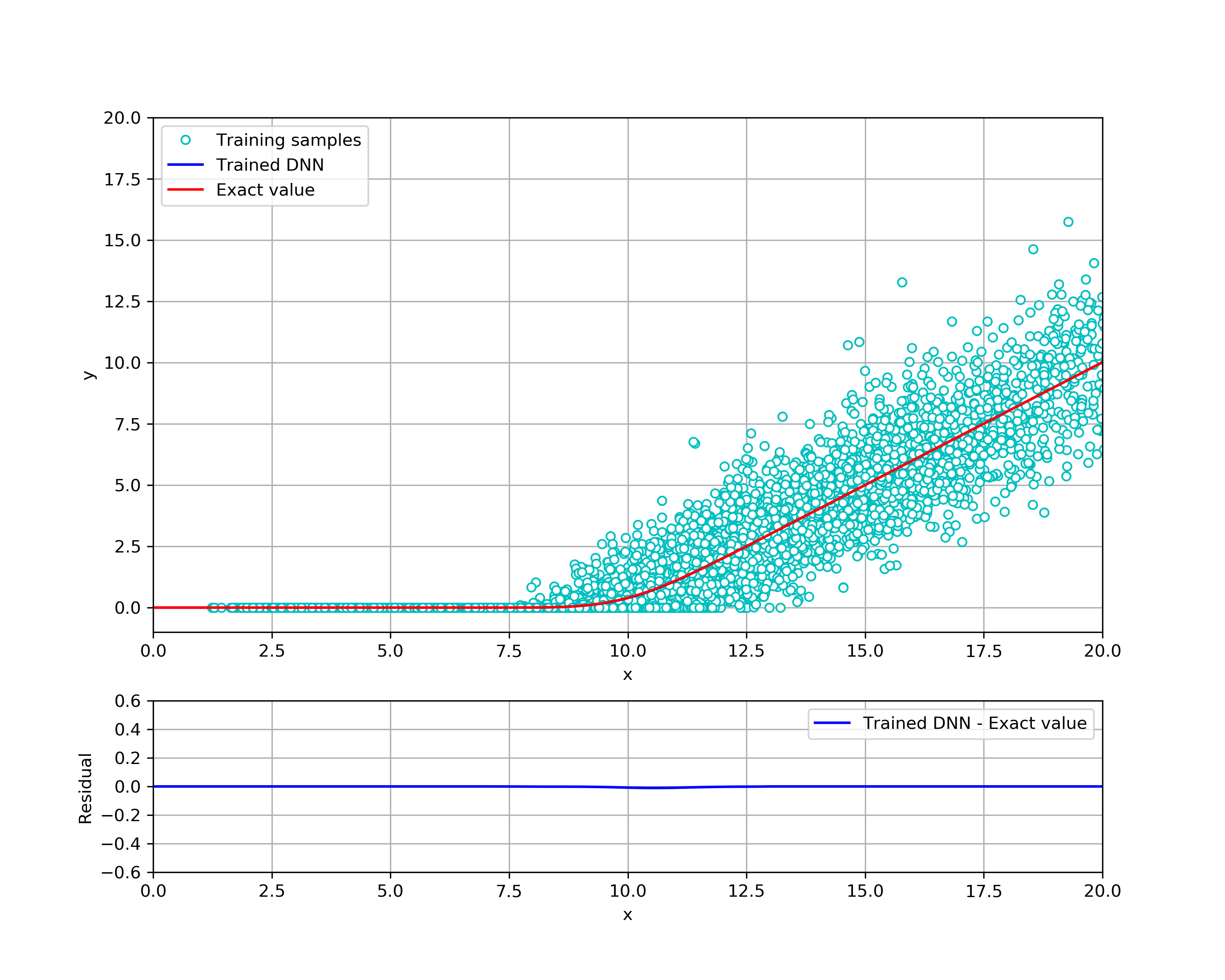} }}%
    \qquad
    \subfloat[\centering Gradient of trained DNN along with true derivative]{{\includegraphics[width=7cm]{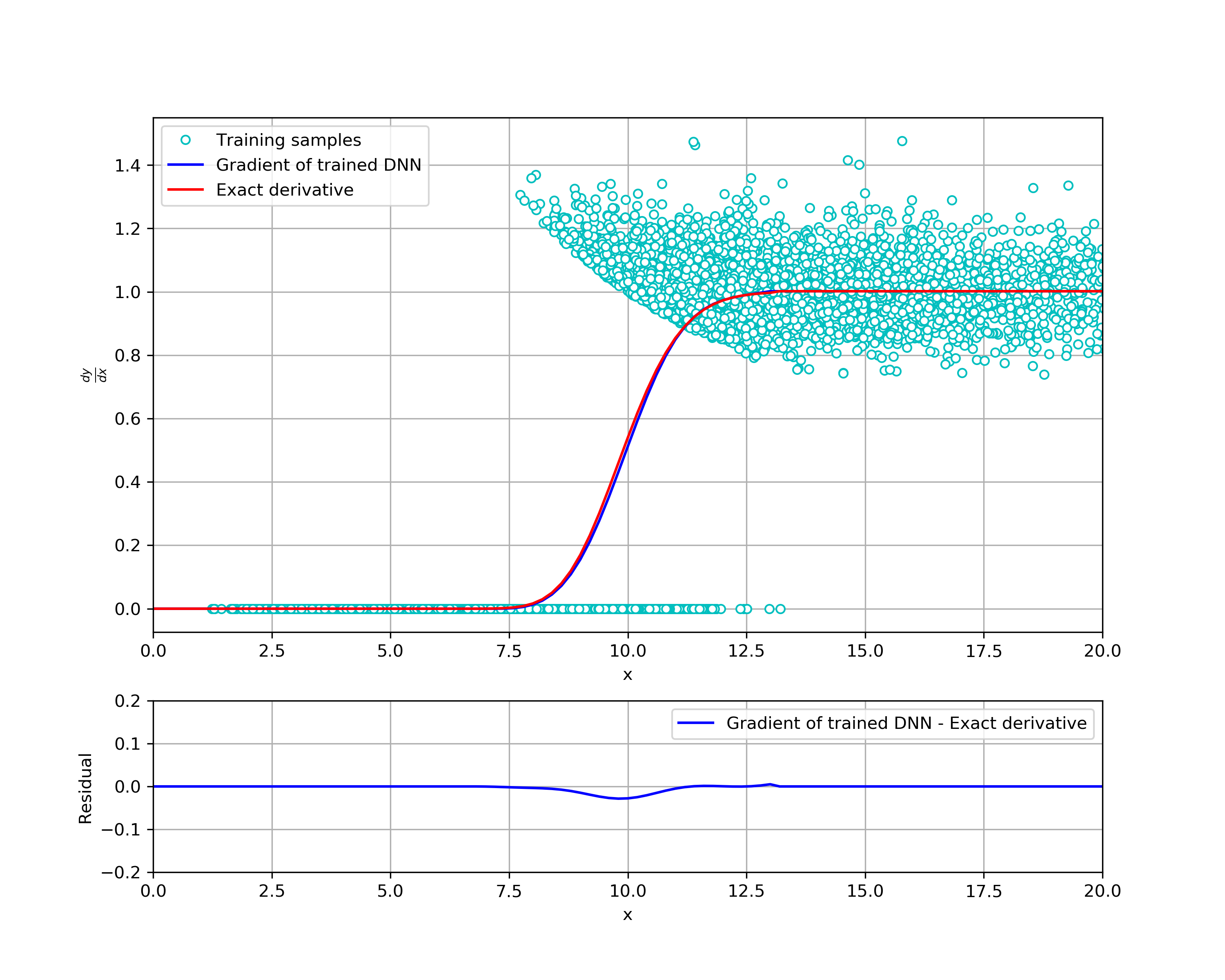} }}%
    \caption{Black-Scholes model regression:  Differential Machine Learning with asymptotic treatment (fixed parameters) where the scatter points represent the samples used in training.}%
    \label{fig:results_regression_bs_dml_fixedasymp}%
\end{figure}

\begin{figure}%
    \centering
    \subfloat[\centering  Difference to true value]{{\includegraphics[width=7cm]{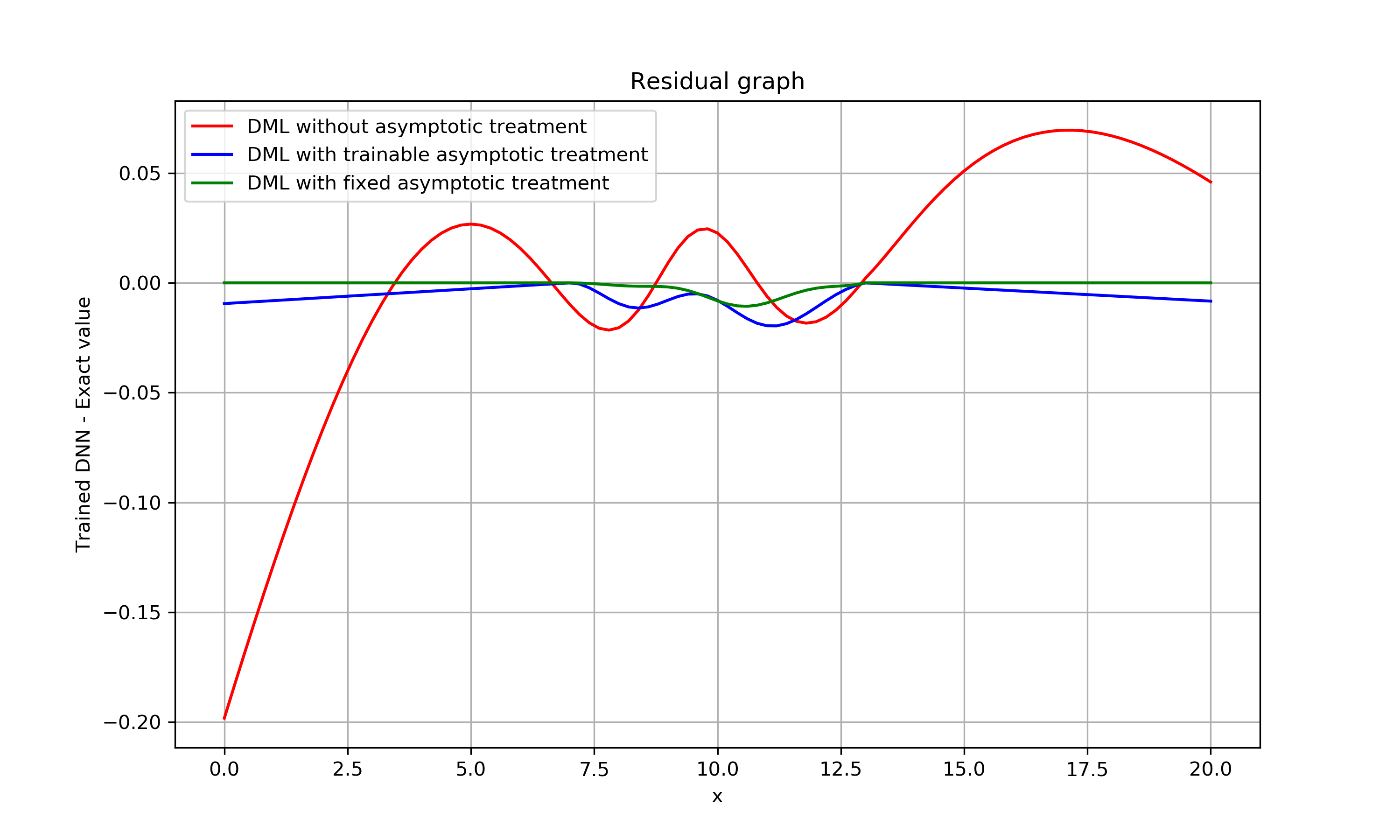} }}%
    \qquad
    \subfloat[\centering  Difference to true derivative]{{\includegraphics[width=7cm]{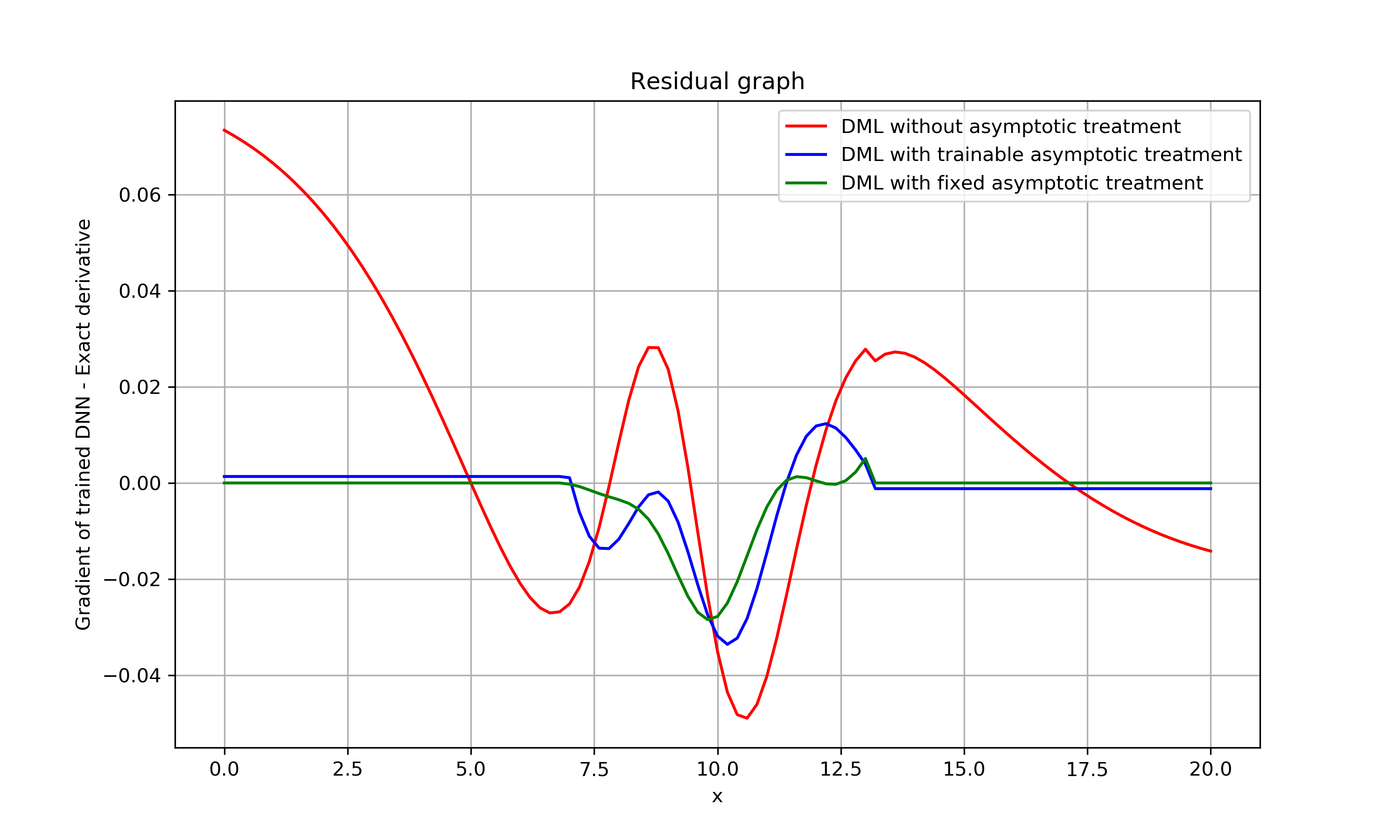} }}%
    \qquad
     \subfloat[\centering Training loss graph (DML loss)]{{\includegraphics[width=7cm]{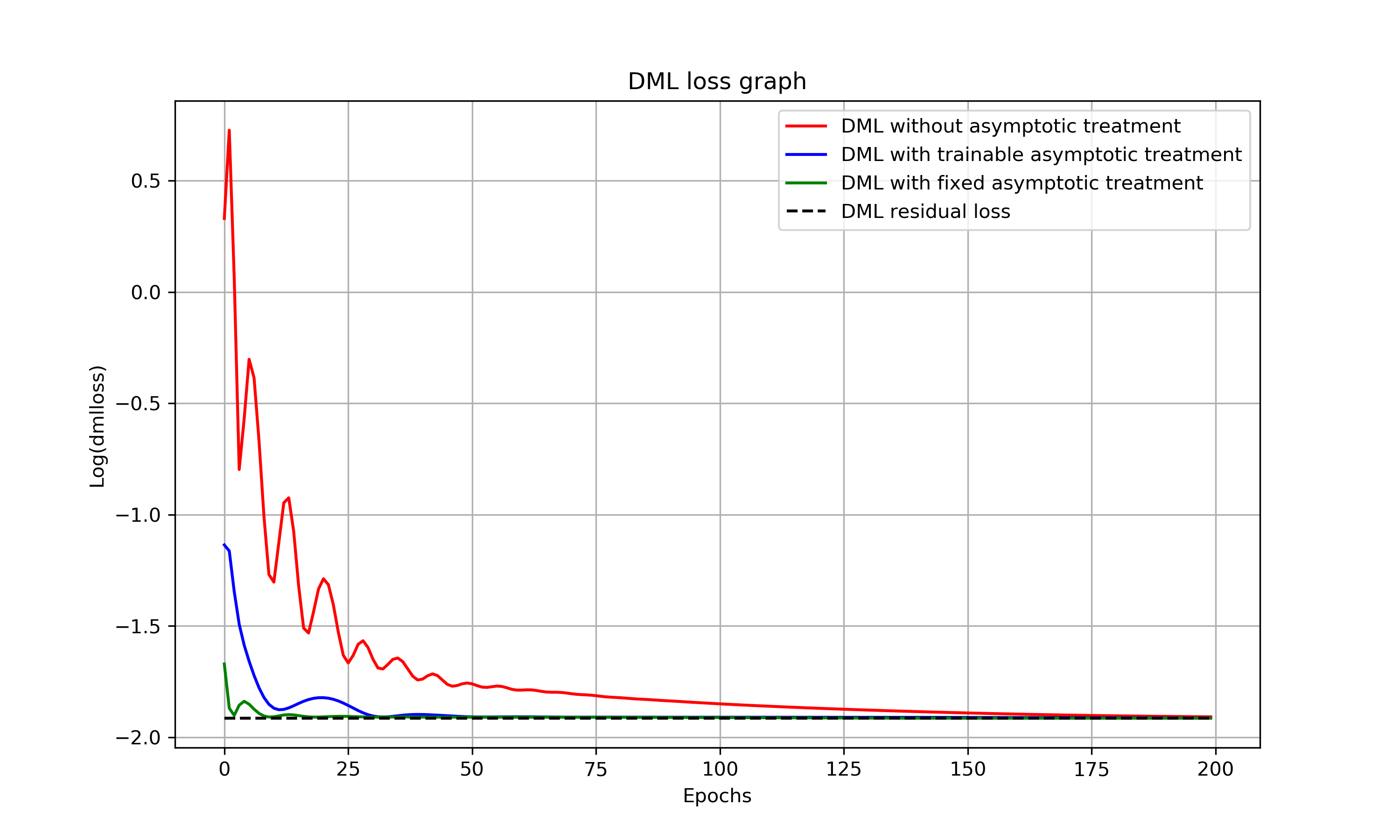} }}%
    \caption{Black-Scholes model regression: Difference and loss graphs for DML without and with asymptotic treatment}%
    \label{fig:lossgraphs_regression_bs}%
\end{figure}

\begin{figure}%
    \centering
    \subfloat[\centering Difference to true value]{{\includegraphics[width=7cm]{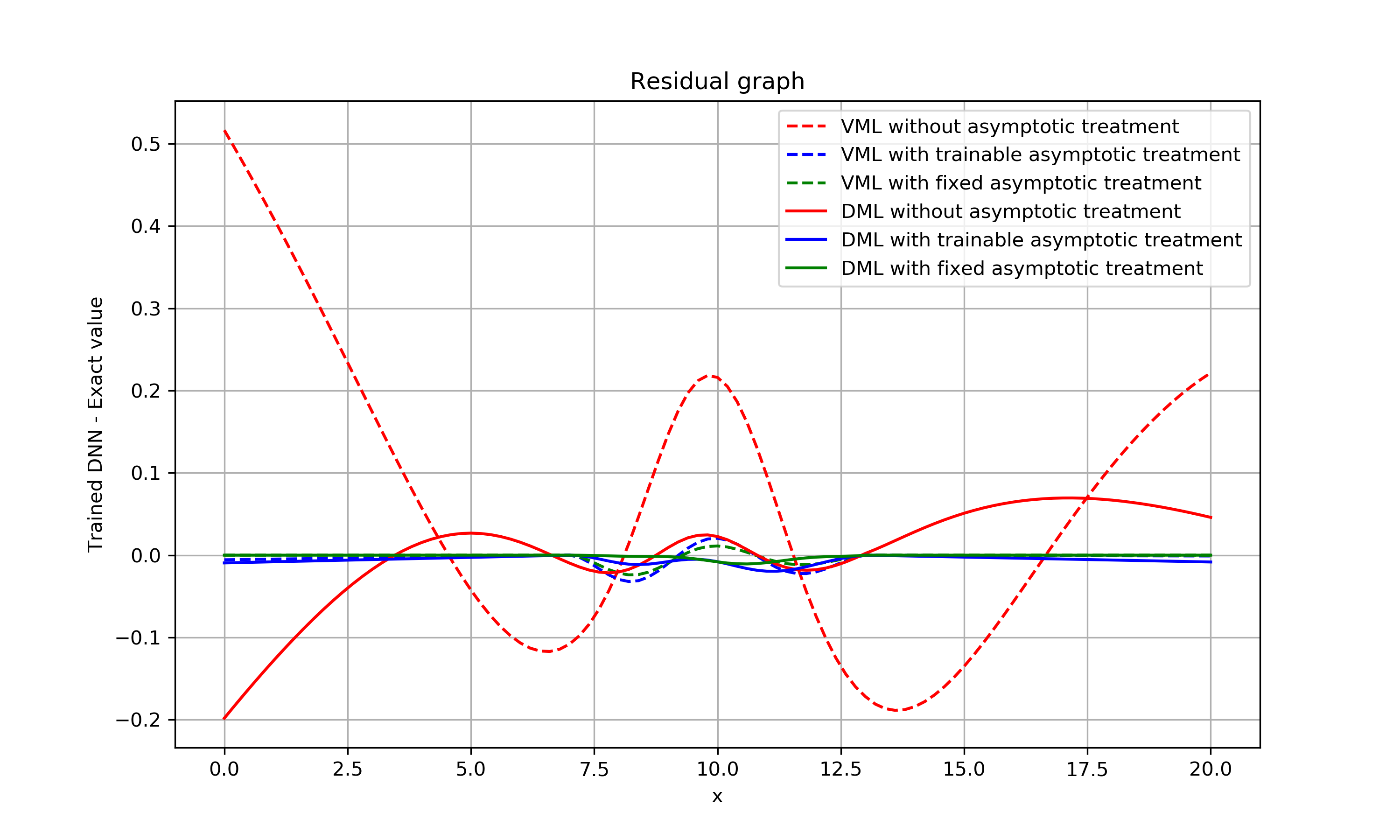} }}%
    \qquad
    \subfloat[\centering Difference to true derivative]{{\includegraphics[width=7cm]{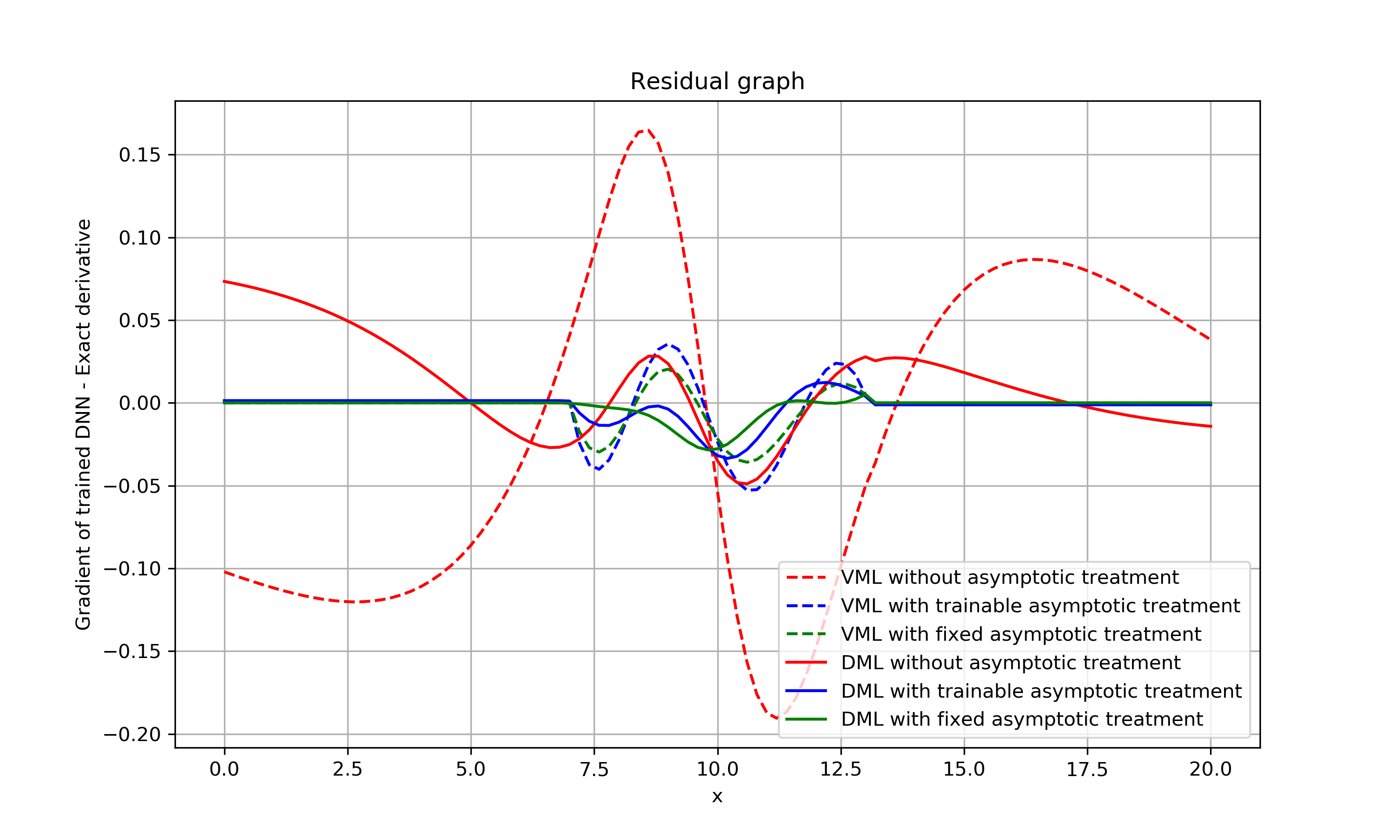} }}%
    \qquad
     \subfloat[\centering Training loss graph (VML loss)]{{\includegraphics[width=7cm]{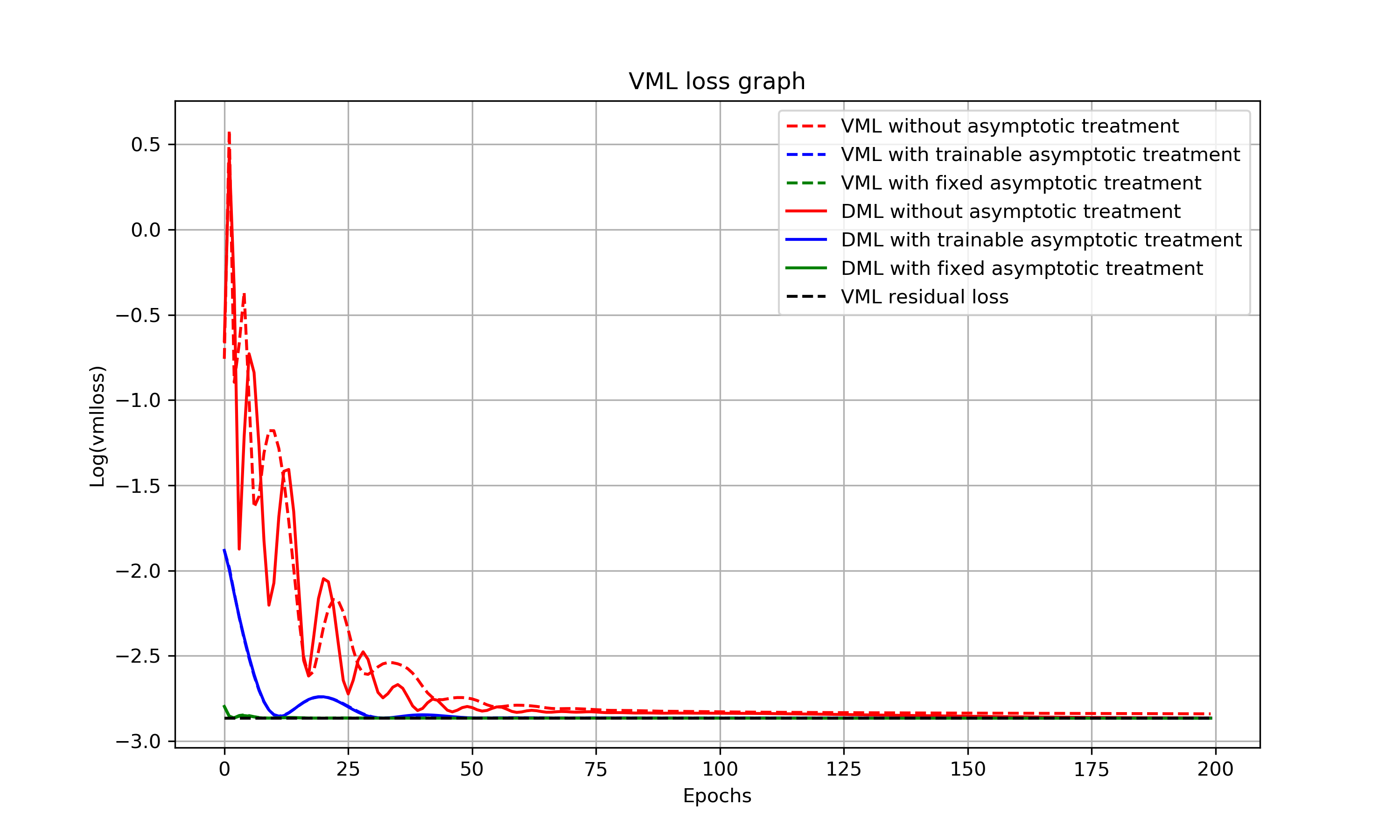} }}%
    \qquad
     \subfloat[\centering Training loss graph (DML loss)]{{\includegraphics[width=7cm]{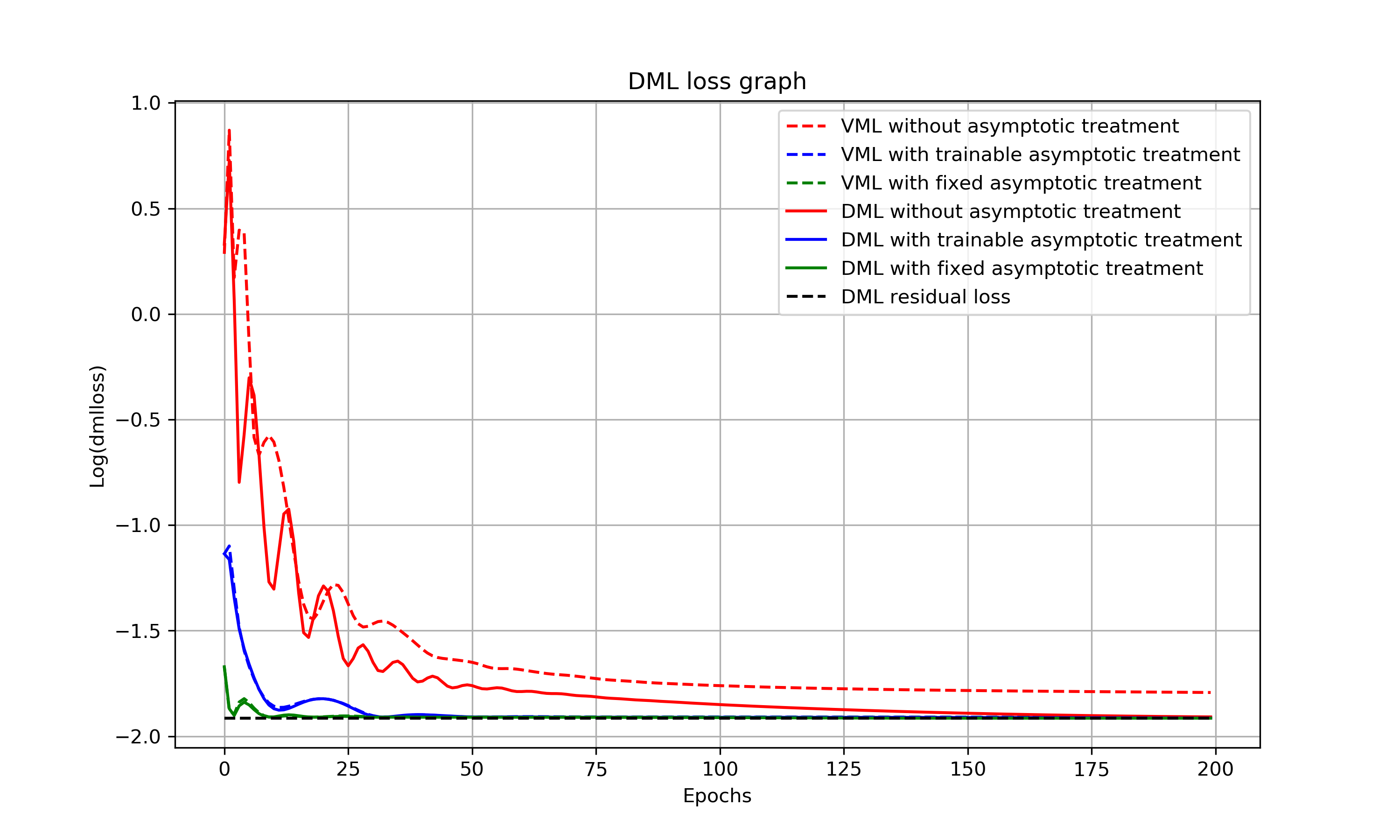} }}%

    \caption{Black-Scholes model regression: Difference and loss graphs for VML and DML (without and with asymptotic treatment)}%
    \label{fig:lossgraphs_regression_bs_all}%
\end{figure}

Fig.~\ref{fig:results_regression_bs_vml} shows the results of the trained deep neural network using VML without asymptotic treatment, 
Fig.~\ref{fig:results_regression_bs_vml_trainasymp} shows the results of the trained deep neural network using VML with asymptotic treatment 
(trainable parameters) and Fig.~\ref{fig:results_regression_bs_vml_fixedasymp} shows the results of the trained
 deep neural network using VML with asymptotic treatment (fixed parameters). 
Fig.~\ref{fig:results_regression_bs_dml} shows the results of the trained deep neural network using DML without asymptotic treatment, 
Fig.~\ref{fig:results_regression_bs_dml_trainasymp} shows the results of the trained deep neural network using DML with asymptotic treatment 
(trainable parameters) and Fig.~\ref{fig:results_regression_bs_dml_fixedasymp} shows the results of the trained
 deep neural network using DML with asymptotic treatment (fixed parameters). The scales for the graphs are kept fixed across different methods to facilitate easier comparison. Inspecting the difference and loss graphs, we conclude that 
 the proposed asymptotic
  treatment also provides better performance in regression problems. The associated difference and loss graphs are also 
  shown in Fig.~\ref{fig:lossgraphs_regression_bs_vml},  Fig.~\ref{fig:lossgraphs_regression_bs} and Fig.~\ref{fig:lossgraphs_regression_bs_all}. 

\clearpage

\begin{figure}%
    \centering
    \subfloat[\centering  Difference to true value]{{\includegraphics[width=7cm]{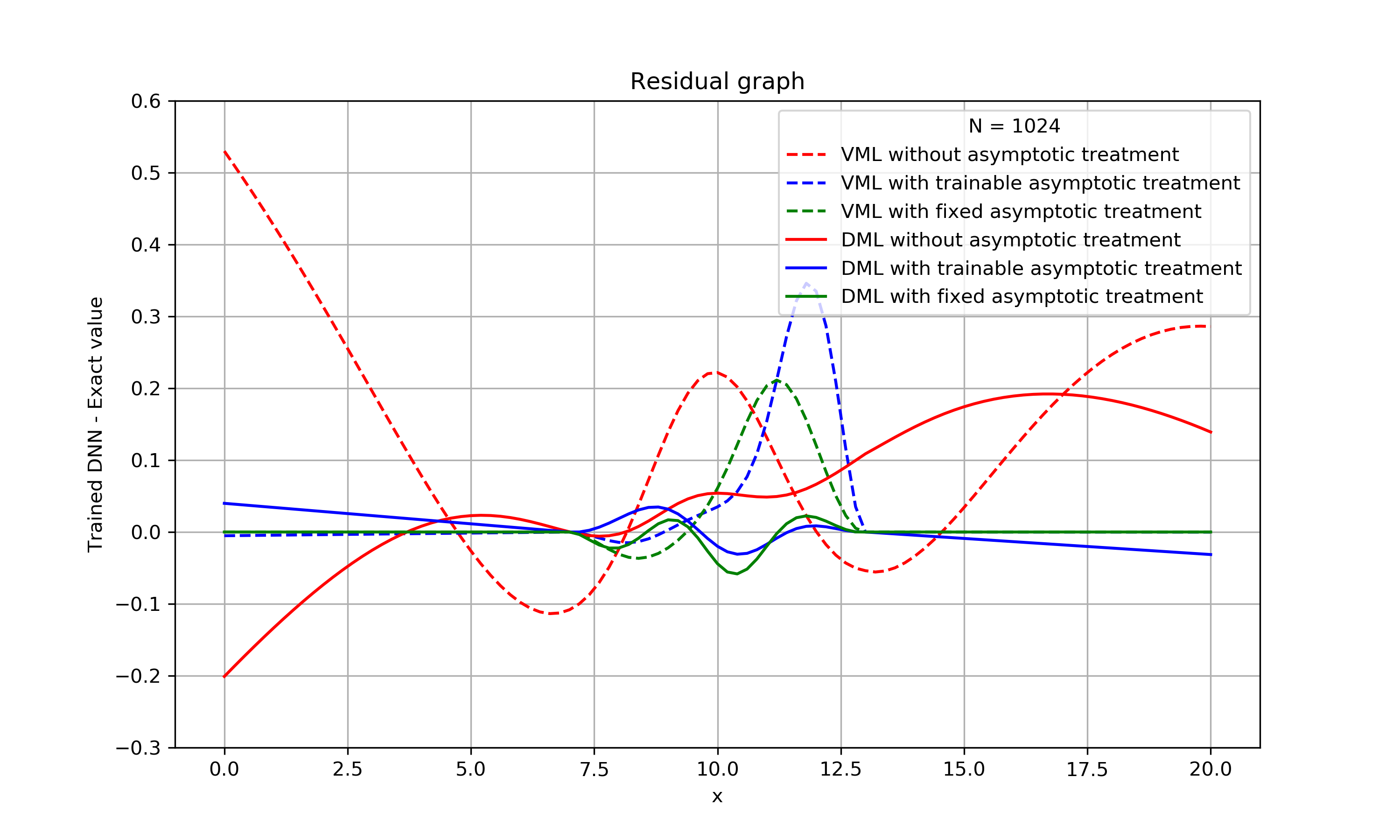} }}%
    \qquad
    \subfloat[\centering  Difference to true derivative]{{\includegraphics[width=7cm]{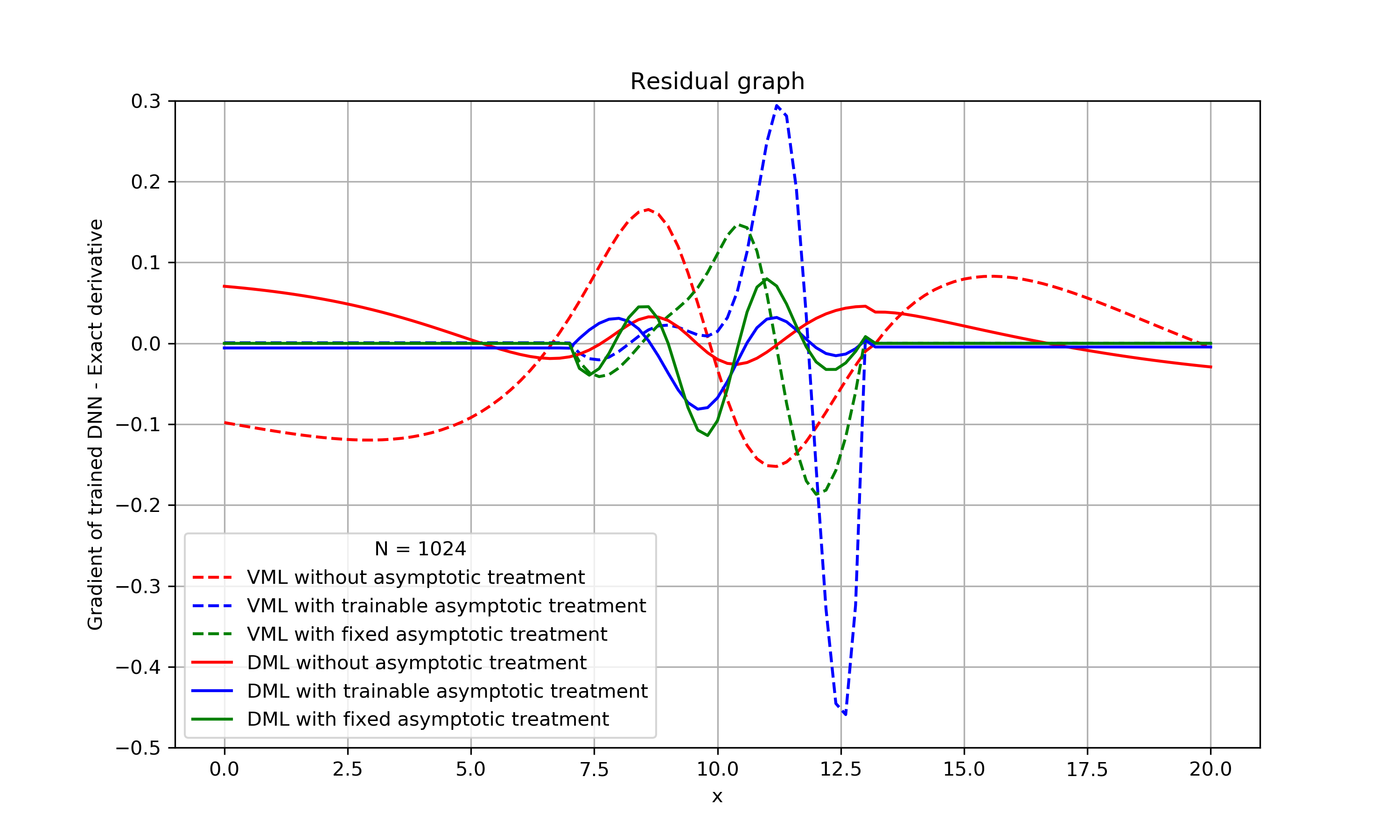} }}%
    \caption{Black-Scholes model regression: Difference graphs for DML and VML (without and with asymptotic treatment) for sample size = $2^{10}$}%
    \label{fig:residual_regression_bs_1024_ex3}%
\end{figure}

\begin{figure}%
    \centering
    \subfloat[\centering  Difference to true value]{{\includegraphics[width=7cm]{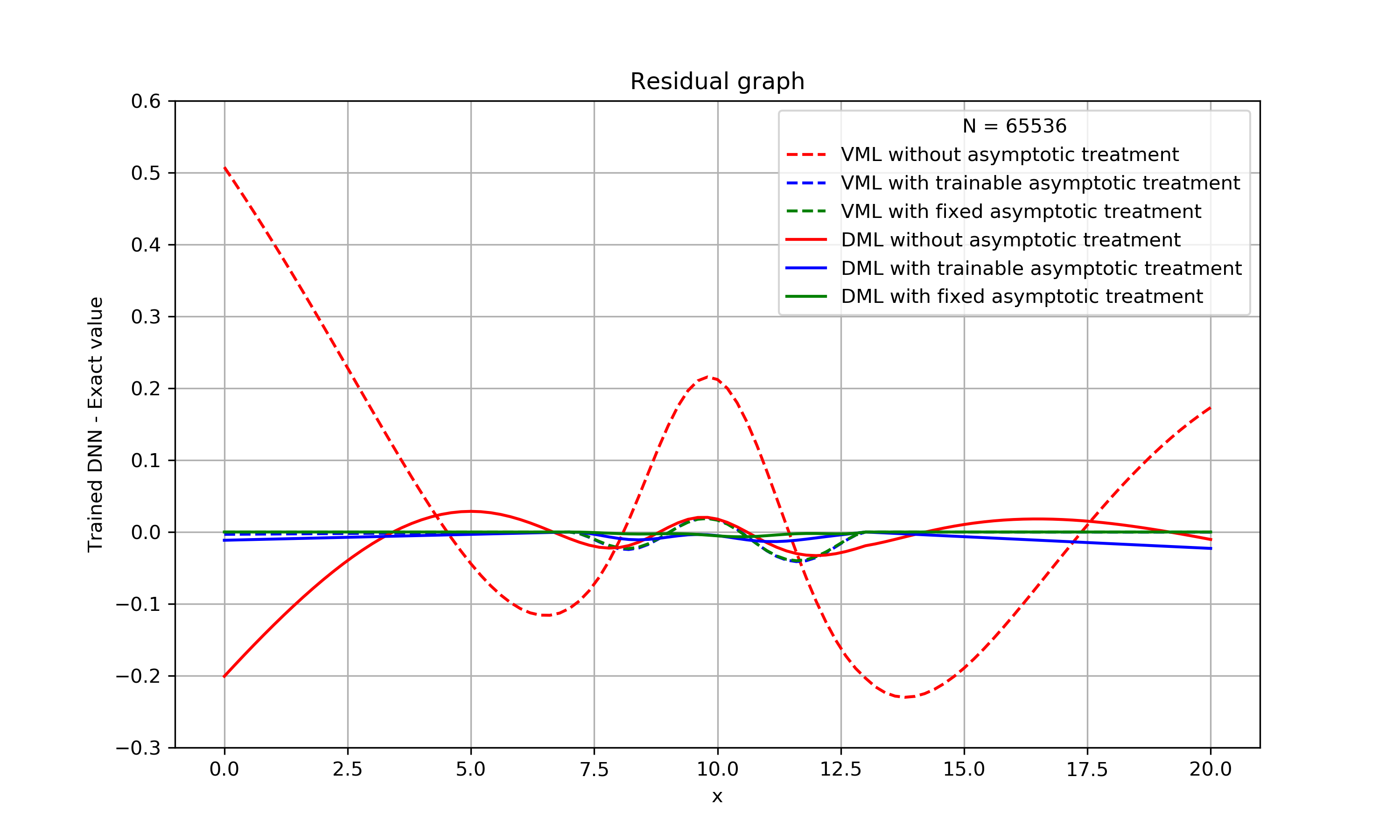} }}%
    \qquad
    \subfloat[\centering  Difference to true derivative]{{\includegraphics[width=7cm]{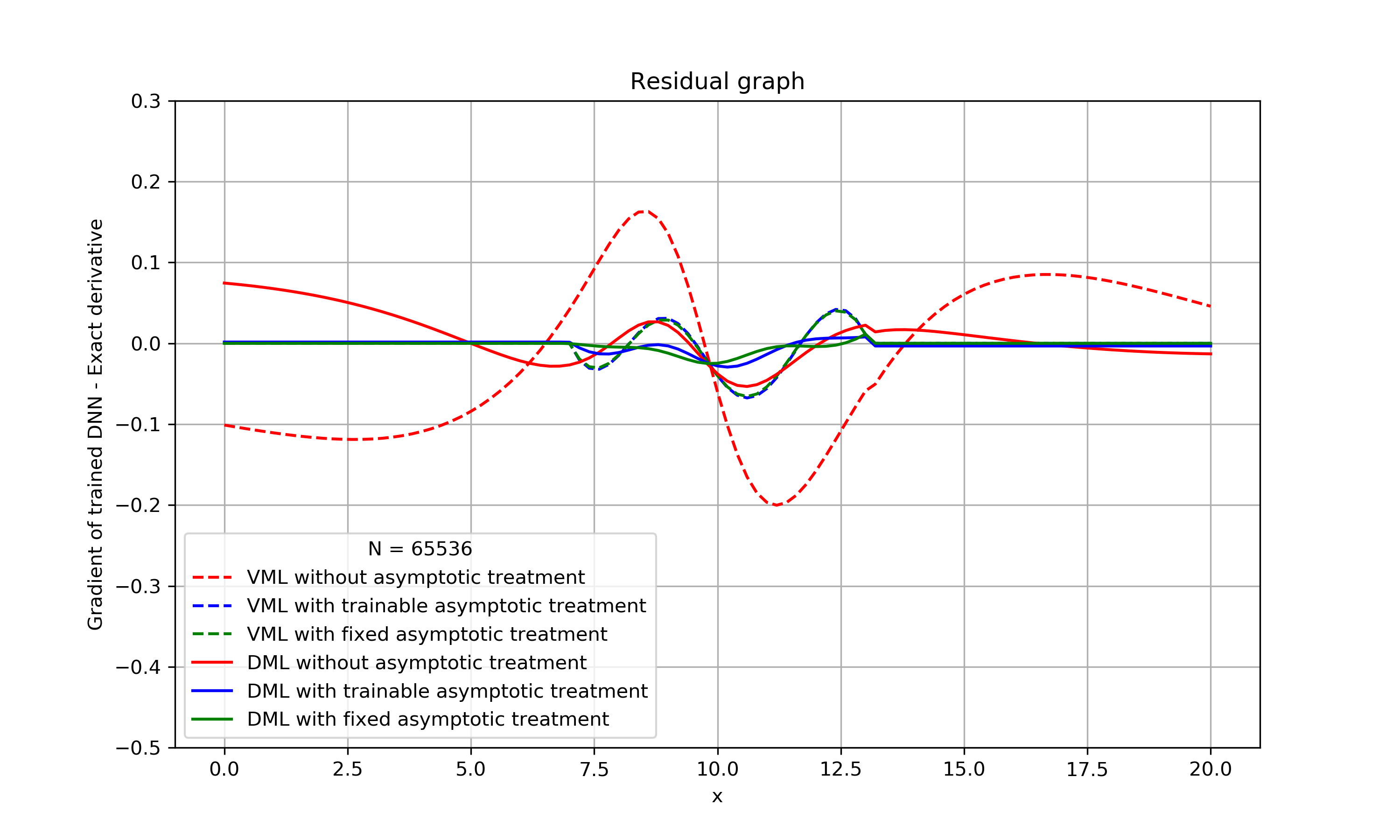} }}%
    \caption{Black-Scholes model regression: Difference graphs for DML and VML (without and with asymptotic treatment) for sample size = $2^{16}$}%
    \label{fig:residual_regression_bs_65536_ex3}%
\end{figure}

We also performed tests on effects of sample size on results using VML and DML (without and with asymptotic treatment). Fig.~\ref{fig:residual_regression_bs_1024_ex3} and 
Fig.~\ref{fig:residual_regression_bs_65536_ex3} shows the results for a sample size of $2^{10}$ and $2^{16}$ respectively. 
It can be observed that the impact of sample size is most prominent when asymptotic treatment is used in both VML and DML with higher sample size resulting in improved approximation. 
The impact is more clearly visible and sizable for the VML approaches (dashed lines) and smaller for the DML approaches (solid lines).

Lastly, we also performed tests with non-zero interest rates $r$ and observed very similar results. Adding $r$ as a drift and in 
the discounting will change the shape and the coefficients slightly but otherwise does not seem to impact the results and we thus 
omit the plots.

\clearpage

\section{Conclusion}
\label{sec:conclusion}

Often, true or approximate asymptotic behavior or forms are known for pricing functions in quantitative finance in asymptotic regimes
and regions, given by known functional forms either with given parameters or with free parameters. Approximating such 
pricing functions with standard DNN through deep learning approaches will often not lead to good approximations of asymptotic behavior. 

While some approaches in the literature enforce asymptotic forms \cite{antonov2020neural,antonov2020deeprisk}, they represent functions 
in the non-asymptotic domain by a linear combination of Gaussian kernels (on top of the extended asymptotic form),
requiring computation of many exponentials during training and evaluation. In this paper, we proposed and tested a simpler form where 
instead of Gaussian kernels, the product of a simple polynomial (which assures proper matching of asymptotic behavior) and an 
unconstrained DNN is used. This approach leads to a simple implementation and good results in our tests, 
on two example function approximation and one example regression problem. It thus seems to be a promising choice 
and candidate for implementation in cases where asymptotic behavior is known. 

In future work, we plan to test further cases with known asymptotic behaviors, including the also encountered exponential forms. 
We intend to work on extensions to multiple dimensions. We plan to apply such asymptotic treatment in the regression setting 
in our PDML approach for parametric pricers and calibration \cite{PDML,caplet1fcheyette2024}, and to the applications of the DML approach for regression 
to XVA \cite{huge2020differential,huge2020differentialrisk}. We will also pursue the application of such constructions in further areas
beyond the function approximation and regression setting, such as in stochastic control approaches \cite{fathi2023comparisonreinforcementlearningdeep,hientzsch2023reinforcementlearningdeepstochastic}, 
reinforcement learning \cite{fathi2023comparisonreinforcementlearningdeep,hientzsch2023reinforcementlearningdeepstochastic}, 
and deepBSDE approaches \cite{hientzsch2019intro,hientzsch2021intro,yuhientzsch2019backward,yu2023backward}.

\section{Acknowledgements and Disclaimer}

The authors thank Vijayan Nair for his support and suggestions regarding research, this work, and its presentation. 
We thank Orcan Ogetbil for his close reading and feedback improving the presentation in this paper.

Any opinions, findings and conclusions or recommendations expressed in this material are those of the authors and do not necessarily reflect the views of 
Wells Fargo Bank, N.A., its parent company, affiliates and subsidiaries.

\bibliographystyle{alpha}

\bibliography{references}

\end{document}